\documentclass[usegraphicx,usenatbib]{mn2e}

\usepackage{times}
\usepackage{graphicx,epsf}

\newcommand{\apj}{ApJ}           
\newcommand{\mnras}{MNRAS}       
\newcommand{\aap}{A\&A}
\newcommand\aaps{{A\&AS}}

\newcommand{\apjs}{ApJS}
\newcommand{\aj}{AJ}
\newcommand{\pasp}{PASP}
\newcommand{\araa}{ARA\&A}

\title[H$\beta$ in low z quasars]
{Detailed characterization of H$\beta$ emission line profile in low
z SDSS quasars}

\author[Zamfir et al.]{S.\ Zamfir$^1$\thanks{E-mail:
zamfi001@crimson.ua.edu}, J.~W.\ Sulentic$^{1,2}$, P.\ Marziani$^3$, D.\ Dultzin$^4$\\
$^1$Department of Physics and Astronomy, University of
Alabama, Box 870324, Tuscaloosa, AL 35487, USA\\
$^2$Instituto de Astrof\'{\i}sica de Andaluc\'{\i}a, CSIC, Apdo. 3004, 18080, Granada, Spain\\
$^3$INAF-Osservatorio Astronomico di Padova, Vicolo
dell'Osservatorio 5, I-35122 Padova, Italy\\
$^4$ Instituto de Astronom\'{\i}a, Universidad Nacional Aut\'{o}noma
de M\'{e}xico, Apdo. Postal 70-264, 04510 M\'{e}xico, D. F.,
M\'{e}xico}

\pagerange{\pageref{firstpage}--\pageref{lastpage}} \pubyear{2009}

\begin{document}
\label{firstpage} \maketitle

\begin{abstract}

We explore the properties of the H$\beta$ emission line profile in a
large, homogeneous and bright sample of N$\sim$ 470 low redshift
quasars extracted from Sloan Digital Sky Survey (DR5). We approach
the investigation from two complementary directions:
composite/median spectra and a set of line diagnostic measures
(asymmetry index, centroid shift and kurtosis) in individual
quasars. The project is developed and presented in the framework of
the so called 4D Eigenvector 1 Parameter Space, with a focus on its
optical dimensions, FWHM(H$\beta$) and the relative strength of
optical FeII (R$_{FeII}$$\equiv$W(FeII4434-4684\AA)/W(H$\beta$)). We
reenforce the conclusion that not all quasars are alike and
spectroscopically they do not distribute randomly about an average
typical optical spectrum. Our results give further support to the
concept of two populations A and B (narrower and broader than 4000
km s$^{-1}$ FWHM(H$\beta$), respectively) that emerged in the
context of 4DE1 space. The broad H$\beta$ profiles in composite
spectra of Population A sources are best described by a Lorentzian
and in Population B by a double Gaussian model. Moreover, high and
low accretion sources (an alternative view of the Population A/B
concept) not only show significant differences in terms of Black
Hole (BH) mass and Eddington ratio L$_{bol}$/L$_{Edd}$, but they also
show distinct properties in terms of line asymmetry, shift and
shapes. We finally suggest that a potential refinement of the 4DE1
space can be provided by separating two populations of quasars at
R$_{FeII}$$\sim$0.50 rather than at FWHM(H$\beta$)=4000 km s$^{-1}$.
Concomitantly, the asymmetry and centroid shift profile measures at
1/4 fractional intensity can be reasonable surrogates for the
FWHM(H$\beta$) dimension of the current 4DE1.

\end{abstract}


\section{Introduction}

\subsection{Current View on the Structure of BLR of AGN}
\subsubsection{Observational considerations}

A defining characteristic of Type 1 Active Galactic Nuclei (AGN) is
the presence of broad emission lines in their optical and UV
spectra. We cannot resolve the AGN central regions at other than
radio (Very Long Baseline Interferometry - milliarcsecond
resolution) or, more recently, infrared (e.g., \citealt{Raban09})
wavelengths. This leaves the structure, geometry, kinematics and
nebular physics of the Broad Line Region (BLR) beyond our direct
reach. It is unlikely that we will spatially resolve the central
regions of more than a few low z AGN in the foreseeable future. We
are left with two practical methods to probe the BLR structure and
kinematics: a) reverberation studies
\citep[e.g.,][]{Koratkar91,Peterson93,Peterson99,Wandel99,Kaspi00,Peterson04,Kaspi07,Bentz09}
and b) single epoch spectroscopy for large numbers of sources. The
former requires dedicated spectroscopic monitoring to achieve
temporal resolution of a relatively small number of sources while
the latter provides kinematic resolution for multitudes of sources.
The time lag between line and continuum flux variations (in the
former approach) provides an estimate of the BLR size. A few dozen
(mostly low z) sources have been monitored with typical BLR radii in
the range from a few light-days to several light-weeks
\citep[e.g.,][]{Peterson93,Kaspi00,Bentz09}. Spectroscopy of large
samples provides a much-needed empirical clarification since,
without a clear understanding of the phenomenology, one can build
models filled with misconceptions about BLR structure/geometry,
kinematics and nebular physics.

We have been exploring the phenomenology of quasars in the context
of a 4D Eigenvector 1 parameter space (4DE1;
\citealt{Sulentic00a,Sulentic00b,Marziani01,Marziani03a,Marziani03b,Sulentic07}),
which serves as a spectroscopic unifier/discriminator for all type 1
AGN. We take it as self-evident that such a parameter space is
needed: 1) to emphasize source differences, 2) to help remove the
degeneracy between the spectroscopic signatures of orientation and
physics and 3) to provide a context within which to interpret
results of multiwavelength studies of individual sources. We adopted
four principal 4DE1 parameters because they were relatively easy to
measure in large numbers of sources and because they showed large
intrinsic dispersion and sometimes major differences between
subtypes. The principal parameters involve measures of: (i) full
width at half-maximum of broad H$\beta$ (FWHM(H$\beta$)), (ii)
equivalent width ratio of optical FeII ($\lambda$4570\AA\ blend) and
broad H$\beta$, R$_{FeII}$ = W(FeII$\lambda$4570\AA)/W(H$\beta$),
(iii) the soft X-ray photon index ($\Gamma$$_{soft}$) and (iv)
CIV$\lambda$1549\AA\ broad-line profile velocity displacement at
half-maximum, c(1/2). Various lines of evidence \citep{Sulentic07}
suggested a change of source properties near FWHM(H$\beta$) = 4000
km s$^{-1}$, which motivated the introduction of the Population A-B
distinction in order to emphasize the differences. This is similar
to the radio-quiet (RQ) vs. radio loud (RL) distinction, but appears
to be more effective (\citealt{Sulentic07}; see also below).

Previous studies \citep{Sulentic03,Zamfir08} indicate that RL
sources show a much more restricted 4DE1 domain space occupation
(FWHM(H$\beta$) versus R$_{FeII}$) than the RQ majority ($\sim$78\%
of all RL quasars belong to Population B domain and $\sim$62\% of RQ
belong to Population A domain). This was in part the motivation for
our Population A-B distinction. A majority ($\approx$ 55-60\%) of
quasars show broad Balmer lines with FWHM(H$\beta$) $<$ 4000 km
s$^{-1}$ \citep{Sulentic00a,Sulentic00b,Zamfir08} and are labeled
Population A. Population A shows: 1) a scarcity of RL sources (3-4\%
at most; \citealt{Zamfir08}), 2) strong/moderate FeII emission, 3) a
soft X-ray excess (perhaps the consequence of a high accretion rate;
e.g., \citealt{Wang03}), 4) high-ionization broad lines (HIL; e.g.,
CIV $\lambda$1549\AA) showing blueshift/asymmetry and 5) low
ionization broad line profiles (LIL) best described by Lorentz fits
\citep{Sulentic02,Bachev04,Sulentic07,Marziani09}.

Population B sources: 1) include the large majority of the RL
sources: $\sim$91\% of all FRII and $\sim$62\% of all core-dominated
RL quasars or, alternatively, $\sim$17\% of all Population B sources
are RL quasars (and this is a conservative\footnote{Assuming that
the true parent population of RL quasars are FRII sources there
should exist for every core-dominated RL source of a certain radio
luminosity an even weaker radio power FRII, if the unification
scenario is valid.} estimate; see \citealt{Zamfir08}), 2) show
weak/moderate FeII emission, 3) do not show a soft X-ray excess, 4)
do not show HIL blueshift/asymmetry and 5) show LIL Balmer lines
best fit with double Gaussian models
\citep{Sulentic02,Bachev04,Sulentic07,Marziani09}. Fundamental
differences between the BLR in Population A and B sources (and
between RL and RQ quasars) appear to be lower electron density
\citep[e.g.,][]{Collin-Souffrin87,Collin-Souffrin88a} in Population
B along with a less flattened cloud distribution
\citep[e.g.,][]{Zheng90,Sulentic00a,Collin06}. Some of the
Population B sources with very broad LIL profiles show double-peaked
structure \citep[e.g.,][]{Eracleous94}, but they are rare.
Nonetheless, there is an underlying idea that a double-peaked
signature is hidden in most profiles. \citet{Murray97} has
emphasized that a disk+wind model favors the regions of low
projected velocity in the disk, which leads to a predominance of
single-peaked profiles. The existence of double-peaked profiles is
not a necessary condition to prove the reality of a disk BLR
geometry \citep[e.g.,][and references therein]{Popovic04}.

Valuable insights are provided by studies of emission lines of
different ionization levels: HIL, e.g., CIV $\lambda$1549\AA\ and
LIL, e.g., Balmer lines, FeII, MgII doublet $\lambda$2800\AA. The
HIL and LIL are probably produced in distinct regions
\citep[e.g.,][]{Collin-Souffrin88b} at least in Population A sources
as inferred from the difference in the time delayed response to
continuum variations \citep[see Section 4.1 of][]{Sulentic00a}.
While CIV is always broader than H$\beta$ (produced at smaller
radii?) in Population A sources, the two lines tend to scale more
closely or, in some cases, H$\beta$ tends to show broader profiles
\citep{Sulentic07} in Population B.

There has been a great deal of disagreement over the meaning of
often poor spectroscopic data for often small samples. Making
progress involves a lot of work obtaining spectra with good S/N and
resolution in order to assemble representative sample. Ideally one
would like be able to observe the same line(s) over a large redshift
range, which raises obvious difficulties that IR spectroscopy has
begun to alleviate \citep[e.g.,][and references
therein]{Sulentic04,Sulentic06,Marziani09}. Or, even better, observe
concomitantly both HIL and LIL for the same source with awareness
that the BLR in high and low luminosity sources may be different as
well \citep{Netzer03,Kaspi07,Netzer07,Marziani09}.

We suggest that the best hope of advancing our understanding of the
AGN phenomenon lies with better empiricism, which is the path to
model improvement. The first step to progress requires a reasonably
large (i.e., representative) quasar sample for which uniform and
good quality spectra exist. We argue that progress cannot come from
studies of indiscriminately averaged quasar spectra, whether their
S/N is high or low. This statement is false only if all quasars are
virtually identical, which we know to not be true.

There is a long series of questions in need of answers. Do RQ and RL
quasars show the same geometry/kinematics? Do optical/UV spectra
offer predictive power on the likelihood of a source being or
becoming RL? How many broad line emitting regions exist in a source
and is it the same number in all sources? Can we distinguish high
and low accretors spectroscopically? How are broad and narrow
emission lines related geometrically and kinematically? How much
variance in nebular physics exists from source-to-source? Do quasars
change spectroscopically from low to high redshift?

\subsubsection{Geometrical and Kinematical Inferences}

Current best guesses about geometry and kinematics involves the idea
that all or much of the broad-line emission arises from a
\textit{flattened cloud distribution} (possibly an accretion disk)
and that kinematics are dominated by \textit{Keplerian motions}.
(e.g., \citealt{Peterson99,Peterson00}). There are empirical
arguments in favor of a flattened geometry, which pertain mostly to
radio-loud (RL) quasars. An observed anticorrelation between the
width of the Balmer lines (FWHM) and the core:lobe radio flux
density ratio R \citep{Wills86} or equivalent alternatives
\citep{Wills95,Corbin97} can be explained by observation of a
flattened BLR at different viewing angles. Similar conclusions could
be drawn from an analysis of the FWHM as a function of radio
spectral index $\alpha$ \citep{Jarvis06} or the Balmer decrement as
a function of R \citep{Jackson91}. In superluminal quasars the
viewing angle $\theta$ can be determined with good accuracy and
there are reported correlations between $\theta$ or other beaming
indicators and Balmer lines FWHM \citep{Rokaki03,Sulentic03} that
favor an axisymmetric BLR dominated by rotation. \citet{Zamfir08}
find that FRII (Fanaroff-Riley II, \citealt{Fanaroff74}) quasars
show both broader H$\beta$ profiles ($\Delta$FWHM $\approx$ 2000 km
s$^{-1}$) and weaker FeII ($\Delta$R$_{FeII}$ $\approx$ 0.2) than
the core/core-jet RL quasars.

\citet{Collin06} offer arguments that inclination effects are
apparent only in sources with narrow profiles - what we call
Population A. Studies in the 4DE1 context suggest that kinematics
and geometry are very different for Population B sources. Even if RL
quasars of Population B show correlations consistent with a
flattened geometry the BLR, it must be different than the one
suggested for Narrow Line Seyfert 1 (NLSy1; \citealt{Osterbrock85})
and other Population A sources; for example if only for the simple
reason that the minimum FWHM observed in face-on RL quasars is at
about 3000 km s$^{-1}$ broader than the narrowest Population A
Balmer profiles $\sim$ 500-600 km s$^{-1}$, whose flattened BLR is
also assumed observed face-on \citep[e.g.,][]{Zhou06,Zamfir08}. The
assumption of Keplerian motion in the BLR (and of FWHM(H$\beta$) as
a virial estimator) finds support from the discovery that the
minimum observed FWHM(H$\beta$) increases systematically with source
luminosity (from $\approx$ 1000 km s$^{-1}$ at logL$_{bol}$(erg
s$^{-1}$) = 44 to $\approx$ 3000 km s$^{-1}$ at logL$_{bol}$(erg
s$^{-1}$) = 48). This trend is easily explained if we are observing
line emission from a Keplerian disk obeying a Kaspi
(\citealt{Kaspi00,Kaspi07}) relation \citep{Marziani09}.

\subsection{The Drivers of Type 1 AGN Diversity}

Numerous studies sought to explain the quasar spectroscopic
diversity in terms of physics (intrinsic to the AGN) coupled with
orientation of the observer relative to the AGN structure and
geometry. Much emphasis was placed on properties like BH mass
\citep[e.g.,][]{Boroson02}, luminosity \citep[e.g.,][]{Marziani09}
and accretion rate \citep[e.g.,][]{Sulentic00b,Kuraszkiewicz00,
Marziani01,Marziani03b} as dominant drivers of the quasar diversity.

From a practical perspective the derivation of these quantities is
complicated by several arguments, of which the most cumbersome are:
1) the nature of the underlying continuum is rather poorly
understood, 2) numerous iron lines (FeII and FeIII) contaminate the
UV-optical-IR window producing also a pseudocontinuum, 3) not all
emitted broad lines are signatures of virialized gas gravitationally
bound to the central mass
\citep[e.g.,][]{Gaskell82,Marziani96,Sulentic07}, 4) broad lines can
show inflections and asymmetries that may indicate a composite
nature of the emitting region \citep[e.g.,][]{Marziani96,Netzer07,
Marziani09}, likely connected to a complex structure and kinematics
of the emitting region even for the same ionic species
\citep[e.g.,][]{Sulentic00b,Netzer07,Marziani09}, 5) physics and
orientation are coupled and spectral information reflects both.

Other contributors have also been suggested or speculated (mainly
for RL/RQ): the BH spin
\citep{Wilson95,Moderski98,Meier01,Volonteri07,delP.Lagos09,Tchekhovskoy09},
the structure of the accretion disk
\citep[e.g.,][]{Ballantyne07,Tchekhovskoy09}, the relation to the
host galaxy \citep[e.g.,][]{Woo05,Ohta07}, the host galaxy
morphology \citep[e.g.][]{Capetti06,Sikora07} and/or its link with
the nucleus (e.g., \citealt{Hamilton08}) or the environment
\citep[e.g.,][]{Kauffmann08}. Some researchers prefer the
orientation (either internal or external) for explaining the
differences between all type 1 quasars \citep{Richards02}, perhaps
to remain more smoothly anchored to the unified model
\citep[e.g.,][]{Urry95}.

Metallicity was also fit into the picture of quasar spectroscopic
diversity, several different correlations being reported:
metallicity-luminosity \citep{Hamann99,Nagao06}, metallicity-BH mass
\citep{Warner03} and metallicity-accretion rate \citep{Shemmer04}.
The strength of the optical FeII BLR emission may be connected with
the abundance/metallicity of quasars \citep{Netzer07}, but all those
results are difficult to relate and interpret in the framework of
the Eigenvector 1 space for the following reasons: (1) BH mass is
typically larger in Population B sources, which also show weaker
FeII emission, (2) accretion rate is typically larger in Population
A sources, which show more prominent/stronger FeII emission and (3)
luminosity is part of Eigenvector 2, formally orthogonal to
Eigenvector 1 space. The Eigenvector 1 has also been interpreted as
an evolutionary or ``age'' sequence \citep[e.g.,][and references
therein]{Grupe04}. Population A sources that are best described by
low BH masses and high Eddington ratios may represent an early stage
in the evolution of AGN, in that sense they may be young.

\subsection{The 4DE1-Based Approach}

The diversity of quasars has been explored from various directions:
RL versus RQ, NLSy1 versus BLSy1 and Population A versus Population
B \citep[e.g.,][]{Boroson02,Marziani03b,Sulentic06}. 4DE1 appears as
a promising tool that organizes the spectral diversity of quasars
maximizing the dispersion of their properties
\citep{Sulentic00a,Sulentic00b,Sulentic07}. We explore quasar
phenomenology in the context of the 4DE1 space which involves
optical, UV and X-ray measures for low redshift sources. Pre-SDSS
work was based on our own Atlas of quasars spectra, usually brighter
than B$\approx$16.5
\citep{Marziani96,Sulentic00a,Sulentic00b,Marziani03a,Marziani03b}.

Empirical evidence \citep[e.g.,][]{Sulentic07} argues against
indiscriminate averaging of spectra. One requires some sort of
context within which to define the averages such as the RQ-RL
populations. As sample sizes and data quality improve it becomes
possible to use ever more refined contexts. The optical plane of the
4DE1 uses two of the most fundamental parameters involving
FWHM(H$\beta$) and R$_{FeII}$. In the case of our bright SDSS DR5
sample a composite average spectrum of all N=469 individual spectra
(see next section) would give us little insight into the nature of
the BLR region in quasars. Averaging spectra with FWHM(H$\beta$)
covering a range from 1000 to 16000 km s$^{-1}$ and R$_{FeII}$ from
0 to 2.0 would be an oversimplification of the quasar phenomenon.
The analogy to indiscriminate averaging of OBAFGKM stellar spectra
is obvious.

We have compared quasar spectra in the population A-B context where
sources are separated at FWHM(H$\beta$)$\simeq$ 4000 km s$^{-1}$
(Figure 1a). In the context of the present study we stress two
important facts:

$\bullet$ NLSy1 sources, customarily defined by FWHM(H$\beta$) $<$
2000 km s$^{-1}$, do not form a distinct class of objects relative
to the bulk of Population A sources\footnote{A recent study
\citet{Nikolajuk09} postulates that ``the division between NLSy1 and
BLSy1 based on the width of H$\beta$ line is less generic than
according to the soft X-ray slope'', which actually is one of the
four adopted dimensions of the 4DE1 space.} \citep{Marziani01} if we
consider their optical/UV spectra \citep{Veron01,Sulentic02}.

$\bullet$ There is a remarkable change in the H$\beta$ line profile
at FWHM(H$\beta$)$\approx$4000 km s$^{-1}$: H$\beta$ is fit by a
Lorentzian function up to FWHM(H$\beta$)$\approx$4000 km s$^{-1}$,
while a double Gaussian is needed for broader sources
(\citealt{Sulentic02,Marziani09}).

It is interesting to note that if one uses the RQ-RL separation to
define this boundary within the optical plane of the 4DE1 space (see
Table 3 and Figure 4 of \citealt{Zamfir08}) then a 2D
Kolmogorov-Smirnov (2D K-S) test yields FWHM(H$\beta$)$\approx$3900
km s$^{-1}$. RQ-RL distinction appears less general and subordinated
to the A/B framework since Population B sources can be both RQ and
RL \citep{Zamfir08}. We are not claiming that the limit at
FWHM(H$\beta$)$\approx$4000 km s$^{-1}$ is absolute or fixed.
Several lines of evidence suggest a dichotomy at
FWHM(H$\beta$)$\approx$4000 km s$^{-1}$
\citep{Sulentic07,Marziani09}; especially impressive is the H$\beta$
profile change mentioned above (and also discussed later in the
paper), although as yet we cannot exclude the possibility that there
is a smooth change of properties along the optical E1 sequence.

The first result listed above suggests that Population A sources are
simply an empirically-motivated extension of the NLSy1s to a
somewhat broader limit. The limit at FWHM(H$\beta$)$\approx$4000 km
s$^{-1}$ is also motivated by the rarity of sources with broader
H$\beta$ if R$_{FeII}$ $>$ 0.5. An underlying physical parameter
could be the accretion rate, and the limit may be dependent on
luminosity (see Figure 11 of \citealt{Marziani09} for the locus of
fixed accretion rate in the FWHM vs. Luminosity plane). In this
case, the effect of orientation may blur the boundary at a critical
accretion rate potentially associated with a BLR structure
transition, creating an almost constant limit at 4000 km s$^{-1}$ at
low and moderate luminosity. No break is observed near
FWHM(H$\beta$)$\approx$4000 km s$^{-1}$ in the 4DE1 optical plane
and it would be truly remarkable if one were present given also
measurement uncertainties. Summing up, several lines of evidence
from previous work mandate the consideration of - at the very least
- a limit at FWHM(H$\beta$)$\approx$4000 km s$^{-1}$, lest gross
errors and misconceptions are introduced in the data analysis.

The paper is organized as follows: \S~2 presents a detailed
characterization of H$\beta$ broad emission line in low redshift
SDSS spectra based on: composite median spectra in two
representative bins defined within the optical plane of the 4DE1
parameter space and a set of diagnostic measures like asymmetry,
kurtosis and centroid shift. It also explores the connection between
these diagnostic measures and the 4ED1 space. In \S~3 we compare
sources of low and high Eddington ratio both in terms of line
measures and composite median spectra. \S~4 and \S~5 are dedicated
to discussion and conclusions. Throughout this paper, we use H$_{o}$
= 70 km s$^{-1}$ Mpc$^{-1}$, $\Omega_{M}$ = 0.3 and
$\Omega_{\Lambda}$ = 0.7.

\section{Detailed Characterization of H$\beta$ Emission Line Profiles
in Low Z Quasars}

\subsection{General Remarks}

The robustness of any systematic study of quasar broad emission
lines is conditioned by two things: a) availability of large,
representative samples and b) the quality of spectra, both in terms
of resolution and signal/noise (S/N) ratio. The situation is greatly
improved thanks to dedicated surveys like the Sloan Digital Sky
Survey (SDSS) that make available to the community large,
homogeneous data-sets. This work involves exploitation of $\sim$ 470
bright low z ($<$0.7) spectra from SDSS (DR5; \citealt{Adelman07}).
No comparable sample exists in terms of uniformity, completeness and
spectroscopic quality. Construction and characteristics of the
sample are explained in \citet{Zamfir08}. Our work focuses chiefly
on the two optical dimensions of the 4DE1 space, with occasional
commentaries relative to the other two based on previous studies.
One goal of this study is to emphasize that composite average
spectra - averaged in a well defined context like 4DE1 space - can
reveal the underlying stable components of the broad lines. We
address or readdress a few
fundamental questions in this paper:\\
$\bullet$ Can we further emphasize quasar diversity employing
measures such as centroid shifts, asymmetry and line kurtosis?\\
$\bullet$ How are these diagnostics tied to and/or incorporated
within the 4DE1 space (see also \citealt{Boroson92})?\\
$\bullet$ Is the distinction between Population A/B more general
than RL versus RQ quasars?\\
$\bullet$ Does it reveal evidence for a Population A/B dichotomy or
continuity across 4DE1?\\
$\bullet$ Can we reveal/infer some kinematics/structural differences
with the aid of the aforementioned line diagnostics?\\

The departure point for our present analysis is the sample of N=477
bright SDSS (DR5) quasars (z$<$0.7) defined in \citet{Zamfir08}. The
sample was defined such that psf g or i-band magnitudes are brighter
than 17.5, representative for the luminosity regime
log[L$_{bol}$(erg s$^{-1}$)]=43-47. The redshift limit was chosen so
that we have a good coverage of the H$\beta$ and its adjacent
spectral regions employed for the underlying continuum definition.
The brightness limit on the apparent magnitude was driven by the
need of good quality spectra, which allows reliable measures on the
broad line profiles. Figures 1 and 2 present the general properties
of the sample in terms of location in the optical plane of the 4DE1
space, other spectral properties of the broad lines, the number of
FRII and core-dominated RL sources along with the estimated radio
loud fraction in each bin, and the S/N (signal to noise) spectral
quality, respectively. For a detailed presentation of the criteria
involved in the construction of our sample please refer to section 2
of \citet{Zamfir08}.

Figure 1a shows the source distribution in the optical plane of 4DE1
whose parameters involve broad H$\beta$, FWHM(H$\beta$), and the
equivalent width ratio of the optical FeII (4570\AA\ blend) and
broad H$\beta$, R$_{FeII}$$\equiv$W(FeII4434-4684\AA)/W(H$\beta$).
Sources show the characteristic ``banana'' shape distribution found
in earlier studies.

\subsection{Spectral Bins in the Optical Plane of the 4DE1}

Bins of $\Delta$FWHM(H$\beta$) = 4000 km s$^{-1}$ and
$\Delta$R$_{FeII}$ = 0.5 are shown in Figure 1 following the
definitions introduced in \citet{Sulentic02}. The width
$\Delta$R$_{FeII}$ = 0.5 basically corresponds to the measurement
uncertainty at $\pm$2$\sigma$ confidence level. For example, if we
consider the center of bin A2 (R$_{FeII}$ = 0.75), data points are
not significantly different from this central value at a confidence
level $\pm$2$\sigma$ since the error of individual measurements is
in the range $\pm$0.20 to $\pm$0.25. A few extreme sources with
FWHM(H$\beta$) $>$ 16000 km s$^{-1}$ or R$_{FeII}$ $>$ 2.0 are not
considered here. The two sources that are located in what would be
bin B3 are ignored at this time as well. The bulk of sources lie
between FWHM(H$\beta$)=1000-12000 km s$^{-1}$ and R$_{FeII}$ $<$
1.5. Our sample does not include a significant population of NLSy1
sources between FWHM(H$\beta$)=500-1000 km s$^{-1}$, which are
classified as ``Galaxy'' in SDSS. Such SDSS sources have been
tabulated by \citet{Zhou06} and n=40 of them were included for
evaluation in \citet{Marziani09}. In our present report we make use
of the FeII optical template provided by \citet{Boroson92} study,
which is not suitable for very narrow broad lines. In only 14
sources (less than 3\%) do we measure a negligible R$_{FeII}$ (see
also \citealt{Dong09}; FeII may be a ubiquitous signature in Type 1
AGN and we speculate that FeII presence may be incorporated in the
empirical definition of Type 1 phenomenology). The median centroid
of the source distribution is located at (R$_{FeII}$,
FWHM(H$\beta$)) = (0.53, 4300 km s$^{-1}$). The distribution shows
no obvious concentration at the centroid values, instead we see a
broader concentration between 0.1-1.0, 1000-6000 km s$^{-1}$. The
source distribution is ``elongated'' (like a banana) and largely
reflects the \textit{intrinsic} source dispersion rather than
measurements errors ($\sim$ 20$\sigma$ in FWHM(H$\beta$) and $\sim$
16$\sigma$ in R$_{FeII}$).

A trend is seen in the sense that narrower FWHM(H$\beta$) sources
show a large dispersion in R$_{FeII}$ and weaker R$_{FeII}$ sources
show a large dispersion in FWHM(H$\beta$). The range of R$_{FeII}$
is similar at FWHM(H$\beta$)= 8000 or 12000 km s$^{-1}$ and the
range of FWHM(H$\beta$) is the same at R$_{FeII}$ = 1.0, 1.5 or 2.0.
Sources with very broad FWHM(H$\beta$) and large R$_{FeII}$ measures
are not present in this sample of quasars and are likely either
nonexistent or obscured. The distribution shows tails at both sides,
i.e., extreme FWHM(H$\beta$) (up to 40,000 km s$^{-1}$,
\citealt{Wang05}) and extreme R$_{FeII}$ (up to 2.5); the tails are
mutually exclusive and they represent about 5\% of the total sample.

Figures 1b and 1c provide median and mean values for key parameters
in each bin: the number of sources and average values of bolometric
luminosity ($L_{bol} \simeq 10\lambda L_{\lambda}$, where $\lambda$
$\equiv$ 5400\AA, see \citealt{Zamfir08}), redshift z, equivalent
width of H$\beta$ and of FeII (4570\AA\ blend), FWHM(H$\beta$) and
R$_{FeII}$. These numbers are representative for quasars of
intermediate (bolometric) luminosity (logL$_{bol}$[erg s$^{-1}$] =
43.5-46.5) at low redshift (z $<$ 0.7). RL sources tend to strongly
prefer the Population B regime. Nonetheless, there is a small
fraction of RL amongst population A sources, but we should also
emphasize that $\sim$40\% of them (7 out of 18) show FWHM(H$\beta$)
in very close to the 4000 km s$^{-1}$ A/B adopted boundary.

Comparing the average (median or mean) equivalent width values
W(H$\beta$) and W(FeII)(Figures 1bc) we see an increase of about
45\% in the former and a decrease of about 50\% in the latter
measure in bin B1 relative to bin A2. R$_{FeII}$ consequently
decreases by a factor 2.8 from A2 to B1. RL quasars strongly
concentrate above FWHM(H$\beta$) $\sim$ 4000 km s$^{-1}$ and below
R$_{FeII}$ $\sim$ 0.50 (see Figure 4 of \citealt{Zamfir08}). Most
exceptions (i.e., Population A RL) show core dominated radio
morphology suggesting that they are viewed near the radio jet axis.
If H$\beta$ emission arises from an accretion disk oriented
perpendicular to the jets' axis then a correction to FWHM(H$\beta$)
would move most core-dominated RL from Population A into the region
of Population B \citep{Marziani03b}. RL sources showing
lobe-dominated FRII and core-dominated radio morphologies show a
clear and significant separation in the 4DE1 optical plane
\citep{Sulentic03,Zamfir08}. This is a hint that the orientation
plays an important role at least for RL quasars and possibly for all
quasars.

A more extreme comparison (relative to Figures 1b-c) could involve
the less populated bins A3, B2 and B1+, each with about 30-40
sources. Average luminosities are very similar. On average W(FeII)
in bin A3 is about 4$\times$ larger than in bin B1+. Average
W(H$\beta$) in bin B1+ is about 1.5$\times$ larger than in bin A3.
There are no RL sources in bin A3, but a large concentration of RL
is found in B1+ (see Figure 4 in \citealt{Zamfir08}). We avoid bin
A1 with the smallest FWHM(H$\beta$) and R$_{FeII}$ measures because
it may involve a mixture of Population A and B sources. In terms of
spectral quality we estimate mean and median ratio S/N $\sim$ 29
(defined in the continuum region 5600-5800 \AA). Figure 2 shows the
(histogram) distribution of pixel S/N ratios for the N=469 SDSS
spectra in our sample.

\subsection{Composite Spectra of Quasars}

Composite average quasar spectra have proven to be a powerful tool
for revealing both commonalities among and differences between
quasars. High S/N spectral composites have served many purposes.
First they allowed a general description of the quasar phenomenon
by: a) providing a reference or cross-correlation template spectrum
by identifying (and measuring) weak emission features (strengths,
widths, shifts, asymmetries, etc.), otherwise hardly observed in
individual spectra \citep[e.g.,][]{Wills80, Veron83, Boyle90,
Francis91, Zheng97, Vandenberk01, Constantin03, Scott04, Glikman06};
b) defining the underlying continuum by isolating wavelength regions
(almost) free of lines \citep[e.g.,][]{Francis91, Vandenberk01,
Constantin03}; c) revealing the omnipresence of FeII blends
throughout the UV and optical domains (see e.g., section 4.2 in
\citealt{Vandenberk01} and references therein); d) evaluating colors
of quasars and/or K(z) corrections \citep{Cristiani90, Glikman06}.

As larger samples of quasars spectra became available, composites
were constructed for sources with very similar properties like: 1)
radio-loudness \citep[e.g.,][]{Cristiani90, Zheng97, Brotherton01,
Telfer02, Bachev04, Glikman06, Metcalf06, deVries06}, 2) radio
core-to-lobe flux density ratio \citep[e.g.,][]{Baker95}, 3) radio
morphology, FRII versus core-dominated, e.g., \citet{deVries06}, 4)
optical continuum slope or relative color \citep{Richards03}, 5)
broad emission line velocity shifts \citep{Richards02, Hu08b}, 6)
CIV $\lambda$1549\AA\ line width \citep{Brotherton94}, 7) emission
line widths and strengths \citep{Sulentic02, Constantin03,
Bachev04}, 8) Black Hole Mass (M$_{BH}$) and Eddington ratio
($L_{bol}/L_{Edd}$) \citep[e.g.,][]{Marziani03b}, 9) variability
\citep{Wilhite05, Pereyra06}, 10) Broad Absorption Lines - BAL
\citep[e.g.,][]{Brotherton01, Richards02, Reichard03} and 11) soft
X-ray brightness \citep[e.g.,][]{Green98}. Such composites
facilitate an emphasis on differences between properties of quasars.
The luminosity and redshift dependence/evolution of quasar
properties was sought and tested with the aid of composite spectra
\citep[e.g.,][]{Cristiani90, Zheng97, Blair00, Croom02, Corbett03,
Yip04, Nagao06, Scott04, Bachev04, Metcalf06, Marziani09}.

An important question involves how to combine individual spectra
into a composite. Moreover, how well does a composite spectrum
preserve the properties of the input components and to what extent
is it representative of the sample from which it has been
constructed? It has been noted that there is no unique best answer
to such questions. While a median composite preserves the relative
fluxes of the emission features, a geometric mean composite
preserves better the shape of the underlying continuum
\citep[e.g.,][]{Vandenberk01}. Additional questions can be related
to how to weight each spectrum's contribution into the composite.
Depending on the number and quality of individual spectra available
one could decide whether a weighted-average (arithmetic mean)
composite \citep[e.g.,][]{Francis91,Glikman06} is more useful than a
composite where all spectra are given equal weight
\citep[e.g.,][]{Constantin03}. The most important thing is the
context within which we define the composites; we consider 4DE1 is
the best approach available.

Given the S/N distribution (Figure 2) of our sample, the homogeneity
of the SDSS spectra and the nature of our spectral analysis we see
the median composites with equal weight from individual components
as most suitable (for a discussion on the relevance of the median
composites see section 4.1 of \citealt{Bachev04}).

\subsection{H$\beta$ Profile Shape}
\subsubsection{Data Analysis}

Low z rest-frame spectra require galaxy decontamination and flux
corrections. They are normalized to the 5100 \AA\ specific flux
prior to being considered for the composite. The full analysis
procedure for a quasar optical spectrum focusing on the H$\beta$
region is outlined in \citet{Zamfir08}. We added an intermediate
step in the present study by defining the underlying continuum as a
simple power-law anchored in the regions 4195-4218 \AA\ and
5600-5800 \AA\. The FeII template used to model FeII emission around
H$\beta$ is based on IZw1 \citep{Boroson92,Marziani03a}. The
template matches the 4750 \AA\ FeII ``blue'' blend significantly
better than it does the FeII blend redward of
[OIII]$\lambda$5008\AA\ (``red'' blend). In many cases the ``red''
blend of FeII may be overestimated. In order to alleviate this
problem we opted for an additional iteration in the reduction
process of the composite spectra: 1) we define the underlying
power-law continuum, 2) subtract the continuum, 3) find the best
match between the FeII template and the spectrum (adjusting the
width and intensity of the template), focusing on the best match for
the blue blend of FeII, 4) subtract this template solution from the
original rest-frame spectrum (available before step 1) and 5)
redefine a power-law continuum using the wavelength regions
4195-4218 \AA\ and 5100-5300\AA. This continuum is slightly steeper
than the initial choice. We now repeat steps 2 and 3. This iteration
significantly improves the match between template and the red blend
of FeII.

From the spectrum thus obtained (continuum and FeII-free) we cut out
the wavelength range bracketed by HeII$\lambda$4687\AA\ and
[OIII]$\lambda$5008\AA\. We simultaneously fit all lines in the
selected interval, allowing for narrow components (NC) in the case
of HeII, H$\beta$, [OIII]$\lambda$4960\AA\ and 5008\AA\ lines. The
[OIII] lines require an additional component, in most cases
blueshifted relative to the quasar's restframe indicated by the NC.
We do not require a match in width between the NC of H$\beta$ and
the core (the narrower of the two) components of the [OIII] lines.
We constrain the [OIII] 5008/4960 flux ratio R=3 allowing for a
maximum 15\% error. All components are modeled with Gaussian
functions in the first stage. Subsequently, we test whether a
Lorentzian function provides an improved fit to the H$\beta$ broad
component (BC) in terms of reduced $\chi^2$ and by visual inspection
of residuals. All other emission lines are modeled with Gaussian
functions. We use the Levenberg-Marquardt
algorithm\footnote{www.nr.com; \citet{Press86}.} to adjust the
parameter values in the iterative procedure.

\subsubsection{Composite Spectra in 4DE1}

Figures 3 and 4 present composite spectra for bins A2 and B1,
respectively. We show both the region (upper) from 3500-7000\AA\
with continuum+line model and an enlarged view (lower) of the
H$\beta$ region from 4600-5100\AA\ with details of the line model.
We focus on bins A2 and B1 as representative of source differences
between Populations A and B. We avoid bin A1 because it likely
contains a mix of both Population A and B as well as because it
shows a range of R$_{FeII}$ values that is too similar to bin B1
sources.

Extreme A bin composites (A3 and A4) tend to have stronger FeII
emission while extreme B bins (B1+, B1++) involve sources with
broader H$\beta$ profiles and weaker FeII emission. The systematic
weakening of FeII between bins B1 and B1++ (see also Figures 1b,c)
may be caused by the increase in FWHM for FeII lines that increases
with FWHM(H$\beta$). As the FeII blends broaden they create a
pseudocontinuum that may lead to systematic underestimation of FeII
strength. However, thanks to the high S/N of our SDSS sample, FeII
is detected for the first time in most sources with FWHM(H$\beta$)
$>$ 8000 km s$^{-1}$, revealing that this effect is not as
influential as feared.

Bins A2 and B1 are well matched in terms of number of sources as
well as mean/median redshift and logL$_{bol}$ (Figures 1-b,c). N=129
and N=131 sources were used to generate the A2 and B1 composites
respectively. The differences between the two composites would be
slightly larger if we included in A2 the NLSy1s sources with
FWHM(H$\beta$) $<$ 1000 km s$^{-1}$ tabulated by \citet{Zhou06}.
However inclusion or exclusion of these sources will make no
significant change in the results. \citet{Marziani03b} show that
sources with FWHM(H$\beta$) $<$ 2000 km s$^{-1}$, often described as
a separate class, are similar in terms of line profile shape (i.e.,
Lorentzian) to those having FWHM(H$\beta$) in the range 2000-4000 km
s$^{-1}$.

The A2 median composite shows high enough S/N to reveal a modest
inflection between the BC and NC of H$\beta$. The NC adopted by the
model shows FWHM $\sim$ 350 km s$^{-1}$, very similar to the core
components of the [OIII] lines. BC shows FWHM(H$\beta$)$\simeq$2600
km s$^{-1}$. The best fit (Figure 3b) to the BC of H$\beta$ profile
is obtained with a Lorentzian function, confirming results of
previous studies (\citealt{Sulentic02,Marziani03b,Marziani09}). The
profiles of NLSy1 sources have long been described as Lorentzian
\citep{Veron01}, but that profile description now extends to
FWHM(H$\beta$)$\approx$4000 km s$^{-1}$. A Gaussian fit does not
provide satisfactory results in light of $\chi$$^{2}$ residuals. A
pair of unshifted Gaussians provides a reasonable fit
\citep{Popovic04,Bon09} since they essentially represent an
approximation of a Lorentzian. A Lorentzian is preferred as a
description of the A2 H$\beta$ profile based on $\chi$$^{2}$. The
broad Lorentzian component shows no shift relative to the quasar's
restframe.

There is a positive residual redward of [OIII]$\lambda$5008\AA. This
residual can be an artifact produced by a mismatch between the FeII
template and the composite spectrum coupled with an oversimplified
definition of the underlying continuum. The red-shelf has been
pointed out in \citet{Marziani03a} (see their section 4.3) and
\citet{Veron02}. The latter study focused on two sources with such a
prominent feature and showed that the main contributors to the
``shelf'' are most probably the HeI lines $\lambda\lambda$ 4922,
5017 \AA\ and is not caused by a broad redshifted component of
H$\beta$ or [OIII] lines. Even if a VBLR is assumed in our case here
and affects or causes the red ``shelf'', its contribution to the
total flux of H$\beta$ is no more than 10\%, which is within the
uncertainty range of the fitting routine.

The median composite for bin B1 (Figure 4) shows striking
differences from the A2 composite, most notably in the shape of the
broad H$\beta$ profile. We see two inflections: 1) between the NC
and the BC and 2) on the red side of the BC. The simplest fit to the
broad component therefore requires a double Gaussian involving a
broad (BC) unshifted and a very broad (VBC) component, redshifted by
$\sim$ 1500 km s$^{-1}$, which also involves about 53\% of the total
H$\beta$ broad line flux. This model is again in agreement with our
earlier studies. Detection and measurement of the VBC requires
spectra with good S/N and moderate resolution - the median composite
suggests that it is very common if not ubiquitous in sources with
FWHM(H$\beta$)$>$ 4000 km s$^{-1}$. It is obvious that some or many
sources show FWHM(H$\beta$)$>$ 4000 km s$^{-1}$ because of this
second component. If we interpret the unshifted BC as the classical
BLR then we might expect it to show a Lorentzian shape. Fitting the
B1 composite with an unshifted Lorentzian plus redshifted Gaussian
yields a poorer $\chi$$^{2}$ and a VBC comprising less than $\sim$
10\% of the total broad line flux, with $\sim$ 6000 km s$^{-1}$
redshift. Past and present results suggest that a double Gaussian is
the best choice. The classical ``unshifted'' BC Gaussian shows a
only small blueshift of $\sim$ -100 km s$^{-1}$ and FWHM of $\sim$
4000 km s$^{-1}$.

The width of the BC component in the bin B1 composite is
significantly less than the median value of FWHM(H$\beta$) $\sim$
5400 km s$^{-1}$ reported for bin B1 in Figure 1b. We conclude that
FWHM measures of broad H$\beta$ that do not correct for the VBC will
lead to serious BH (Black Hole) mass overestimates. At the same
time, for the bin A2 there is no difference between the median value
reported in Figure 1b and the measured FWHM(H$\beta$) in the
composite median spectrum. This suggests that no VBC correction is
required for Population A sources making their H$\beta$-based BH
mass estimates more reliable. The bin A2-B1 comparison suggests
fundamental structural and kinematic differences between Population
A and B sources (see also \citealt{Sulentic02}) most likely coupled
with non-trivial effects of orientation.

\subsubsection{Diversity of H$\beta$ Profiles of Individual Spectra}

We have argued that indiscriminate averaging of spectra will obscure
the most physically fundamental differences in a source sample. The
purpose of this section is to go to the other extreme and examine
source spectra individually. Figure 5 shows a selection of
individual H$\beta$ profiles in order to illustrate that all
previously known (e.g., \citealt{Stirpe90}) diversity is present in
the SDSS sample. The A2 and B1 median composite spectra presented
and discussed earlier (and also discussed in \citealt{Sulentic02})
do not show the variety of features observed in individual,
single-epoch spectra shown in Figure 5. This justifies the
interpretation that average spectra represent the underlying
commonality, more stable components in the H$\beta$ profiles of
different sources. The individual spectra are, to some extent,
equivalent to the results of spectroscopic monitoring of a single
source. Since many of the features do not show up in the composite
they presumably represent relatively short-lived, transient and
therefore lees important changes in the profiles.

We want to go beyond the optical plane of the 4DE1 space because we
find that sources inside bins like A2 and B1, while showing similar
FWHM(H$\beta$) and R$_{FeII}$, also show considerable profile
diversity implying that the two measures, width and relative
strength of FeII emission, are not enough to characterize the
diversity. For each spectrum shown in Figure 5 we also indicate the
bin in the optical plane of the 4DE1 space, the bolometric
luminosity, spectral continuum S/N (as defined earlier) along with
measures of centroid shifts, asymmetry and line shape (see next
section). This set of 32 spectra is assumed representative in terms
of S/N ratio, luminosity and occupation of the 4DE1 optical plane.
In addition to the already stated purpose of presenting the great
variance of the Balmer H$\beta$ line properties amongst our low
z/intermeidate luminosity sample, this subset serves also the
purpose of estimating typical uncertainties for the line diagnostic
measures described in detail in the next section.

\subsubsection{Higher Order Moments of H$\beta$ Profiles as Diagnostics of the BLR Physics}

In order to get a deeper insight into the hypothesized existence of
two quasar populations A and B
\citep[e.g.,][]{Sulentic07,Marziani09} we examine the sample
distributions of higher order moments of the H$\beta$ profiles. We
focus on centroid shifts, asymmetry and kurtosis (line shape). The
H$\beta$ profile is fit with a high order polynomial making no
assumptions about different components that might be present. Higher
order moments of the broad lines are affected by larger
uncertainties than the line width measures, nonetheless, they offer
the possibility to further constrain physical models of the BLR. We
benefit in this analysis from the uniformity of SDSS spectra in S/N
and resolution.

For quantitative definitions of the parameters involved in our study
we consider those proposed in \citet{Marziani96}. For clarity we
reproduce them here in the following order: Asymmetry Index A.I.,
Kurtosis Index K.I. and centroid shift $c(x)$:
$$A.I.\left(\frac{1}{4}\right)=\frac{v_{r,R}(1/4)+v_{r,B}(1/4)-2v_{r,peak}}{v_{r,R}(1/4)-v_{r,B}(1/4)}~~~~~(1)$$
$$K.I.=\frac{v_{r,R}(3/4)-v_{r,B}(3/4)}{v_{r,R}(1/4)-v_{r,B}(1/4)}~~~~~(2)$$
$$c(x)=\frac{v_{r,R}(x)+v_{r,B}(x)}{2}~~~~~(3)$$

The v$_{r,R}$ and v$_{r,B}$ refer to the velocity shift on the red
and blue wing, respectively (at the specified hight of the profile),
calculated relative to the restframe of the quasar. By ``peak'' we
mean 9/10 fractional intensity. The centroid shift will refer to
$x=1/4$ fractional intensity (what we call the line ``base'') or
$x=9/10$ (the line ``peak''). We point out that all these measures
avoid the region below 1/4 fractional intensity due to the larger
uncertainties associated with the definition of line wings and
underlying continuum plus FeII pseudo-continuum.

\subsubsection{Higher Order Moments and the Optical Plane of the 4DE1
Parameter Space}

It is important to identify any kind of relation between 4DE1 and
higher order diagnostic parameters that characterize the profile of
the H$\beta$ profile. Figure 6 shows plots of FWHM(H$\beta$) versus
the newly defined diagnostic measures: (a) vs. asymmetry index
A.I.(1/4), (b) vs. kurtosis line shape measure K.I., (c) vs.
normalized base shift c(1/4)/FW025M and (d) vs. normalized peak
shift c(9/10)/FW0.90M). Whenever shown, the two vertical lines
indicate the 2$\sigma$ uncertainty lever on either side of zero. We
also indicate the two populations A and B defined in terms of
FWHM(H$\beta$) less than and larger than 4000 km s$^{-1}$,
respectively. In each panel we also display the FRII radio sources
with distinct symbols (blue solid). A few commentaries may be
worthwhile: 1) There seems to be a trend in panel (a) in the sense
that lower widths profiles are more symmetric, but sources with
broader profiles show a numerical excess of (larger than the
2$\sigma$ uncertainty) of red-asymmetric basis; 2) Panel (b) shows
no clear trend, but it indicates that the kurtosis values of broader
H$\beta$ profiles cover more uniformly a wider range of vales than
the narrower profiles; 3) Panel (c) reveals bears some similarity to
panel (a), with a subtle difference; at lower widths there seems to
be a slight excess of blueshifted ``basis'', and the slight excess
of redshifted basis is still apparent for broader profiles; 4) there
is clear excess of sources with blueshifted ``peaks'' over the whole
range of FWHM. FRII sources do not seem to have any special behavior
when contrasted against the other Population B sources.

The centroid shift measures are normalized in each case to the width
of the profile at that same fractional intensity. We distinguish in
each case Population A and B sources, as defined in the context of
the 4DE1 parameter space. The vertical dotted lines indicate the
2$\sigma$ median uncertainty derived for the subset of 32 sources
shown in Figure 5. Individual source profile uncertainties (for the
32 spectra) are estimated allowing for $\pm$ 5\% variation in
fractional intensity and then propagating the errors into the final
diagnostic measure. Thus, the typical errors are:
$2\sigma(A.I.(1/4)) \simeq 0.16$, $2\sigma(c(1/4)/FW0.25M(H\beta))
\simeq 0.07$, $2\sigma (K.I.) \simeq 0.092$, $2\sigma
(c(9/10)/FW0.90M(H\beta)) \simeq 0.22$, 2$\sigma$ (c(1/4)) $\simeq$
520 km s$^{-1}$ and 2$\sigma$ (c(9/10)) $\simeq$ 320 km s$^{-1}$.

Figures 7a-d show the histogram distributions of the asymmetry index
A.I.(1/4), the centroid shift at 1/4 and 9/10 fractional intensities
c(1/4) and c(9/10), respectively and the kurtosis measure K.I. The
width of the bins in Figure 7 (and 8) are chosen such that they are
approximately equal to 1$\sigma$ typical (i.e., average)
uncertainty. In Table 1 we present the average (mean and median)
values of the parameters shown in Figure 7. We make the following
comments relative to Figure 7: i) Population B sources show an
excess (outside 2$\sigma$) of red asymmetric profiles and profiles
red-shifted at the base, ii) Population A sources are more symmetric
and they show a slight excess of blue-shifted profiles at 1/4 from
zero-level, iii) there is an excess of blue-shifted peaks in both
Population A and B sources and iv) Population B sources show an
extended tail of large kurtosis values, while Population A quasars
show a more symmetric distribution about K.I.=0.35, with a clear
preference for lower K.I. values. A one dimensional K-S test
indicates a 4\% probability of null hypothesis when comparing the
normalized ``peak'' shift distributions (lower left panel of Figure
7). For all other measures shown in Figure 7 a 1D K-S test shows a
null-hypothesis probability less than 10$^{-4}$. The numbers in
Table 1 appear to confirm the aforementioned observations i) and
ii).

\begin{table*}
\caption{Average values for line asymmetry, kurtosis, base and peak
shift for the two populations A and B, relative to Figure 7.}
\tabcolsep=3pt
\begin{tabular}{lcc}
\hline

Parameter & Population A (N=260) & Population B (N=209)  \\
          & mean (SD) ; median & mean (SD) ; median \\

\hline\hline

A.I.(1/4) & 0.002 (0.118) ; 0.001 & 0.096 (0.155) ; 0.083  \\

\hline

K.I. & 0.356 (0.063) ; 0.350 & 0.383 (0.092) ; 0.374  \\

\hline

c(1/4)/FW0.25M & -0.011 (0.062) ; -0.010 & 0.029 (0.076) ; 0.024  \\

\hline

c(9/10)/FW0.90M & -0.064 (0.277); -0.042 & -0.102 (0.300) ; -0.080 \\
\hline

\end{tabular}

\end{table*}

Figure 8 shows a comparison of ``base'' and ``peak'' shift for
Populations A and B, this time without normalization to the line
width at the corresponding fractional intensity as in Figure 7. We
are aware that the un-normalized centroid shift values are less
meaningful than the normalized ones. Nonetheless, the two panels in
Figure 8 seem to show: 1) the two populations are significantly
different (Population A sources being distributed mostly within
$\pm$2$\sigma$, while Population B quasars showing large excesses of
blueshifted peaks and redshifted basis outside the 2$\sigma$ range
and 2) the measured shifts are quite large, ``base'' redshifts over
1500-2000 km s$^{-1}$ and ``peak'' blueshifts in excess of -1000 km
s$^{-1}$ being observed.

We tested whether the observed distributions are spurious effects of
the S/N. For this task we reconstructed the histogram distributions
shown in Figure 7 restricting the sample to sources of S/N $<$ 15
($\sim$ 20\% of our whole sample; see Figure 2). All distributions
presented above are confirmed and are not caused by the spectral
S/N. We also tested if the distributions presented in Figure 7 are
driven by the preferential addition of a subset of SDSS quasars
selected with some restriction on their radio properties
\citep{Zamfir08}. For this reason, we isolated only the RQ sources
of our sample (i.e., N=388 sources with logL$_{21cm}$ $<$ 31.6 erg
s$^{-1}$ Hz$^{-1}$.). Although we don't show the new histograms
here, the distributions are qualitatively very similar to those in
Figure 7, thus we have clear indication that the population A/B
concept is more general than the RL/RQ distinction.

The pioneering study of \citet{Boroson92} incorporated measures of
asymmetry, line shape and shift in their set of quasar spectral
measures. They reported a significant correlation between R$_{FeII}$
and the asymmetry of H$\beta$ (see their Figure 5). Sources with
strong FeII emission tend to show a flux excess on the blue wing of
H$\beta$ and the weaker FeII emitters tend to be show more
red-asymmetric H$\beta$ profiles. The larger sample under
investigation here reenforces their result (Figures 9ab), the trend
being clearly illustrated in diagrams that combine A.I.(1/4),
c(1/4)/FW0.25M and R$_{FeII}$.

Figure 9 presents the most interesting plots involving pairs of
measures from the following set: R$_{FeII}$, A.I.(1/4),
c(1/4)/FW0.25M, K.I., L$_{bol}$/L$_{Edd}$. The plots involving the
peak shift are not shown because they do not illustrate any
particularly special aspects beyond the already mentioned ones
(related to the histograms in Figure 7).

Figures 9ab (log-linear plots) show the correlations between 4DE1
parameter R$_{FeII}$ and the asymmetry and ``base'' line shift
measures. We also identify sources showing FRII radio morphology
symbols (dark blue). The FRII quasars also serve as an independently
(radio) defined subsample of Population B sources. Both panels show
a trend from blue asymmetries/shifts at large R$_{FeII}$ values to
red asymmetries/shifts at lower R$_{FeII}$ values. A Pearson
correlation test applied show to the whole sample yields a
coefficient of r$_{p}$ = - 0.50 (with a probability of not being
correlated P $\sim$ 0) for panel (a) and r$_{p}$ = - 0.58 (P $\sim$
0) for panel (b). The bisector fit \citep{Isobe90} for (a)
corresponds to the linear relation A.I.(1/4)=
-0.54logR$_{FeII}$-0.15 and for (b) c(1/4)/FW0.25M =
-0.27logR$_{FeII}$-0.09.

It is interesting to point out that in both cases the linear
regression indicates that when the c(1/4) or A.I.(1/4) yield a null
value, the corresponding R$_{FeII}$ $\sim$ 0.5. The majority of FRII
sources lie below the bisector fits and show positive A.I.(1/4) and
c(1/4) values. This indicates that Population A sources are less
asymmetric and less redshifted (at profile ``base'') than Population
B quasars (see also Figure 20 and related discussion). A 2D K-S test
for RL vs. non-RL quasars within the optical plane of the 4DE1 space
\citep{Zamfir08} indicates the same R$_{FeII}$ $\sim$ 0.50 as a
relevant boundary.

Figure 9c shows the plot of R$_{FeII}$ versus the line shape K.I.
measure. We see a complex distribution, quite different from Figures
9ab. We see different trends on either side of R$_{FeII}$$\sim$0.5.

The parent population of RL quasars (FRII sources) seems to prefer
almost exclusively the region to the left of R$_{FeII}$ = 0.50.
Thus, we may already have several lines of evidence that seem to
suggest R$_{FeII}$ = 0.50 as a viable alternative to the original
FWHM(H$\beta$) = 4000 km s$^{-1}$ Population A-B boundary.

Previous quasar studies
\citep{Sulentic00b,Kuraszkiewicz00,Marziani01,Marziani03b} indicated
that the principal physical driver of the variance in 4DE1 was the
Eddington ratio L$_{bol}$/L$_{Edd}$. In Figure 9d we show the
correlation between logR$_{FeII}$ and log(L$_{bol}$/L$_{Edd}$);
L$_{Edd}$ $\approx$ 1.3 $\times$ 10$^{38}$ (M$_{BH}$/M$_{\odot}$)
erg s$^{-1}$ and
$$M_{BH}(M_{\odot}) \approx 5.48 \times 10^{6}
\left[\frac{\lambda L_{\lambda}(5100\AA)}{10^{44}erg
s^{-1}}\right]^{0.67}\left[\frac{FWHM(H\beta_{BC})}{1000 km
s^{-1}}\right]^{2}~~~~~(4)$$ (see \citealt{Sulentic06}).

A Pearson test yields r$_{p}$ = 0.56 (P $\sim$ 0) and a Spearman
correlation test yields r$_{s}$ = 0.59 (P $\sim$ 0). We show the
bisector fit (correlation coefficient $\rho$ = 0.58). The linear fit
log(L$_{bol}$/L$_{Edd}$) = 1.47logR$_{FeII}$-0.32 indicates that
R$_{FeII}$ = 0.50 corresponds to an Eddington ratio
L$_{bol}$/L$_{Edd}$ = 0.18. FeII emission increases in strength
relative to H$\beta$ as the Eddington ratio increases, i.e., higher
accretion rate. We are now in a position to test whether the base
asymmetry and centroid shift measures A.I.(1/4) and c(1/4)/FW0.25M
are correlated with the accretion rate measured by
L$_{bol}$/L$_{Edd}$. Figures 9e-f show the two line diagnostic
measures versus log(L$_{bol}$/L$_{Edd}$). For the plot involving the
A.I.(1/4) (panel e) a Pearson test gives r$_{p}$= - 0.39 and a
Spearman test gives r$_{s}$ = - 0.43 (P $\sim$ 0 in each test). For
panel (f) r$_{p}$ = - 0.28 (P $\sim$ 7 $\times$ 10$^{-10}$) and
r$_{s}$= - 0.32 (P $\sim$ 6 $\times$ 10$^{-13}$). The FRII sources
prefer the low accretion regime. It appears that both the A.I. and
the normalized c(1/4) measures are intimately connected to the 4DE1
parameter space. Nonetheless, their inclusion in this space
definition may not necessarily lead to an extension of the number of
dimensions. One of the two indices may be regarded as a meaningful
surrogate for FWHM parameter for example, although the problem
requires further investigation.

Figure 9(g) shows the plot of the Eddington ratio vs. the line shape
measure K.I.; the large majority of FRII sources are low accreting
sources and their kurtosis measures spread over a wide range of
values. It also seems to indicate that high Eddington ratios and and
large kurtosis values are mutually exclusive.

\section{Low versus High L$_{bol}$/L$_{Edd}$ Sources}

Several recent articles
\citep{Hubeny00,Marziani03b,Collin06,Bonning07,Kelly08,Hu08a,Marconi09}
came from various directions to the conclusion that around some
critical L$_{bol}$/L$_{Edd}$ $\sim$ 0.2$\pm$0.1 the
properties/structure of the active nucleus may change fundamentally,
thus providing support for the notion that two distinct populations
of quasars exist \citep{Collin06,Sulentic07}. While the concept of
two Populations A and B is deeply anchored into a long series of
empirical studies (see section 3 and Table 5 of \citealt{Sulentic07}
for a summary), the nominal boundary at 4000 km s$^{-1}$
FWHM(H$\beta$) may not have a very intuitive or immediate physical
meaning or may even appear arbitrary. However, it is solely intended
as a critical empirical measure that could indicate a fundamental
change in the structure, geometry and kinematics of the BLR.

It has been suggested (as was emphasized earlier) that the Eddington
ratio L$_{bol}$/L$_{Edd}$ is the main physical driver of the
Eigenvector 1 parameter space variance. Our sample reenforces these
arguments (e.g., Figure 9d). Figure 10 compares the histogram
distributions of bolometric luminosity L$_{bol}$, black hole mass
M$_{BH}$ and Eddington ratio L$_{bol}$/L$_{Edd}$ for populations A
and B. Taken at face value they suggest that the two populations
have very different distributions of M$_{BH}$ and Eddington ratio,
even though they show similar luminosity distributions (see Table
2). Obviously, given that the two populations have similar
luminosity distributions, the statistical difference in BH mass and
Eddington ratio is a natural and direct consequence of the line
width alone. Populations A and B being defined based on the
FWHM(H$\beta$). A 1D K-S test in each case indicates a 1.2\%
null-hypothesis probability for the bolometric luminosity
distribution of the two populations and a probability less than
10$^{-4}$ for the other two parameters. However, although the AGN
central mass and the Eddington ratio may be more physically
meaningful quantities, FWHM is a more direct and robust measure,
much less affected by uncertainties and derived with fewer
assumptions.

\begin{table*}
\caption{Average values for bolometric luminosity, BH mass and
Eddington ratio for the two populations A and B, relative to Figure
10.} \tabcolsep=3pt
\begin{tabular}{lcc}
\hline

Parameter & Population A (N=260) & Population B (N=209)  \\
          & mean (SD) ; median & mean (SD) ; median \\

\hline\hline

log[L$_{bol}$(erg s$^{-1}$)] & 45.4 (0.9) ; 45.6 & 45.6 (0.8) ; 45.8  \\

\hline

log[M$_{BH}$(M$_{\odot}$)] & 7.8 (0.7) ; 7.9 & 8.7 (0.6) ; 8.8  \\

\hline

L$_{bol}$/L$_{Edd}$ & 0.36 (0.28) ; 0.28 & 0.07 (0.05) ; 0.06  \\

\hline
\end{tabular}
\end{table*}

We decided to construct two test samples which would be
representative of high and low accretors; all sources with
$L_{bol}/L_{Edd} > 0.3$ (N=123) and with $L_{bol}/L_{Edd} < 0.1$
(N=177) are assumed to be high and low accretion quasars,
respectively. This distinction turns out to be representative of the
previously adopted populations A and B. All high L$_{bol}$/L$_{Edd}$
in our definition are part of population A and 87\% of the low
L$_{bol}$/L$_{Edd}$ are population B, only 13\% spilling over the
4000 km s$^{-1}$ boundary into population A. Obviously we are aware
that this kind of distinction would automatically introduce some
bias, i.e., the low accretors would prefer larger BH masses and less
luminous sources and the high accretors would populate more the high
luminosity and low BH mass regime.

We point out a few remarkable results in Figures 11a-d. Panel (a)
shows source luminosity as a function of the normalized centroid
shift at line ``base'' (1/4 fractional intensity). The bias
indicated previously is clearly apparent. We may however focus on
the region logL$_{bol}$[erg s$^{-1}$] $>$ 45.5, where both high and
low accretor sources are well represented. There is a clear
separation of the two samples in terms of base shift, with high
accretors showing a preference for blueshifts and the low accretors
for redshifts. Inside the $\pm2\sigma$ uncertainty interval there is
some overlap while outside (most significant ``base'' line shifts)
the separation is almost total. A few FRII sources show blueshifted
profiles at 1/4 base and very few high accretors spill over the
+0.07 boundary into the redshifted region.

Panel (b) shows the normalized ``peak'' shift (at 9/10 fractional
intensity) versus the asymmetry index A.I.(1/4). The separation is
again quite striking: high accretors tend to show unshifted peaks
with an excess of blue-asymmetric bases, while the low accretors
distribute more toward blueshifted peaks and red-asymmetric bases.
The simplest accretion disk models predict redshifted profile bases
and blueshifted peaks (\citealt{Chen89,Sulentic90}, but see also
\citealt{Laor91} for X-ray line model predictions). We may find here
support for the idea that low accreting sources produce their broad
emission lines (the LIL/Balmer components) in a disk-like geometry.
A 2D Kolmogorov-Smirnov test applied for the case of Figure 11b
gives a null-hypothesis probability P $\sim$ 3 $\times$ 10$^{-11}$.

One of the clearest separations between the two test samples is
illustrated in panel (c), where we plot the line shape measure
kurtosis K.I. in terms of normalized ``base'' shift. High accretors
are almost exclusively confined to K.I. $<$ 0.4. The clear excess of
blueshifted and redshifted bases is evident at low values of K.I. A
2D Kolmogorov-Smirnov test applied for the two samples gives a
null-hypothesis P $\sim$ 2 $\times$ 10$^{-17}$. Panel (d) should be
examined in conjunction with the previously presented Figure 9c.
Figure 12d is now restricted to high and low accretors only, as we
defined them in this section. It is clear now that the two test
samples show rather different trends on either side of  $R_{FeII}
\sim 0.50$. There is a trend extending along the K.I. axis (low
accretors) and another one expanding mostly along the R$_{FeII}$
axis (high accretors).

\subsection{Composite Spectra of Low and High Accretors}

Starting from Figure 11a we construct median composite spectra for
two groups of sources. The first group (N=31) is defined by the low
accreting sources with logL$_{bol}$[erg s$^{-1}$] $>$ 45.5 and
normalized c(1/4) larger than 0.07 (which is the estimated typical
2$\sigma$ uncertainty of c(1/4)/FW02.5M). The second one (N=26) is
drawn with the same luminosity constraint from the highly accreting
sources having the normalized c(1/4) less than -0.07. Focusing on a
common luminosity regime implies that we are basically emphasizing a
difference in BH mass. Thus, the composites reflect also the
influence of the BH masses that are on average different by 1dex in
the two groups (low accretors, mean and median
log[M$_{BH}$(M$_{\odot}$)] are both 9.3, standard deviation of the
mean 0.3); high accretors, mean and median
log[M$_{BH}$(M$_{\odot}$)] are both 8.2, standard deviation of the
mean 0.4). We present in Figures 12-13 the composite spectra for the
two cases. We also show our choice for the underlying power law
continuum and the best solution of the FeII template that matches
the optical FeII emission, with a focus on the region 4100-5150\AA.

In Figure 12 we analyze the H$\beta$ and adjacent emission lines in
the range $\sim$ 4600-5150\AA\ for the low accretors, after we have
removed the contaminating FeII in this region and the continuum
model. We fit simultaneously the lines following the approach
explained/outlined in earlier in \S\S~2.4. We show two
possibilities, first Gauss+Gauss for BLR+VBLR contributions and
second Lorentz+Gauss for BLR+VBLR. The $\chi^2$ is very similar in
the two cases. The BLR component in either case shows a blueshift of
$\sim$ -500 km $s^{-1}$ relative to the adopted quasar restframe,
based on NC. The VBLR on the other hand shows a big difference. In
the Gauss+Gauss case it has a contribution of $\sim$ 70\% to the
whole Balmer BLR+VBLR emission and shows a redshift of $\sim$ +2500
km s$^{-1}$. In the case Lorentz+Gauss the VBLR contributes less to
the total broad line emission ($\sim$ 30\%) and is redshifted by
$\sim$ +4000 km s$^{-1}$. The FWHM of the BLR component is close to
5000 km s$^{-1}$ in each case, much less than the FWHM for the
combined BLR+VBLR of $\sim$ 7600 km s$^{-1}$.

Figure 13 presents the H$\beta$ profile extracted from the composite
of the luminous and highly accreting sources. Because we see an
inflection along the blue wing of H$\beta$ (indicated in the figure)
we are attempting a two-component fit (Lorentz+Gauss and
Gauss+Gauss). This time we isolate only H$\beta$ because it is
narrow enough so that after FeII subtraction its wings go down to
the defined underlying continuum. We also assumed that a narrow
component may be present. The solution where the ``classical'' BLR
is modeled with a Lorentzian function provides a lower $\chi^2$
(better by a factor 1.6 relative to the Gaussian choice). The
blueshift of the second component is about -1000 km s$^{-1}$ and its
contributes $\sim$ 38\% to the total broad H$\beta$ emission. In the
case of Gauss+Gauss the second component is blueshifted by $\sim$
-600 km s$^{-1}$ and has a larger contribution$\sim62\%$. We note
that in the fitting procedure presented in Figure 13, even though we
did not constrain the components to have zero-level baseline, the
returned solution is still reasonable.

\section{Discussion}

We presented a dual approach in exploring the BLR properties in low
z quasars: using composite/median spectra and defining a diagnostic
set of line profile measures in individual sources.

The first approach allowed us to reveal that the adopted concept of
two populations A and B is supported by large differences in the
line profiles of the H$\beta$ emission lines (Figures 3 and 4).
Comparison of two representative bins defined in the context of 4DE1
clearly illustrates that Population B sources typically require an
additional redshifted component to describe the total H$\beta$
profile, while in Population A sources H$\beta$ lines are more
symmetric and are best described by a single component. Individual
H$\beta$ profiles on the other hand show a large diversity
displaying shoulders and bumps (Figure 5), which are not seen in
composite/median spectra. This may be telling us that there is a
stable emission component signature (seen in composites) and a
variable one (reflected in the large differences between individual
and average spectra).

One of the most important results explored is that low accreting
sources (representative of the so-called Population B) show a
typical red asymmetric and redshifted Balmer profile at 1/4
fractional intensity, while the highly accreting sources (found in
population A) typically show a blue asymmetric and blueshifted
H$\beta$ at the profile's base.

\subsection{Very Broad Line Region}

The red asymmetry in Population B can be linked to the proposed
existence of a distinct emitting region, the so-called VBLR
\citep[e.g.,][]{Marziani09}. In that study it is shown that the VBLR
contribution to the total Balmer broad emission scales with the
source luminosity (see also \citealt{Netzer07}) and becomes
increasingly dominant at higher redshift. \citet{Marziani09} also
suggest that the asymmetry index A.I.(1/4) correlates with the BH
mass for the intermediate redshift sources (more luminous than the
SDSS sample). We plot in Figure 14 the asymmetry index A.I.(1/4) and
the normalized c(1/4) centroid shift versus BH mass (panels (a) and
(b), respectively), indicating the two populations A and B
(separated at FWHM(H$\beta$) = 4000 km s$^{-1}$) with distinct
symbols. All together (A+B) our low z sample does not show a
correlation between the asymmetry/shift and BH mass. However, there
may be two distinct trends for the two populations A and B
separately. For Figure 14a statistical tests Pearson and Spearman
indicate for Population A r$_{P}$ $\simeq$ -0.14 (P$\sim$ 0.02) and
r$_{S}$ $\simeq$ -0.17 (P$\sim$ 0.006) and for Population B r$_{P}$
$\simeq$ 0.28 (P$\sim$ 3.6 $\times$ 10$^{-5}$), r$_{S}$ $\simeq$
0.29 (P$\sim$ 1.9 $\times$ 10$^{-5}$). For Figure 14b we get for
Population A: r$_{P}$ $\simeq$ -0.08 (P$\sim$ 0.21) and r$_{S}$
$\simeq$ -0.09 (P$\sim$ 0.15); for Population B: r$_{P}$ $\simeq$
0.30 (P$\sim$ 8.4 $\times$ 10$^{-6}$), r$_{S}$ $\simeq$ 0.35
(P$\sim$ 2.3 $\times$ 10$^{-7}$). The statistical tests seem to
indicate that only Population B sources may show weak correlations
between asymmetry/centroid shift at 1/4 fractional intensity and the
BH mass.

If there is indeed a distinct component present in population B
sources then the magnitude of its redshift, typically 1000-2000 km
s$^{-1}$ is unlikely to be explained by gravitational redshift
(\citealt{Marziani09}, but see also \citealt{Corbin95}). In Figure
15 we consider the subset of N=249 sources for which we measured a
positive centroid shift (redshift) at 1/4 fractional intensity
(which could be interpreted as a signature of infalling material).
The predicted gravitational redshift \citep{Popovic95}:
$$\Delta\lambda\approx\lambda_{o}\frac{GM}{c^{2}R}~~~~~(5)$$
is much lower than the measured value for virtually all sources.
Herein $\lambda_{o}\equiv4862.7\AA$ is the vacuum wavelength of
Balmer H$\beta$, c is the vacuum speed of light and G is the
gravitational constant and the ratio of the BH mass (M) and the BLR
radius (R) for a given source is estimated as in \citet{Sulentic06}
(their formulas 4 and 7):
$$\frac{M}{R} \approx 1.564 \cdot 10^{23} \times \left(\frac{FWHM(H\beta)}{1000 km~s^{-1}}\right)^{2} (g~cm^{-1})~~~~~(6)$$

A pertinent argument in favor of a separate and distinct emitting
region is the existence of clear inflections on the red wings of the
H$\beta$ profiles in many individual sources (see for example the
case of B2 1721+34 in \citealt{Corbin97}). In Figure 16 we show an
interesting case, SDSS J230443.47-084108.6, for which we have
available recorded spectra at two different epochs, about eight
years apart. The top panel shows the spectrum of the H$\beta$ region
observed in May 1993 \citep{Marziani03a} and the bottom panels show
the H$\beta$ and H$\alpha$ spectral regions of the same source,
provided by SDSS, observed in December of 2001. Not only do we see a
significant change in the functional shape of the Balmer profiles,
but we note the dramatic change on the red side of these lines. In
the SDSS spectrum we see a clear inflection on the red wing of both
H$\beta$ and H$\alpha$ (see also the detailed monitoring study and
complex behavior of NGC 4151 as presented in
\citealt{Shapovalova09}). In many other cases an inflection between
the two components BLR and VBLR may or may not clearly show up in
the line profiles. The two components may be structured differently
and apparently they do not respond in tandem to the variations in
the continuum (see the interesting case of PG1416-129 in
\citealt{Sulentic00c}), nor do they respond similarly to luminosity
variations \citep{Marziani09}. The search for a clear inflection on
the H$\beta$ red wing is further complicated by two additional
emission features: i) optical FeII has a nontrivial contribution in
this region especially due to its multiplet m42 components and ii)
possible contamination by HeI broad permitted emission lines that
may contribute to the build-up of a red-shelf for both H$\beta$ and
H$\alpha$ \citep{Veron02}, as mentioned earlier.

However, the presence/absence of inflections is not necessarily
offering us a definitive answer for or against multiple components
of the broad lines. The inflections are sensitive to kinematics
and/or orientation.

\subsection{Disc Wind}

The blue asymmetry/blue-shifted signature in high accretors
(Population A sources) could indicate an additional wind
contribution. In such highly accreting sources the presence of
radiation driven winds is very likely
\citep[e.g.,][]{King03,Komossa08}. We note that this potential
second contribution on the blue wing of H$\beta$ becomes more
clearly inflected relative to the classical BLR when the [OIII]
lines are weak and show a significant blueshift, too (see the
examples in Figure 5, where the [OIII] shift is indicated and see
also the more extreme case shown in Figure 17, where the [OIII]
blueshift is $\approx$ 1500 km s$^{-1}$ and the inflection on the
blue wing of H$\beta$ is unquestionable). A kinematic linkage
between the [OIII] lines and the HIL line CIV blueshift was
suggested by \citet{Zamanov02}. The same kind of linkage may exist
between the [OIII] lines and the LIL Balmer lines. This idea is
further reenforced by our Figure 18. We show that there is a
correlation between the shift of CIV (rest frame $\lambda$1549\AA)
at 1/2 fractional intensity and the normalized shift of H$\beta$ at
1/4 fractional intensity (see also similar comparisons in
\citealt{Marziani96}). The data in the plot represent the common set
of sources of the current sample of SDSS quasars and the sample
explored spectroscopically in \citet{Sulentic07}. Although the size
of the overlapping sample is rather small (n=25) there is an
important observation: there is no source showing a CIV redshift and
an H$\beta$ blueshift. We also note that a comparison between the
Atlas presented in \citet{Marziani03a} for the H$\beta$ and the
sample presented in \citet{Sulentic07} for CIV reveals an overlap of
n=87 sources. 69 of those sources ($\sim$ 80\%) show a shift of CIV
that is larger than the shift in the Balmer line (at both 1/4 and
1/2 fractional intensity). This kind of empirical evidence suggests
that the blue asymmetry/shift of H$\beta$ might be produced by the
same mechanism that is responsible for CIV shift, and the simplest
interpretation may be a wind/outflow
\citep[e.g.,][]{Murray95,Kollatschny03,Young07}.

The shift of quasar broad emission lines (relative to their
restframe) has been intensively studied
\citep[e.g.,][]{Gaskell82,Wilkes84,Carswell91,Tytler92,Corbin95,McIntosh99,Vandenberk01,Richards02}.
It has been reported that the amplitude of the shift is correlated
with the ionization potential
\citep[e.g.,][]{McIntosh99,Vandenberk01}. The aforementioned
comparison between the shift of the H$\beta$ and CIV (based on the
Atlas of \citealt{Marziani03a} and the sample of quasars from
\citealt{Sulentic07}) supports it.

\subsubsection{Other Arguments for the Wind Scenario}

\citet{Kallman86} argued that when HILs are systematically
blueshifted relative to the quasar's restframe by thousands of km
s$^{-1}$ it becomes more likely that absorption lines with z$_{abs}$
$>$ z$_{em}$ will exist. Let us examine the concrete case of quasar
PG 1700+518, which shows both low and high ionization absorptions
\citep{Laor02,Pettini85} and a clearly blue-asymmetric H$\beta$
profile \citep{Marziani03a}. \citet{Young07} offer arguments in
favor of a strong rotating wind in this source. This scattering wind
rises nearly vertically from the disk over a range of radii
approximately coincident with the Balmer emitting region. So, the
assertion of \citet{Kallman86} could be extended to the blueshifted
components of LILs like H$\beta$, i.e., when sources show large
blue-asymmetries in the Balmer emission lines, the BAL signatures in
MgII $\lambda$2800\AA\ or CIV $\lambda$1549\AA\ could be present as
well. In the case of PG 1700+518 it is also possible that we see the
BLR disk more face-on than equatorially \citep{Richards02} and the
blue-asymmetry is formed by scattering in a wind outflow. Another
similar case studied in detail is IRAS 07598+6508
\citep{Lipari94,Veron06}. This source shows both MgII and CIV BAL
signatures and an inflected blue wing of the H$\beta$ emission line.
Both quasars mentioned here have been interpreted as examples of
nuclear young/starburst (IRAS 07598+6508) or poststarburst (PG
1700+518) by \citet{Lipari94}.

Exploiting the analogy between microquasars, stellar-scale
phenomenal and and quasars, galactic-scale phenomena, one can also
interpret the recent study \citet{Neilsen09} as supportive for the
idea that the presence of disk winds inhibits the formation of
jetted radio structures. This may have implications for the fact
that FRII quasars and sources where disk wind signatures are more
evident occupy mutually exclusive regions in the optical plane of
the 4DE1 parameter space.

\subsection{Other Scenarios for the Shifted Broad Lines}

\citet{Richards02} offer a different view on the matter of the CIV
blueshift. They propose that the blueshift is an apparent effect
produced by obscuration of the redshifted photons. In the proposed
scenario the largest blueshifts (which may not be real, but rather
due to the lack of flux in the attenuated red wing) are explained by
an equatorial viewing angle of the line emitting region(s). An
argument of \citet{Richards02} that favors an equatorial view of the
BLR when we observe increasingly larger blueshifts of the HIL like
CIV is that the composite spectrum of the most blueshifted CIV
spectra is most similar to that of BAL quasars. This similarity
would indicate that the red wing of CIV is attenuated by obscuring
material. Another argument raised in favor of obscuration coupled
with orientation (external or internal) is the finding that the
strength (equivalent width) of the CIV line declines with increasing
blueshift. This is difficult to reconcile with the FeII emission
that is likely produced in a high density/column density environment
(flattened geometry; e.g., \citealt{Collin-Souffrin88a}). We do
observe that the average strength of the H$\beta$ broad emission
line declines by a factor two from A1 to A4 (Figures 1b-c) and the
H$\beta$ blue asymmetric and blueshifted base is typical for bins
A3/A4 (lower equivalent width W(H$\beta$)), however, this can be
explained by a quenching of the Balmer emission in high electron
density environments \citep{Gaskell00}, which favor the FeII
emission (as we see in Figures 1b-c, W(FeII) increases by the same
factor two from bin A1 through bin A4). Largest blueshifts
correspond to narrowest and strongest FeII, and we suggested that
FeII decreases from face-on because of a sort of ``limb darkening''
\citep{Marziani96}, which supports a wind scenario plus disk
obscuration, but a different orientation dependence from the one
suggested by \citet{Richards02}: the largest blueshifts are observed
when the source is viewed pole-on.

\citet{Richards02} advocate the ubiquity of the CIV blueshift. This
last result is not confirmed by the most recent study of
\citet{Sulentic07}, where only Population A quasars show such a
spectral signature. Population B sources (where the majority of RL
quasars belongs) show unshifted and symmetric CIV profiles. Most
likely the bins of least CIV blueshift in \citet{Richards02}, which
also have the largest radio-detected fraction and the largest CIV
equivalent width (see their Table 2), must have a significant
overlap with the Population B sources. It is difficult to see how
attenuation of the CIV red wing, according to the orientation
dependence suggested by \citet{Richards02}, can leave a symmetric
profile in so many quasars: population B sources account for $\sim$
60\% of our CIV sample \citep{Sulentic07}.

An alternative to the the wind interpretation of blueshifts was
recently proposed by \citet{Gaskell08}. They advocate that the
blueshift of HIL like CIV is produced by Rayleigh and/or electron
scattering off of inflowing material of the BLR. Their model
predicts that the relative blueshifts of different lines in the same
AGN should be proportional to the inverse square root of the radius
where they are produced. The results we indicated in the previous
paragraph, that the H$\beta$ blueshift is almost always less than
that of CIV are consistent with the prediction of \citet{Gaskell08}
in the sense that CIV is probably emitted at a smaller radius within
the BLR relative to the Balmer emitting region. This is
qualitatively supported also by the fact that HIL CIV line shows
broader FWHM profiles than H$\beta$ in Population A sources (but not
in B; see \citealt{Sulentic07}). Moreover \citet{Gaskell08} predict
the amplitude of the blueshift is proportional to the infall
velocity. Spectroscopic signatures of the inflow have been indicated
(see section 2.1 of \citealt{Gaskell08}, see also
\citealt{Bentz08}), although very recent reports
\citep{Denney09,Shapovalova09} indicate evidence for a very diverse
behavior of the BLR, which can show signatures of outflowing,
inflowing and virialized gas motion in different sources.

If the scattering effect depends on the wavelength only, then
low-ionization lines in UV should show a blueshift comparable to
that of the UV high-ionization lines. Published data
\citet{Tytler92} as well as a tentative analysis on several HST
spectra (Negrete et al. in preparation) do not support this
suggestion. The OI1304\AA\, for example, does not seem to show the
large blueshift of CIV and seems to be more consistent with the
quasar rest-frame as derived from the narrow low-ionization emission
lines.

Recently \citet{Hu08a,Hu08b} propose that inflow may be a ubiquitous
phenomenon as revealed by their measured (red-) shift of the FeII
optical lines. However, \citet{Hu08a,Hu08b} report that the infall
is increasingly pronounced as the accretion rate (or Eddington
ratio) decreases. This works against the model proposed by
\citet{Gaskell08} because the largest blueshifts are typically
observed in highly accreting sources (members of the Population A).
The \citet{Hu08a,Hu08b} recent papers propose that there is an
Intermediate Line Region (ILR) component in H$\beta$ that is
systematically redshifted and narrower than a so-called VBLR
rest-frame component. This is a rather different view compared to
the proposed two-component model of \citet{Sulentic02} or
\citet{Marziani09}, where the VBLR shows a significant and
systematic redshift and the classical BLR signature is basically at
the quasar's rest frame. A two-component model employing the
VBLR/ILR terminology is proposed by \citep{Popovic04}, where the
emission of the line wings comes from a disk and the line core from
a distinct emitting region that has spherical geometry. The two
emitting regions may be linked by a wind. One must be cautious about
different definitions and labels of the broad line emitting regions
used by different studies; the same name may be assigned to
different structural components.

We are not yet convinced of the \emph{ubiquity} of the FeII redshift
proposed by \citet{Hu08a,Hu08b}. Although we don't dismiss that in
sources like SDSS J094603.94+013923.6 or SDSS J101912.57+635802.7
(both shown in our Figure 5) the FeII could be redshifted, we also
encounter sources like SDSS J001224.02-102226.5\footnote{A similar
case SDSS J153636.22+044127.0 was found in SDSS DR7 by
\citet{Boroson09} and interpreted as a possible signature of a
sub-parsec binary BH system. although a number of other articles
propose alternative explanations.} or SDSS J094756.00+535000.3 that
can be difficult to fit in the scenario they propose. And we also
find an excess of blueshifted peaks in both populations A and B,
which does not support the idea of a redshifted ILR in virtually all
quasars. This is further confirmed by the composite spectrum shown
in our Figure 4 for bin B2, which should show a significantly
redshifted FeII signature, which we do not see. Moreover, we
selected the (apparently) brightest sources in their sample (psf i
or psf g $<$17.5, just as we constrained our SDSS low z sample;
\citealt{Zamfir08}). We found about 50 such sources with a claimed
FeII redshift larger than 1000 km s$^{-1}$. The authors state that
their sources are bright enough that the host galaxy plays a
negligible role. We did find sources where prominent absorption
lines indicate a host galaxy contamination, e.g., SDSS
J110538.99+020257.3. This raises the question whether the mismatch
between the FeII rest-frame template and the spectral lines is
severely affected by the dips created by an ignored host galaxy
signature. The problem has profound implications toward
understanding the link between the 4DE1 space (through R$_{FeII}$)
and its proposed main driver L$_{bol}$/L$_{Edd}$ (see the very
recent study of \citealt{Ferland09}) and thus requires further
scrutiny.

\subsection{The Relative Fraction of Population A/B}

\citet{Shen08} compile BH masses for 60,000+ Sloan quasars within
redshift z=4.5, starting from the catalog of \citet{Schneider07},
which imposes a faint limit of -22 for the absolute i-band
magnitude. In order to compare to our sample, we selected from
\citet{Shen08} (using the VizieR catalogue database,
\citealt{Ochsenbein00}) only the quasars with z$<$0.7. There are
12,003 listed sources, of which 8,979 have reported measurements of
the H$\beta$ line. We were unable to understand why $\sim$ 25\% of
the sample has no reported measures of H$\beta$. They are neither
fainter, nor noisier/more problematic spectra than the rest. The
other major criticism is the relative proportion of sources that we
would call A/B in the light of our adopted 4000 km s$^{-1}$
boundary. They report 71\% sources with FWHM(H$\beta$) $>$ 4000 km
s$^{-1}$ (of those 8,979 measured). We tested whether and how this
proportion is tied to the S/N reported by their study. We found that
as the S/N decreases there is an increased fraction of Population A
sources (i.e., 19\% for S/N$>$20, 26\% for S/N$>$15-20, 30\% for
S/N$>$5-15). All previous major studies of quasars in the context of
Eigenvector 1 space
\citep[e.g.,][]{Boroson92,Sulentic00a,Sulentic00b,Marziani03a}
report a slight numerical dominance of Population A sources (at low
z), i.e., FWHM(H$\beta$) $<$ 4000 km s$^{-1}$. Our main concern is
their adopted H$\beta$ fitting routine, i.e., a single Gaussian for
all profiles. We now know that Population A spectra are best
described by Lorentz-profiles, while for Population B sources is a
double-Gaussian model works better, due to their pronounced red
asymmetry \citep{Sulentic02,Netzer07,Marziani09}. A single Gaussian
fit would overestimate the FWHM measure in both Population A and B
sources; for instance a single Gauss fit to the A2 or B1 bin median
composites spectra (discussed earlier in \S\S~2.4) would
overestimate the FWHM of the broad H$_{beta}$ by 20-25 \%. Another
serious problem with their analysis is the inclusion of a large
number of spectra of sources that are either in a high-continuum
phase are the emission lines are intrinsically very weak (e.g., SDSS
J083353.88+422401.8, SDSS J083148.87+042939.0), where the broad
lines are almost missing. For such cases they report hugely
overestimated values of the FWHM(H$\beta$).

\section{Conclusions}

We have shown that the Population A/B concept proposed in the
context of the 4DE1 Parameter Space holds robustly when other line
diagnostic measures like centroid shifts, asymmetry and kurtosis are
considered. Taken at face value the spectral measures tell us that
populations A/B show a great deal of difference in terms of BH mass
and Eddington ratios (Figure 10). This may be indicating that the
numerous observed dissimilarities between A and B are indeed driven
by large differences in the structure and kinematics of the central
engine. Nonetheless, large differences in terms of BH mass and
Eddington ratio may be reduced. On one hand it was pointed out that
NLSy1 sources may not harbor such small BH masses as inferred from
their narrow emission lines. They might not be super-Eddington
accretors if proper correction for radiation pressure
(\citealt{Marconi08,Marconi09}, but see also \citealt{Netzer09}),
orientation \citep{Marziani03b,Decarli08} and geometry
\citep[e.g.,][]{Collin06} are made. On the other hand, if the
H$\beta$ profiles that show a large red asymmetry, which is probably
caused by a distinct VBLR redshifted contribution, are properly
corrected for such ``unvirialized'' contributions, then the
estimated BH mass gets smaller and thus the measured Eddington ratio
gets higher. The large variance of the 4DE1 space would be
significantly reduced from both sides of the accretion rate
sequence. As a consequence, the FWHM measure, which is an important
optical dimension within 4DE1, may be replaced with A.I.(1/4) or the
normalized c(1/4)/FW025M centroid shift. They correlate at least
equally well with R$_{FeII}$ (Figure 9). Nonetheless, incorporating
these two new measures in the 4DE1 is not an easy task, as they also
manifest (at least in Population B) a sensitivity to luminosity
\citep{Netzer07,Marziani09}, which was indicated as part of the
Eigenvector 2 space \citep{Boroson92}, formally orthogonal to
Eigenvector 1. Moreover, the Population B H$\beta$
asymmetry/centroid shift at the base (1/4 fractional intensity)
appears weakly correlated with the BH mass (Figure 14; see also
\citealt{Marziani09}).

We also presented several arguments that two populations of quasars
may be redefined based on an R$_{FeII}$ nominal boundary of 0.50.
This could be seen as an alternative empirical definition of some
equivalent Populations A and B. The lines of evidence are as
follows: 1) R$_{FeII}$=0.50 seems a critical value that best
isolates RL quasars from the rest of quasars (recall Figure 4 from
\citealt{Zamfir08}), 2) around this measured value of R$_{FeII}$ the
centroid shift and the asymmetry of H$\beta$ change sign (Figures
9a-b), 3) a plot of the line shape kurtosis index K.I. versus
R$_{FeII}$ shows two different trends on either side of R$_{FeII}$
$\sim$ 0.50 (Figure 9c), likely separating low from high accreting
sources (Figure 12d) and 4) a tight correlation between R$_{FeII}$
and the Eddington ratio indicates that R$_{FeII}$ $\sim$ 0.50
roughly corresponds to $L_{bol}/L_{Edd}\sim0.18$ (Figure 9d),
consistent with several other studies that suggest a critical change
in the structure/geometry/kinematics in the quasar ``engine'' around
that value. The red and blue H$\beta$ base asymmetries observed at
the opposite ends of the Eddington ratio sequence may have very
different origins.

We also find that our measures for low accretors (typically members
of Population B) seem to match the predictions of accretion disk
models, i.e., blueshifted peaks c(9/10) and redshifted/red
asymmetric bases c(1/4) and A.I.(1/4) (Figure 12b). This allows us
to suggest a typical flattened BLR geometry for all Population B
quasars, not only for the RL quasars, whose majority is seen in
Population B subspace. We emphasized that our conclusions are not
driven by radio-loudness. Focusing our investigation on the
radio-quiet subsample the results are similar to what we get using
the whole sample of N=469 quasars.

Although the sample of FRII sources is relatively small (N=44), we
find a very interesting correlation between the core:lobe flux
density ratio, which may be seen as an orientation indicator (based
on the 1.4GHz FIRST survey measures) and the redshift of the
H$\beta$ base (Figure 19). The more core-dominated FRII sources show
a more pronounced redshift at 1/4 fractional intensity (see also
Figures 9a-b and related commentaries). We may just speculate that
the excess of redshifted photons could be a signature of an inflow
feeding the central mass and allowing for a jetted
burst\footnote{See the interesting case of the microquasar
GRS1915+105 presented by \citealt{Neilsen09}. Their study suggests
that accretion disk winds prevent the jetted outbursts and
viceversa.}. A larger sample should be investigated.

\section*{Acknowledgments}

Funding for the SDSS and SDSS-II has been provided by the Alfred P.
Sloan Foundation, the Participating Institutions, the National
Science Foundation, the U.S. Department of Energy, the National
Aeronautics and Space Administration, the Japanese Monbukagakusho,
the Max Planck Society, and the Higher Education Funding Council for
England. The SDSS Web Site is http://www.sdss.org/. The SDSS is
managed by the Astrophysical Research Consortium for the
Participating Institutions. The Participating Institutions are the
American Museum of Natural History, Astrophysical Institute Potsdam,
University of Basel, University of Cambridge, Case Western Reserve
University, University of Chicago, Drexel University, Fermilab, the
Institute for Advanced Study, the Japan Participation Group, Johns
Hopkins University, the Joint Institute for Nuclear Astrophysics,
the Kavli Institute for Particle Astrophysics and Cosmology, the
Korean Scientist Group, the Chinese Academy of Sciences (LAMOST),
Los Alamos National Laboratory, the Max-Planck-Institute for
Astronomy (MPIA), the Max-Planck-Institute for Astrophysics (MPA),
New Mexico State University, Ohio State University, University of
Pittsburgh, University of Portsmouth, Princeton University, the
United States Naval Observatory, and the University of Washington.

This research has made use of the NASA/IPAC Extragalactic Database
(NED) which is operated by the Jet Propulsion Laboratory, California
Institute of Technology, under contract with the National
Aeronautics and Space Administration. This research has made use of
the VizieR catalogue access tool, CDS, Strasbourg, France.

\clearpage

\begin{figure*}
\centering
\includegraphics[width=1\columnwidth,clip=true]{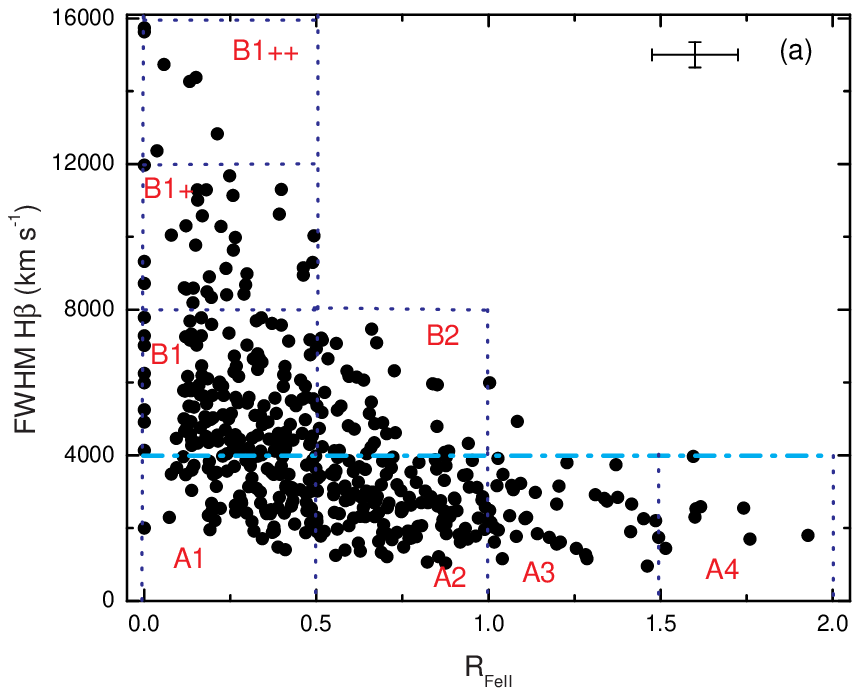}\\
\includegraphics[width=1\columnwidth,clip=true]{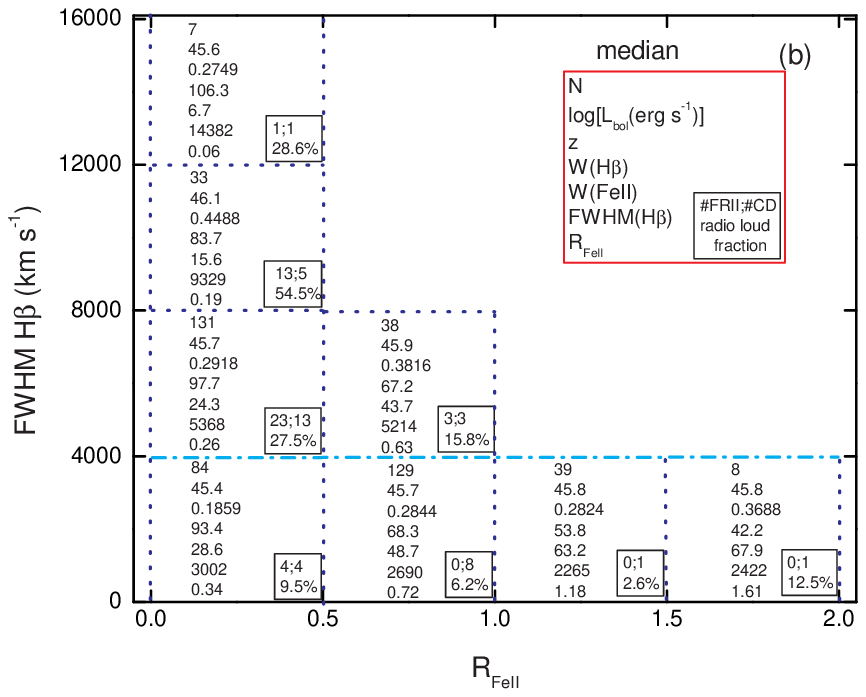}\\
\includegraphics[width=1\columnwidth,clip=true]{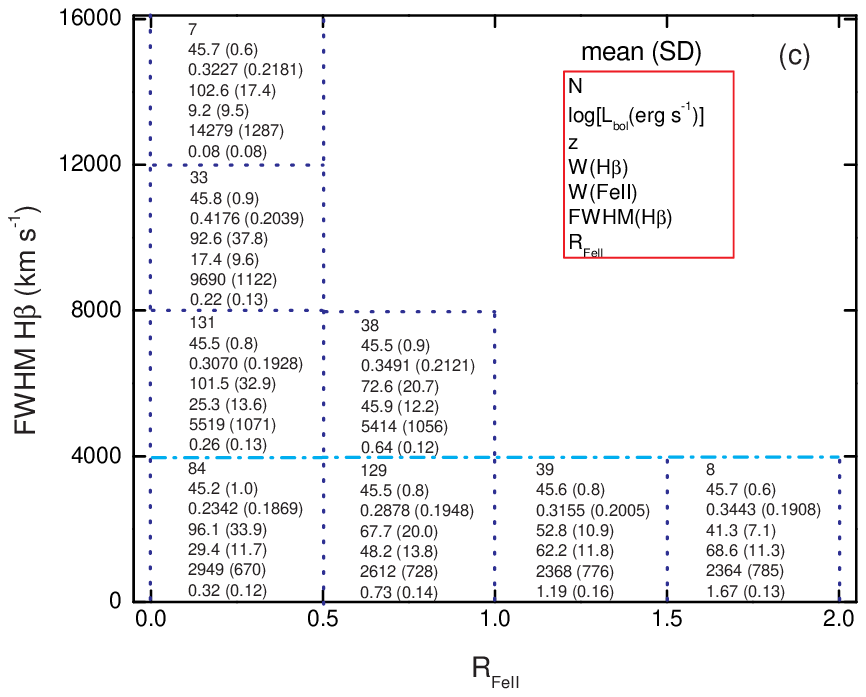}
\caption {(a) The optical plane of the 4D Eigenvector 1 Parameter
Space. The quadrants indicate the bins defined by \citet{Sulentic02}
and adopted here for composite spectra. Typical 2$\sigma$ error bars
are shown in the upper right corner. The horizontal dash-dot line
indicates the nominal boundary between Populations A and B. The
relative numerical contribution of from each population is 55\% in A
and 45\% in B. (b) The number of sources considered in each bin
along with a few (median) relevant measures in each bin are
indicated. The square in the upper right is the legend of the
numbers shown in each bin. Inside each bin (lower-right square) we
report the number of FRII and core-dominated RL sources along with
the estimated radio-loud fraction. (c) Similar to (b), this time the
numbers represent mean and standard deviations (in the parenthesis
in each case).}
\end{figure*}
\clearpage

\begin{figure*}
\centering
\includegraphics[width=1.8\columnwidth,clip=true]{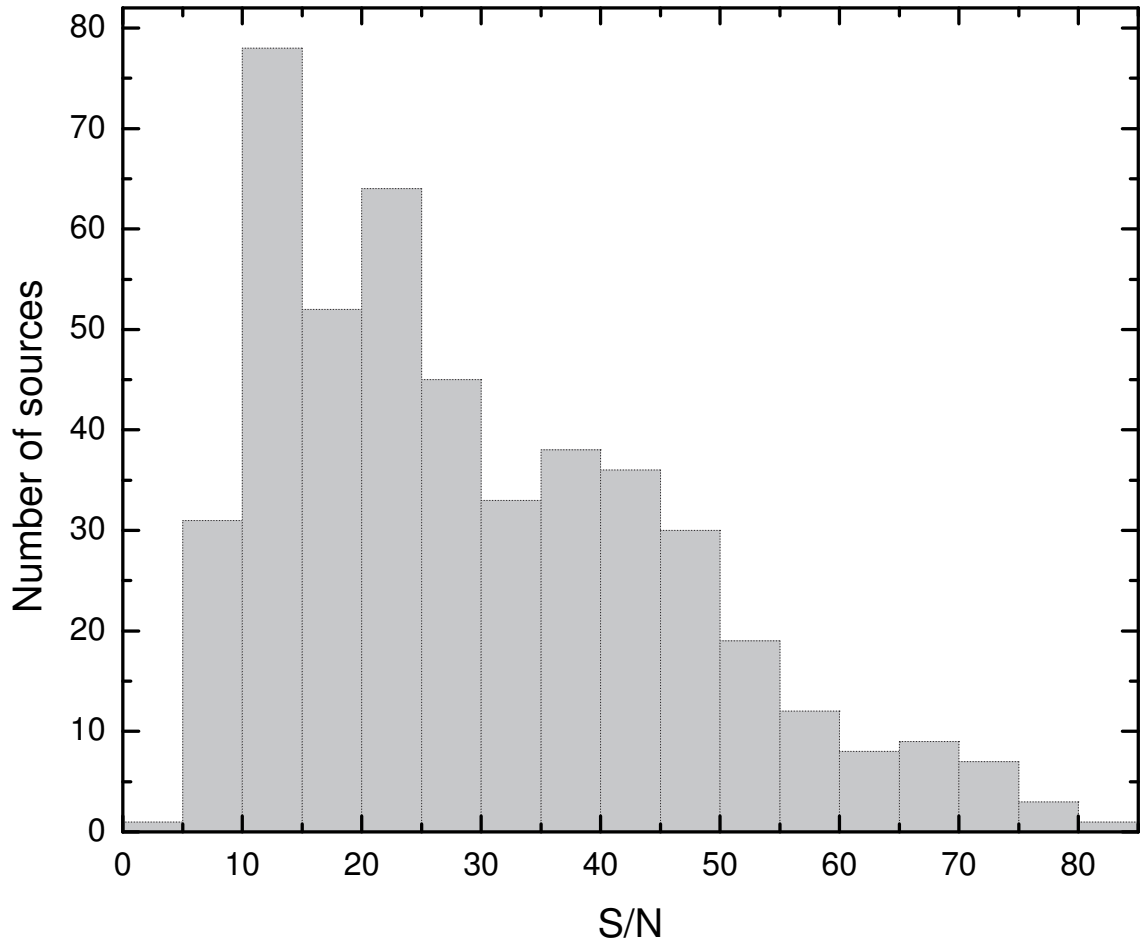}
\caption{Distribution of the S/N measure (per pixel) in the
5600-5800 \AA\ region for our sample of N=469 sources.}
\end{figure*}
\clearpage

\begin{figure*}
\centering
\includegraphics[width=1.62\columnwidth,clip=true]{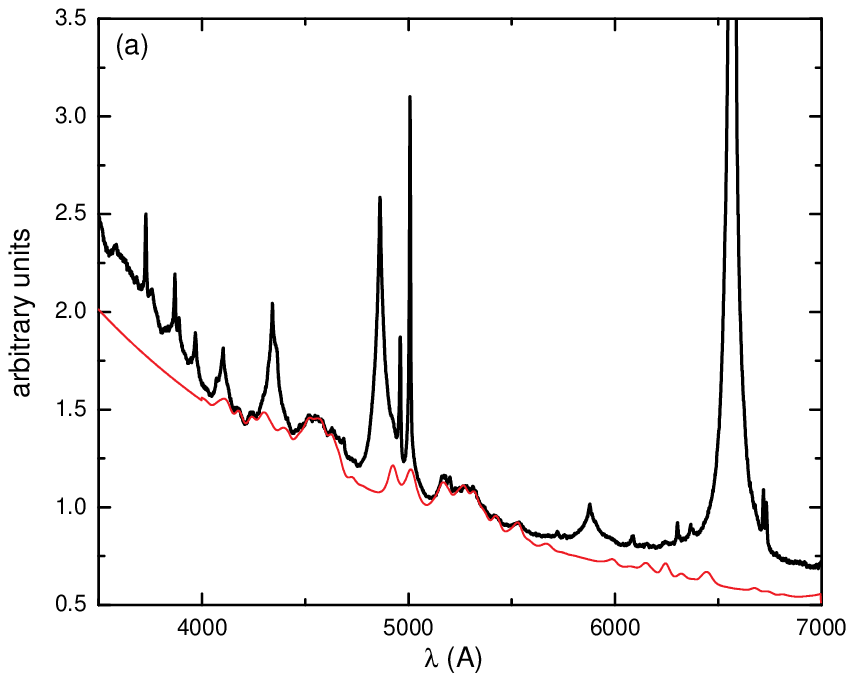}
\vspace*{0.5cm} \hspace*{0.3cm}
\includegraphics[width=1.46\columnwidth,clip=true]{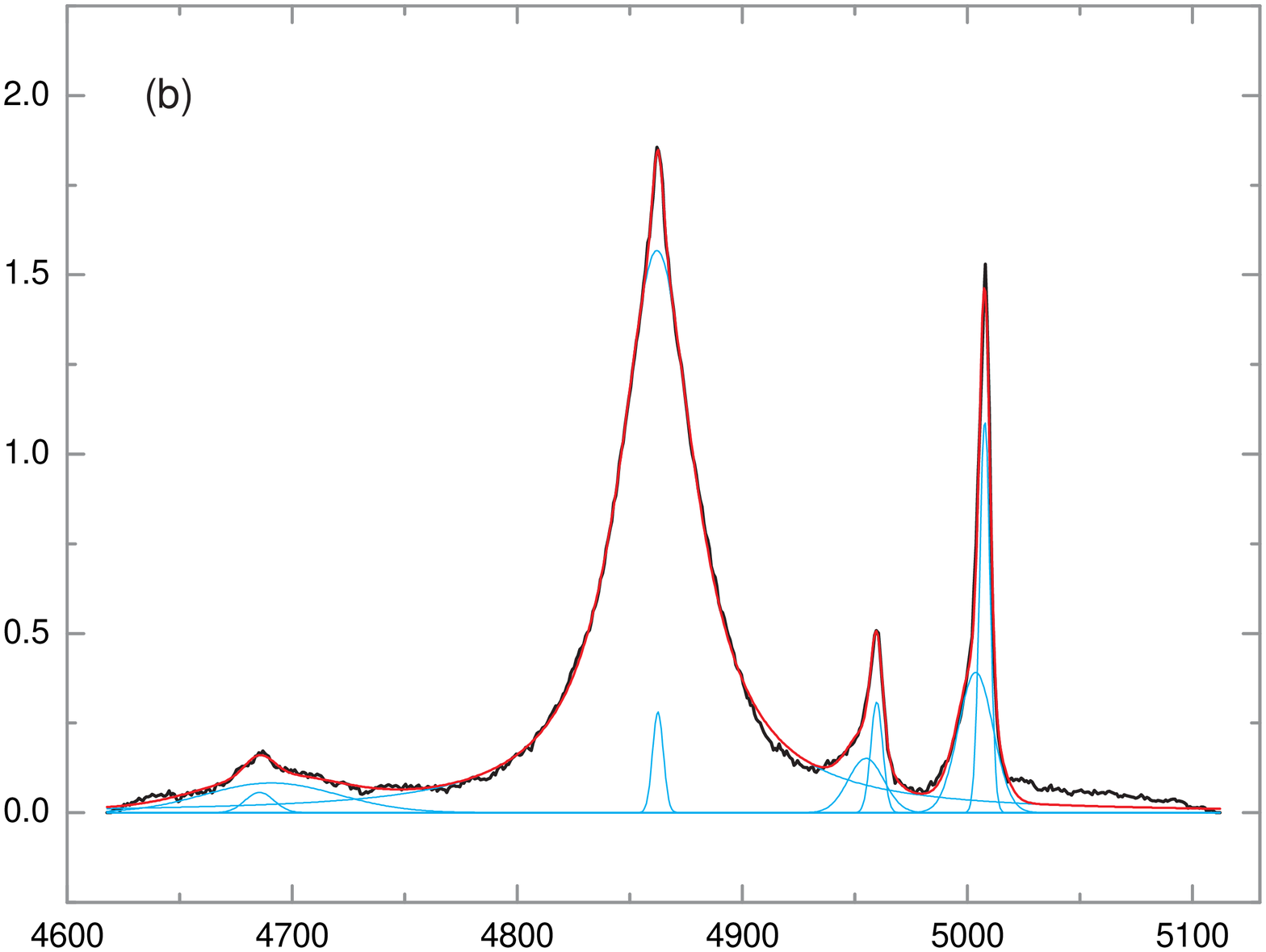}
\caption{(a) Median composite spectrum of A2 bin sources and the
best model of FeII emission around H$\beta$. (b) A simultaneous fit
(our best solution) of H$\beta$ and adjacent lines in the selected
wavelength interval of the composite.}
\end{figure*}
\clearpage

\begin{figure*}
\centering
\includegraphics[width=1.62\columnwidth,clip=true]{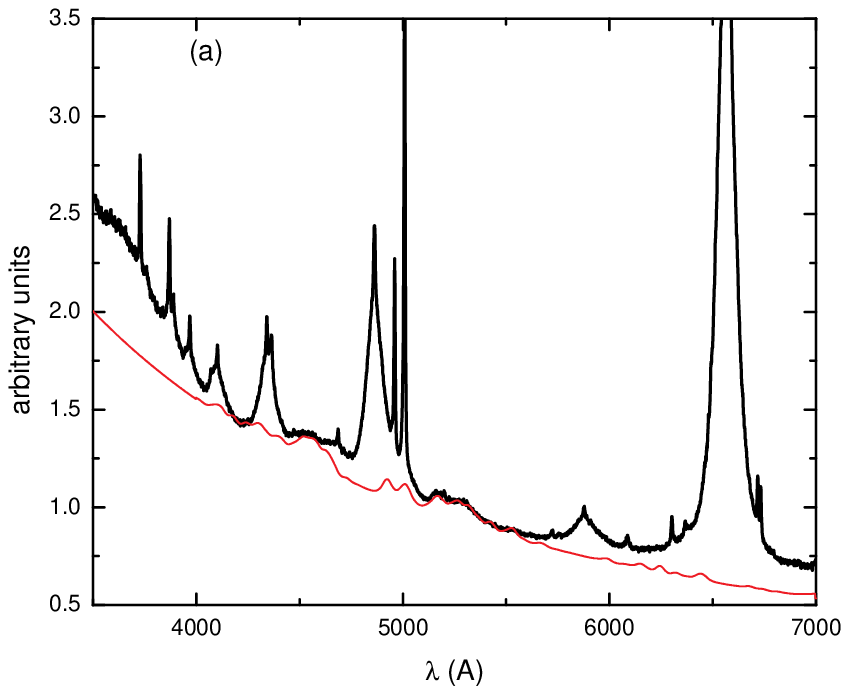}
\vspace*{0.5cm} \hspace*{0.3cm}
\includegraphics[width=1.49\columnwidth,clip=true]{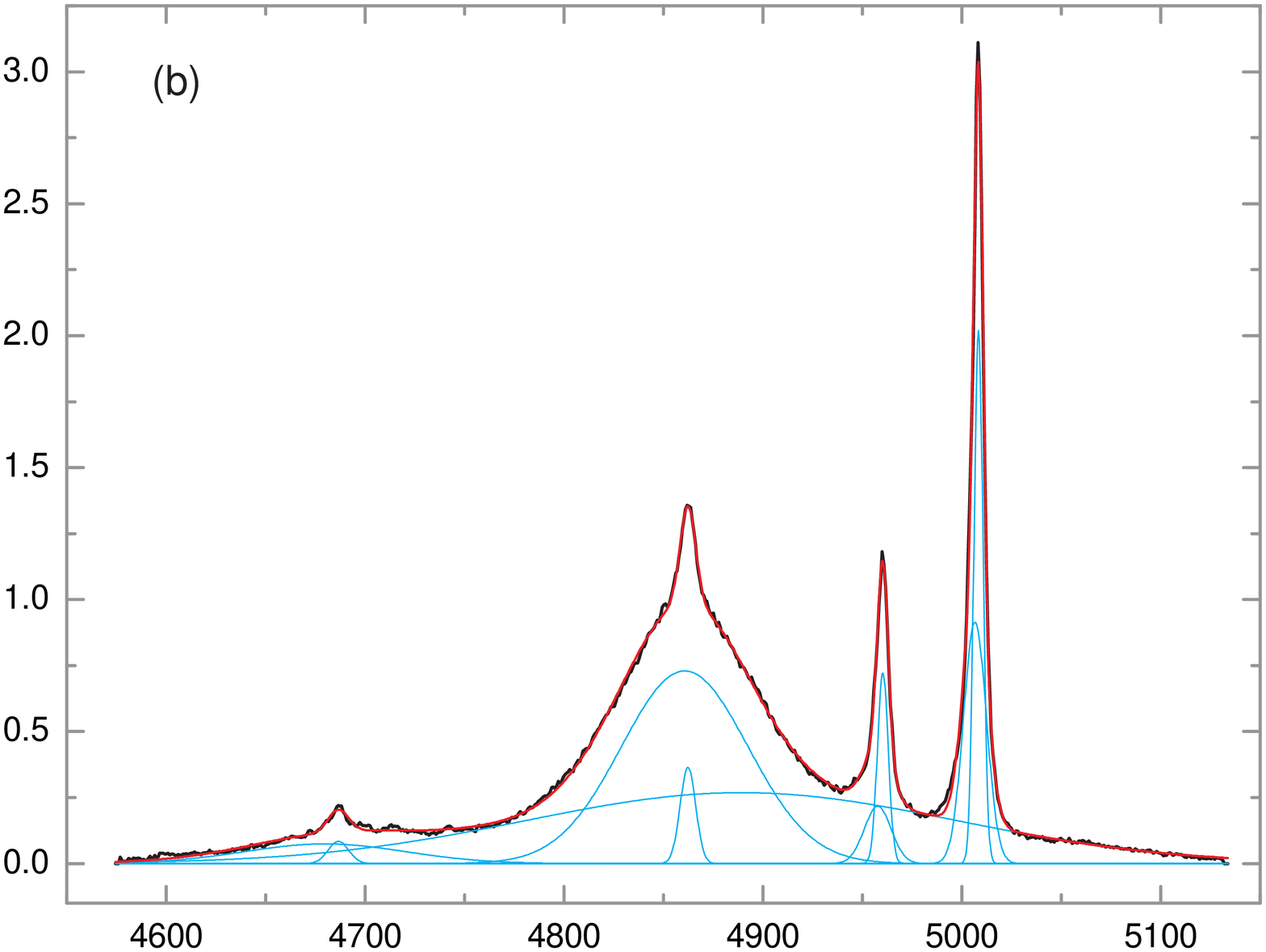}
\caption{(a) Median composite spectrum of B1 bin sources and the
best model of FeII emission around H$\beta$. (b) A simultaneous fit
(our best solution) of H$\beta$ and adjacent lines in the selected
wavelength interval of the composite.}
\end{figure*}
\clearpage

\begin{figure*}
\centering
\includegraphics[width=0.9\columnwidth,clip=true]{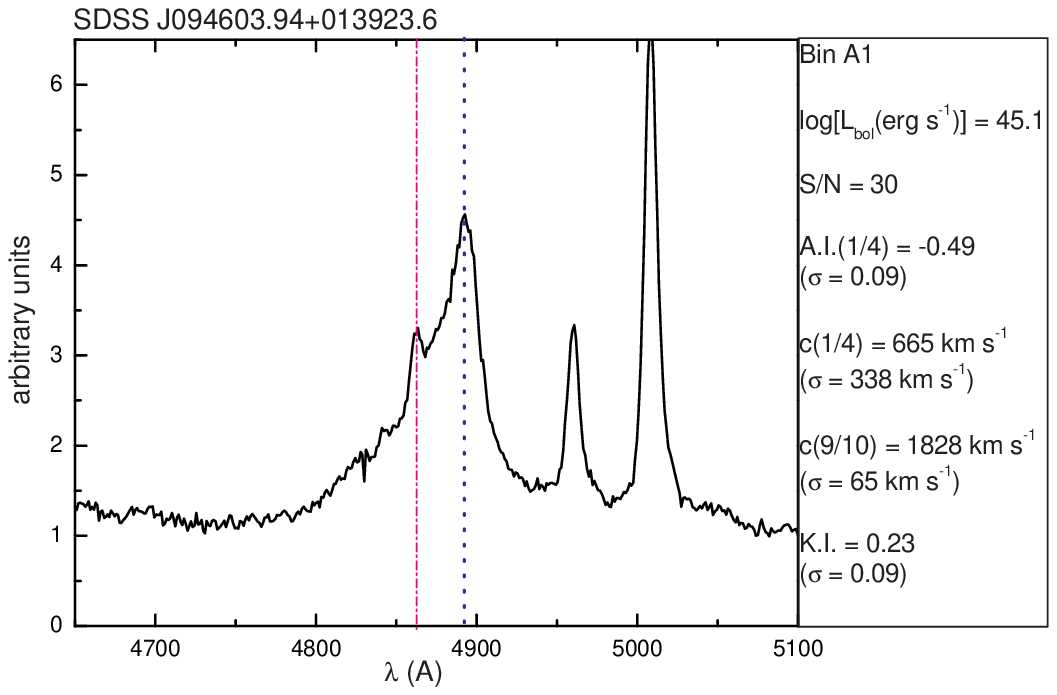}
\includegraphics[width=0.9\columnwidth,clip=true]{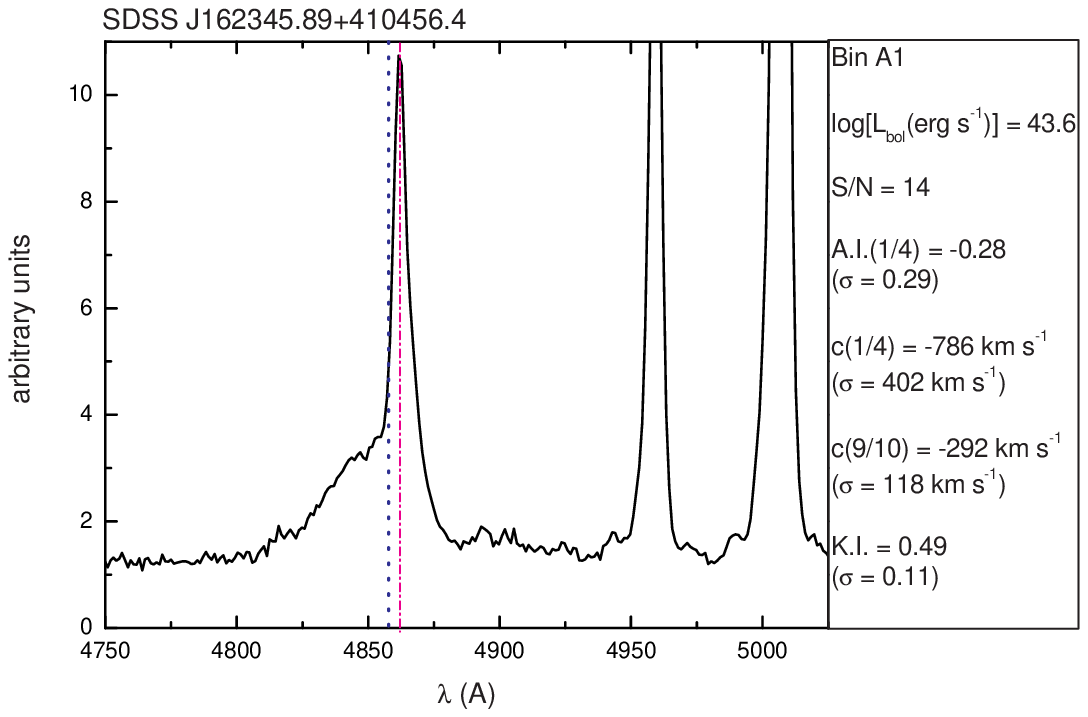}
\includegraphics[width=0.9\columnwidth,clip=true]{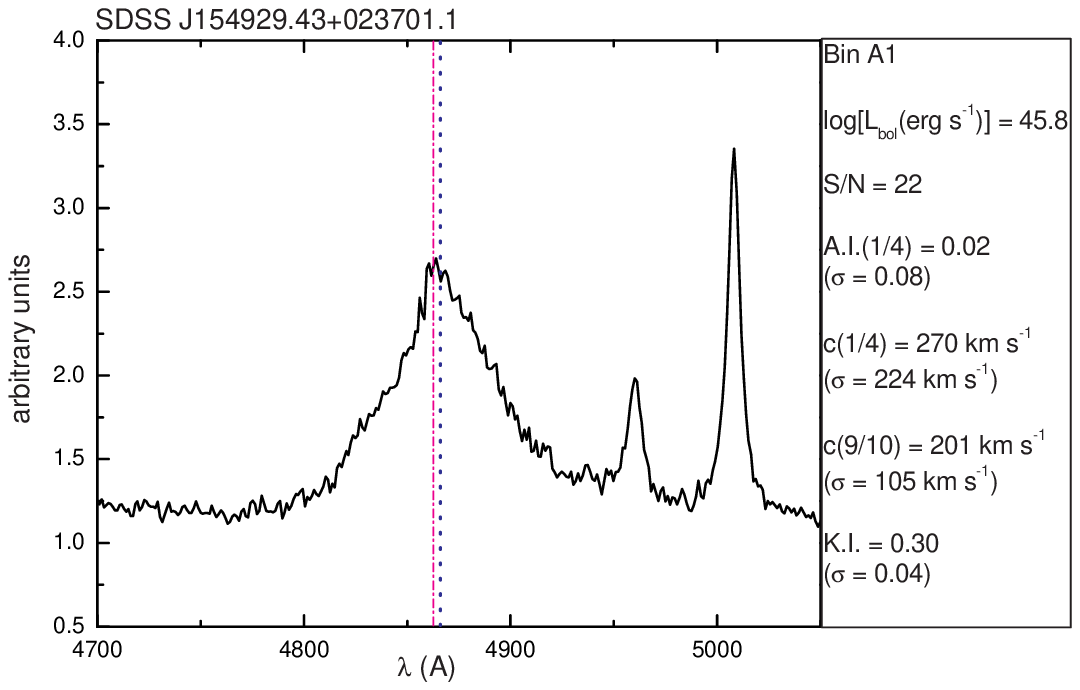}
\includegraphics[width=0.9\columnwidth,clip=true]{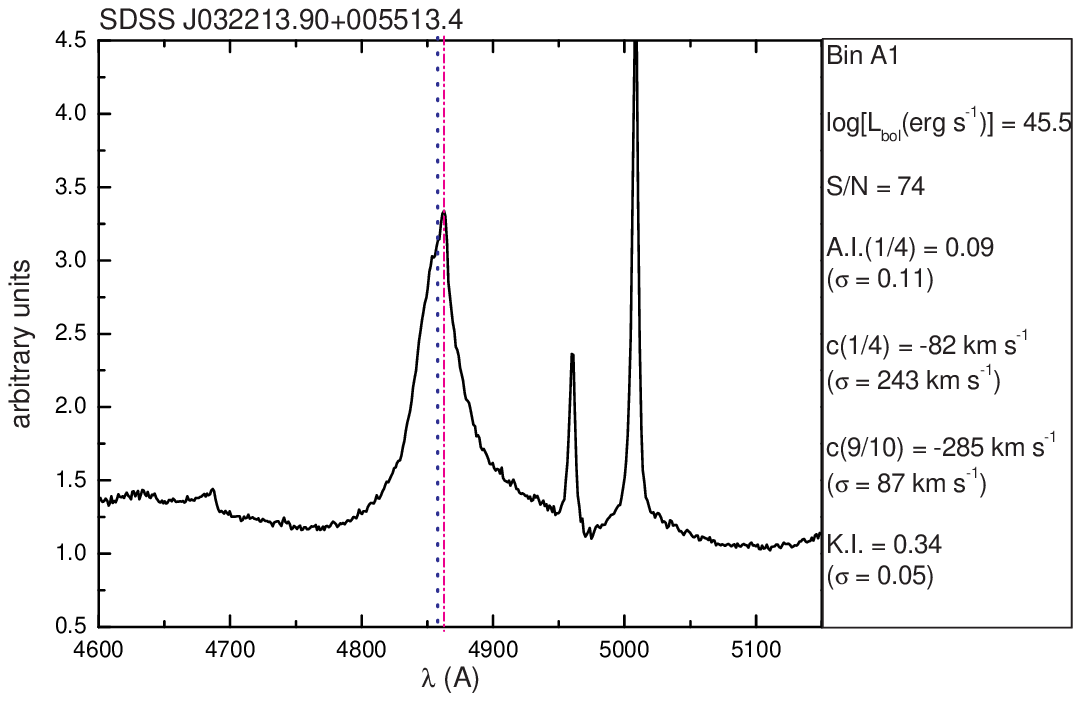}
\includegraphics[width=0.9\columnwidth,clip=true]{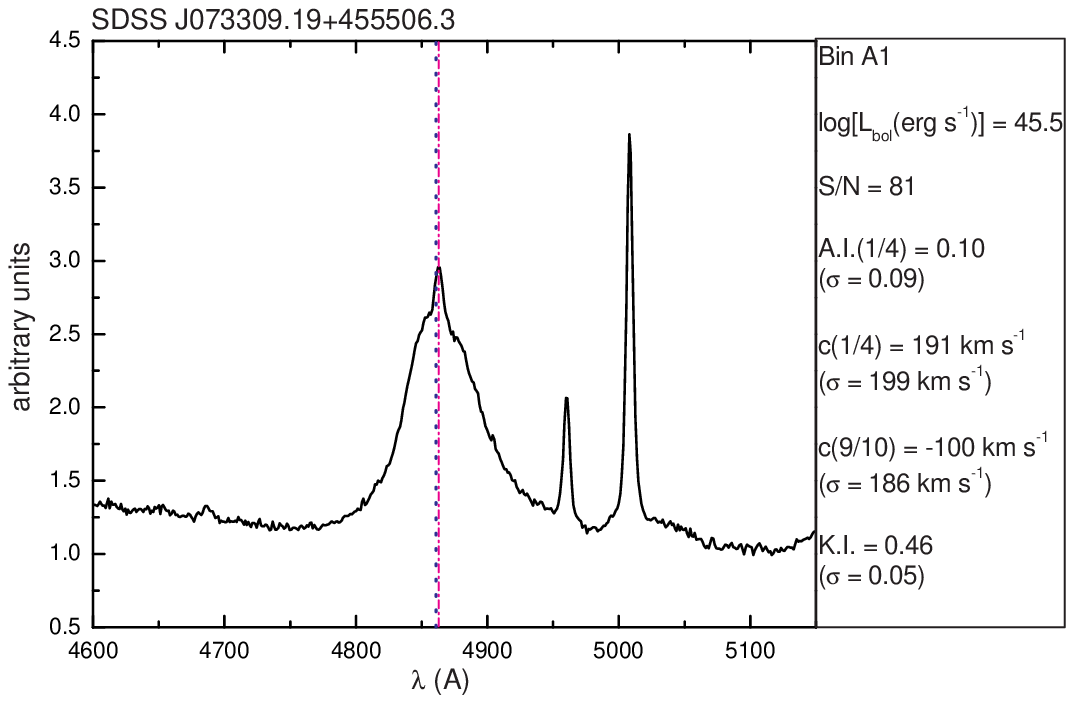}
\includegraphics[width=0.9\columnwidth,clip=true]{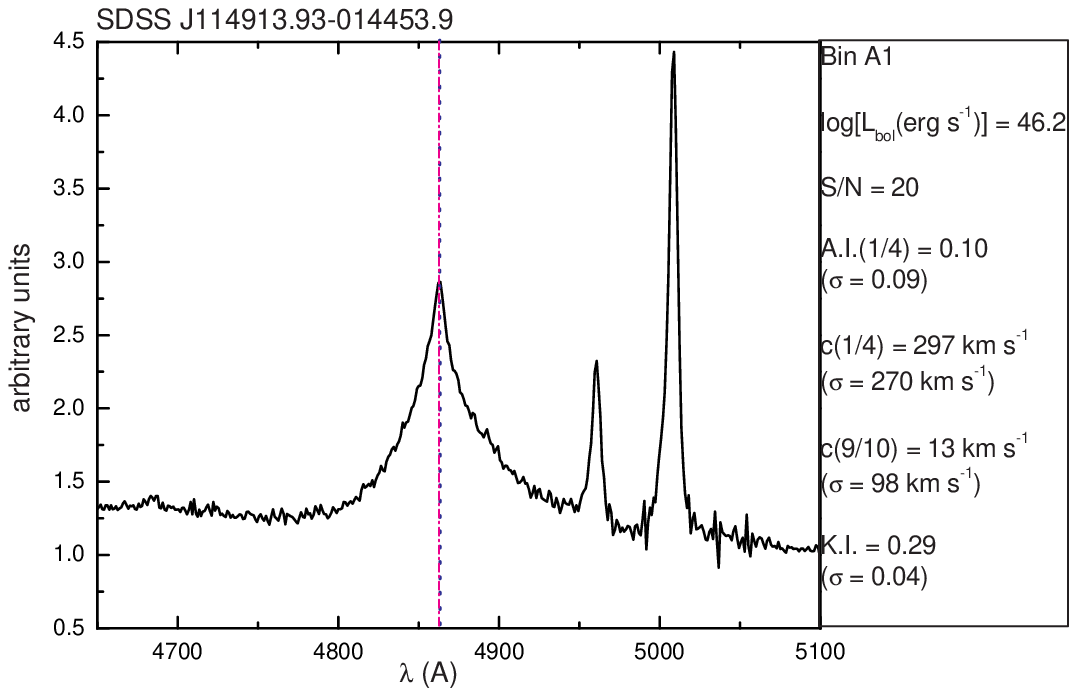}
\includegraphics[width=0.9\columnwidth,clip=true]{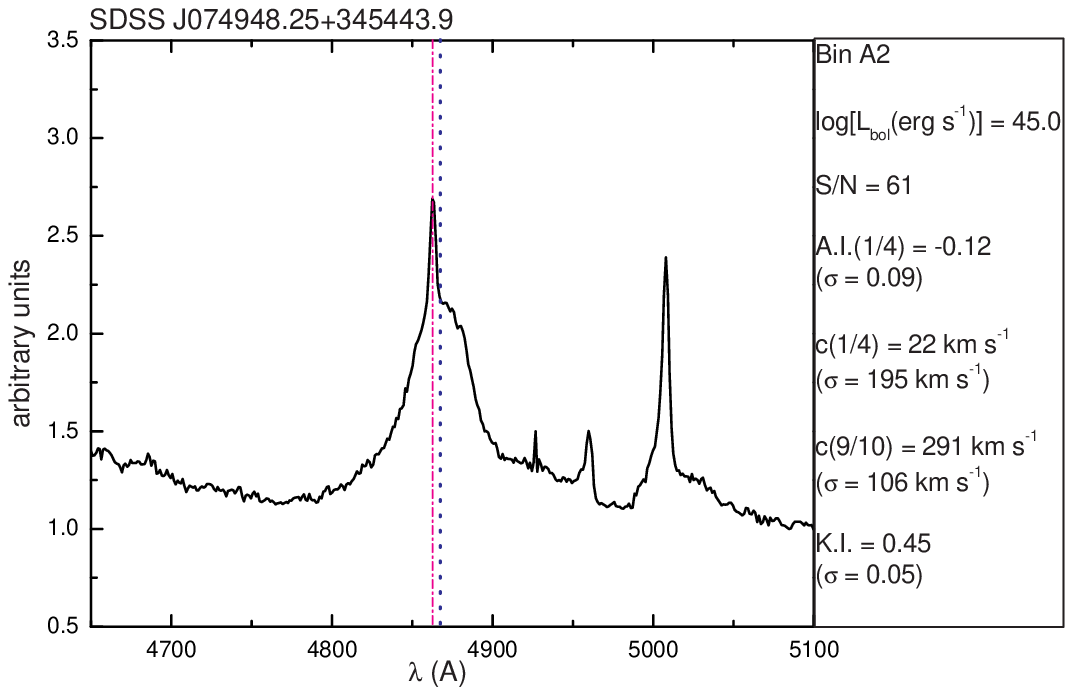}
\includegraphics[width=0.9\columnwidth,clip=true]{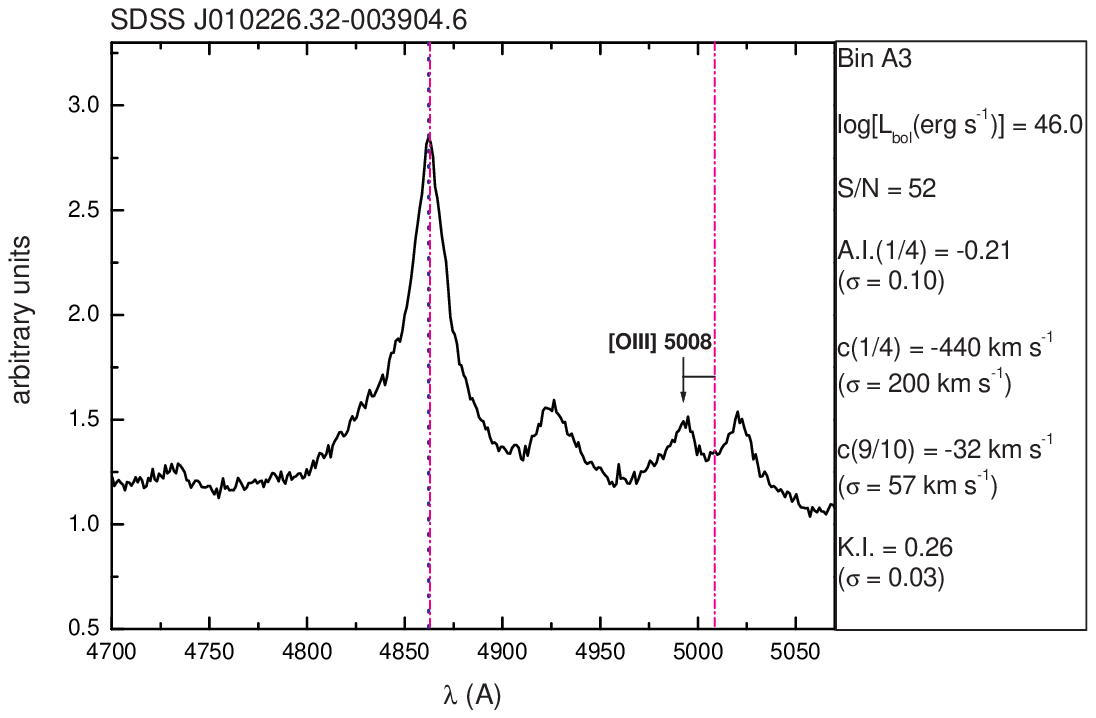}
\caption{A selection of H$\beta$ profiles meant to illustrate the
diversity of asymmetries, shapes and shifts in individual sources. A
magenta dash-dot vertical line in each case indicates the rest-frame
defined by H$\beta$ NC at $\lambda$4862.7\AA\ (Note that we refer
our measures to vacuum rest-frame wavelength values). A blue dotted
vertical line indicates the peak of the broad H$\beta$ defined at
9/10 fractional intensity. In a few panels a third vertical line
indicates the rest-frame [OIII] $\lambda$5008.2\AA\ and the actual
blueshifted position of that line. The displayed spectra still
include the FeII contribution, but the H$\beta$ measures employed in
the paper are performed on FeII-``cleaned'' profiles. Panels are
grouped by bins and displayed in order of increasing asymmetry index
A.I.(1/4) within each bin. In each case we also provide a set of
numbers: bin, bolometric luminosity, S/N measures in the continuum
(see Figure 2), asymmetry and kurtosis indices, centroid shift at
1/4 and 9/10 fractional intensity.}
\end{figure*}
\clearpage

\begin{figure*}
\centering\setcounter{figure}{4}
\includegraphics[width=0.9\columnwidth,clip=true]{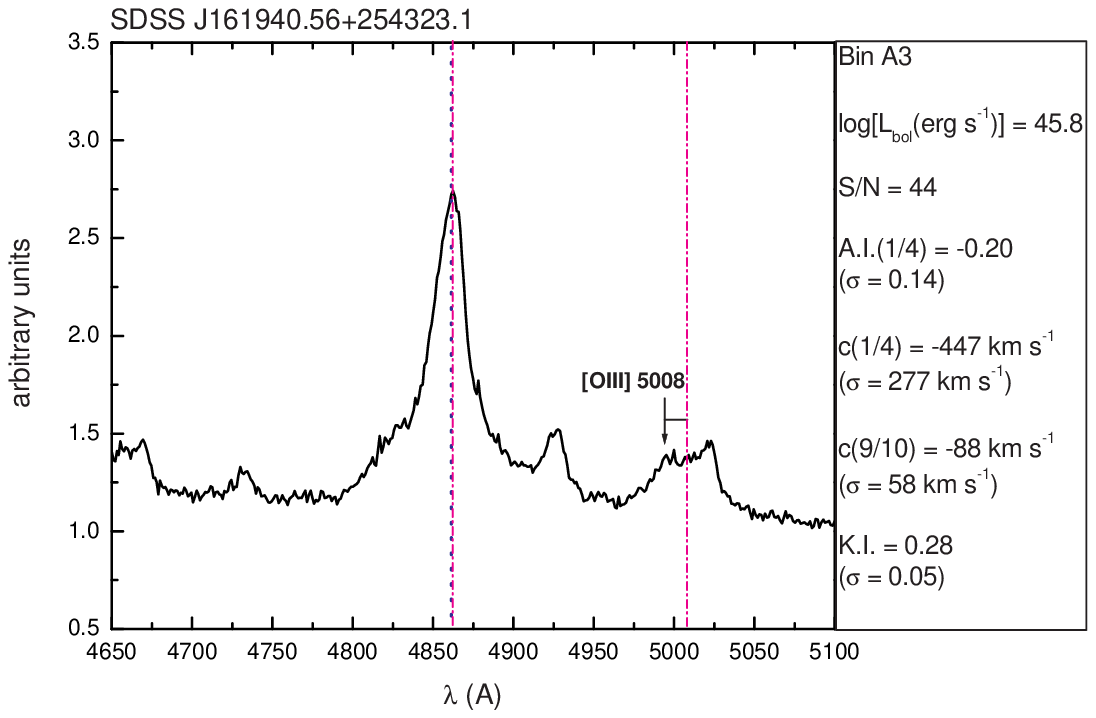}
\includegraphics[width=0.9\columnwidth,clip=true]{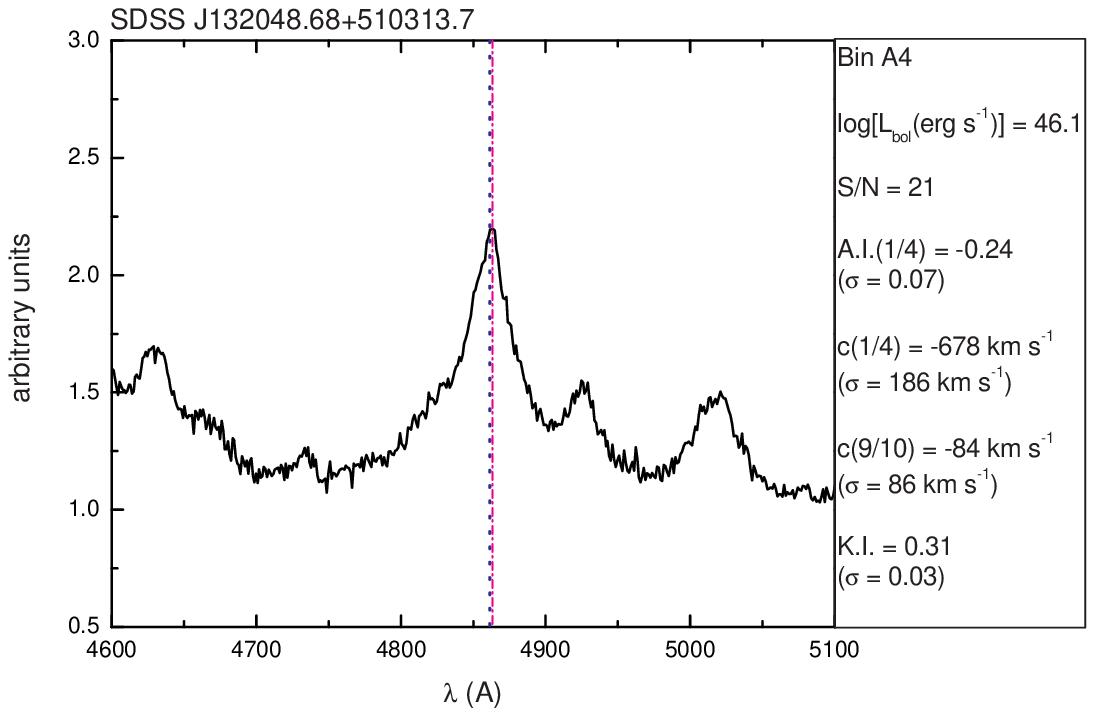}
\includegraphics[width=0.9\columnwidth,clip=true]{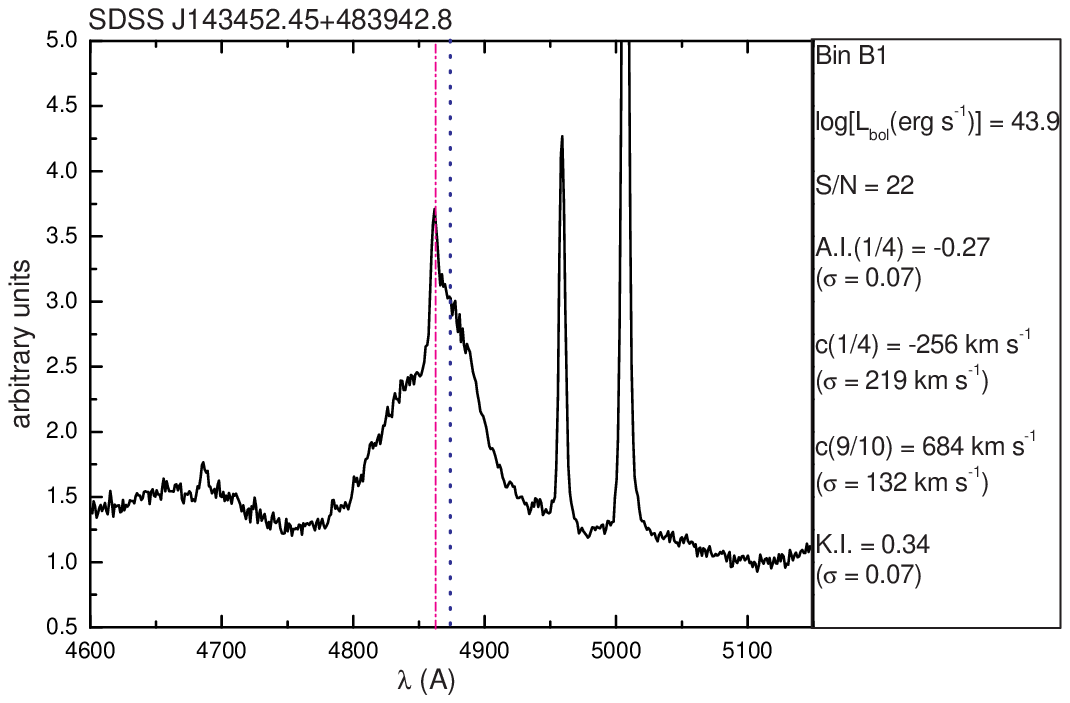}
\includegraphics[width=0.9\columnwidth,clip=true]{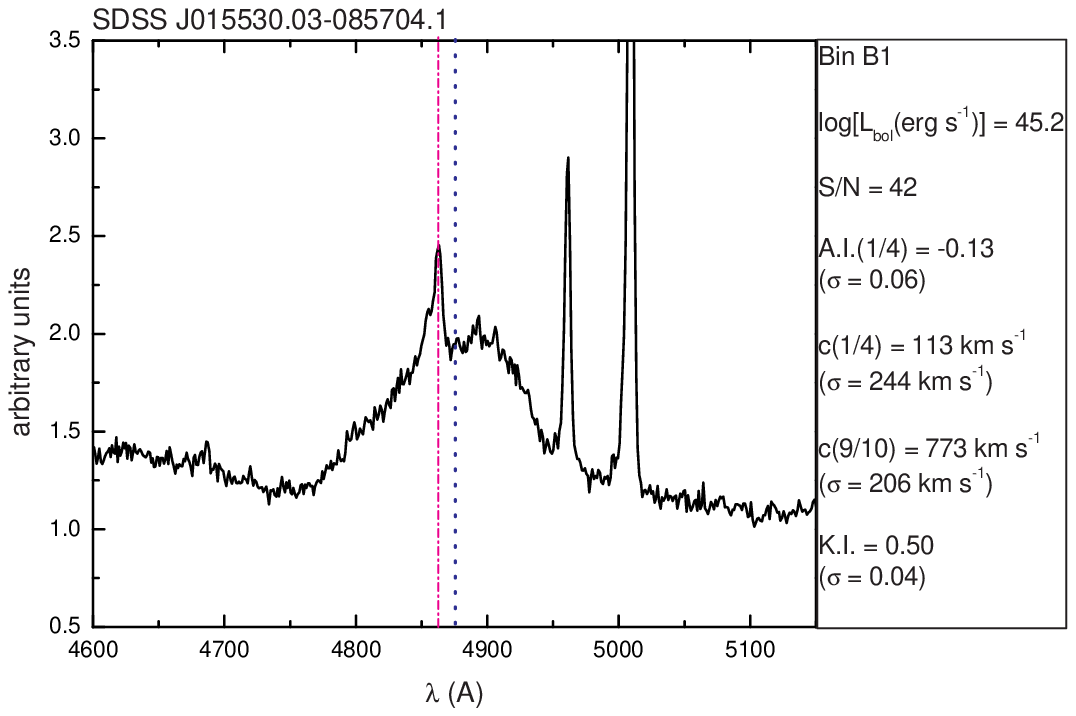}
\includegraphics[width=0.9\columnwidth,clip=true]{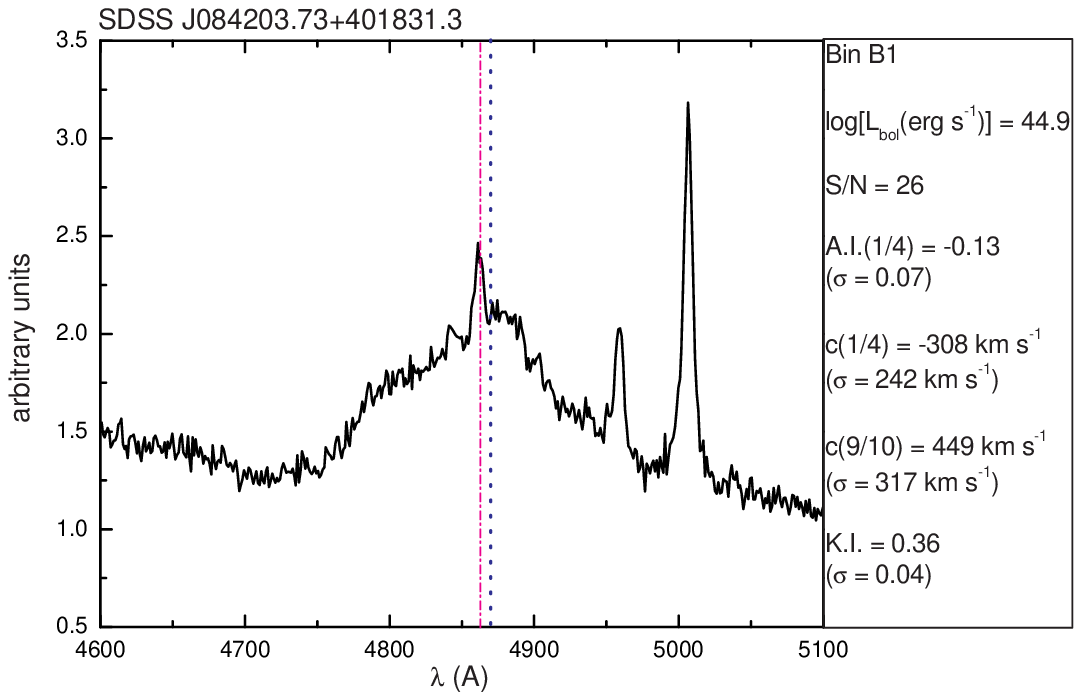}
\includegraphics[width=0.9\columnwidth,clip=true]{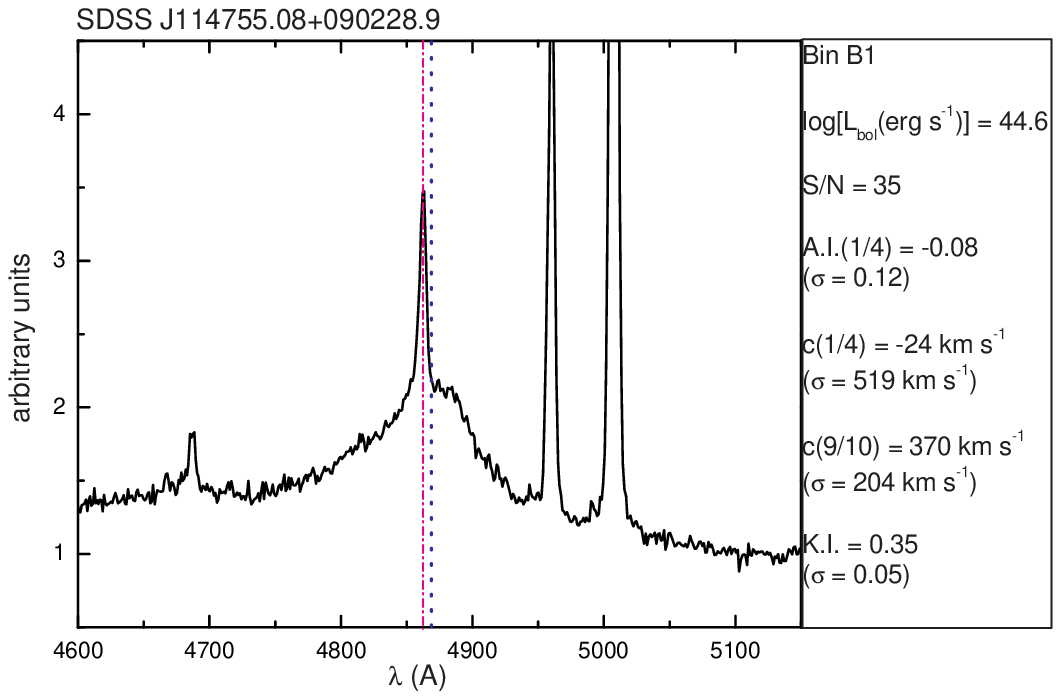}
\includegraphics[width=0.9\columnwidth,clip=true]{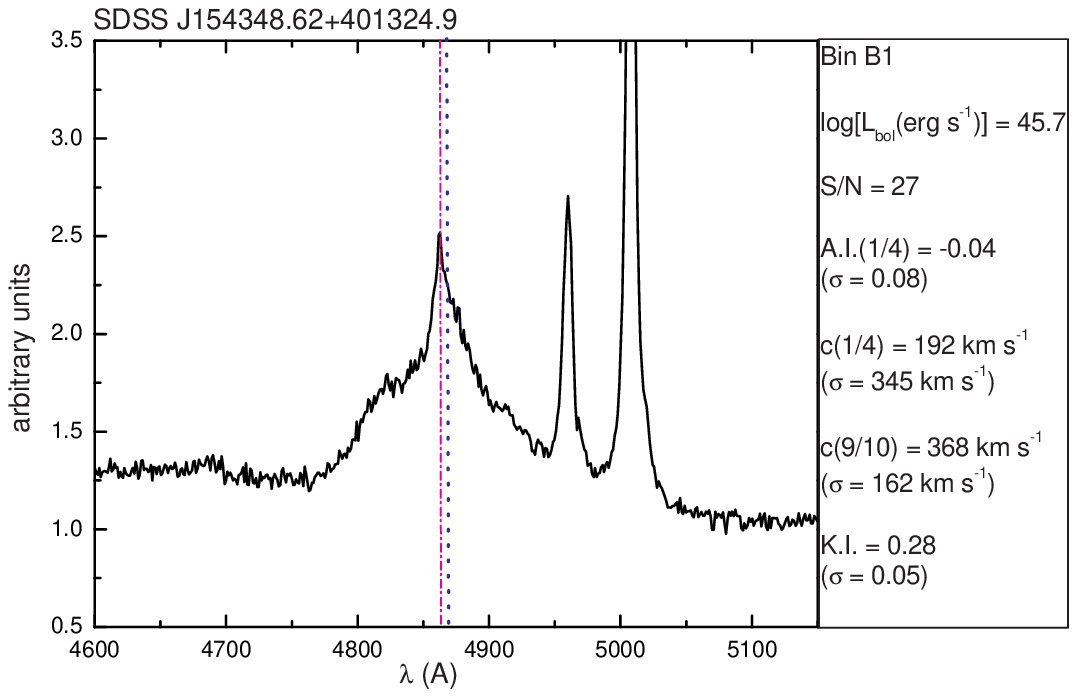}
\includegraphics[width=0.9\columnwidth,clip=true]{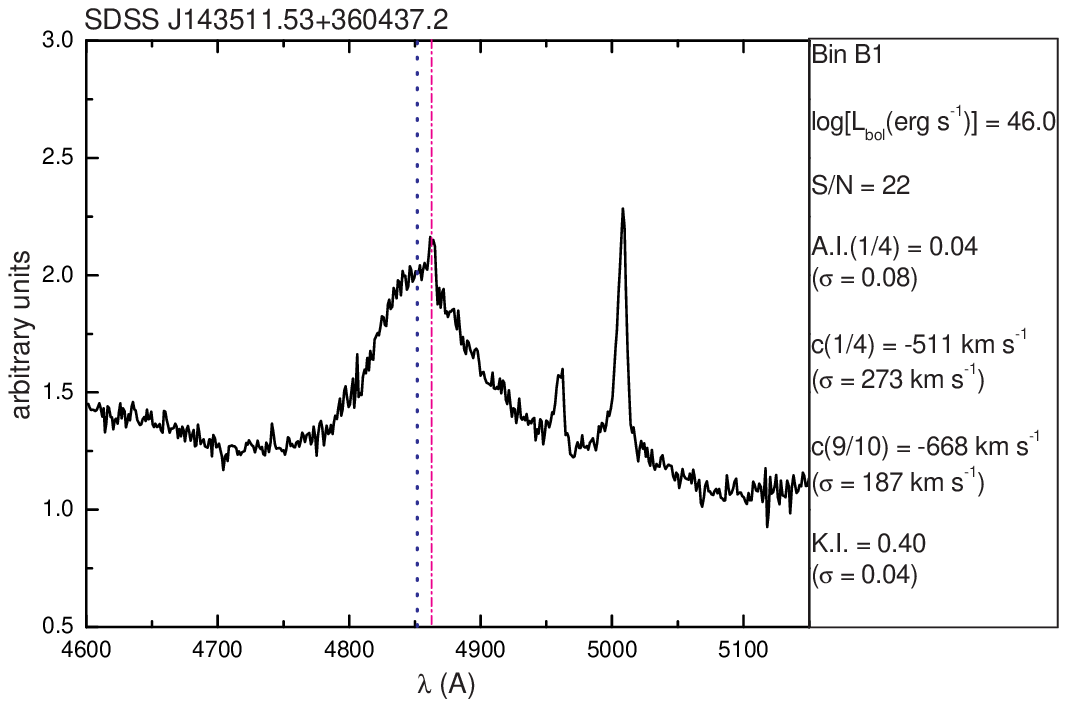}
\caption{Cont'd.}
\end{figure*}
\clearpage

\begin{figure*}
\centering\setcounter{figure}{4}
\includegraphics[width=0.9\columnwidth,clip=true]{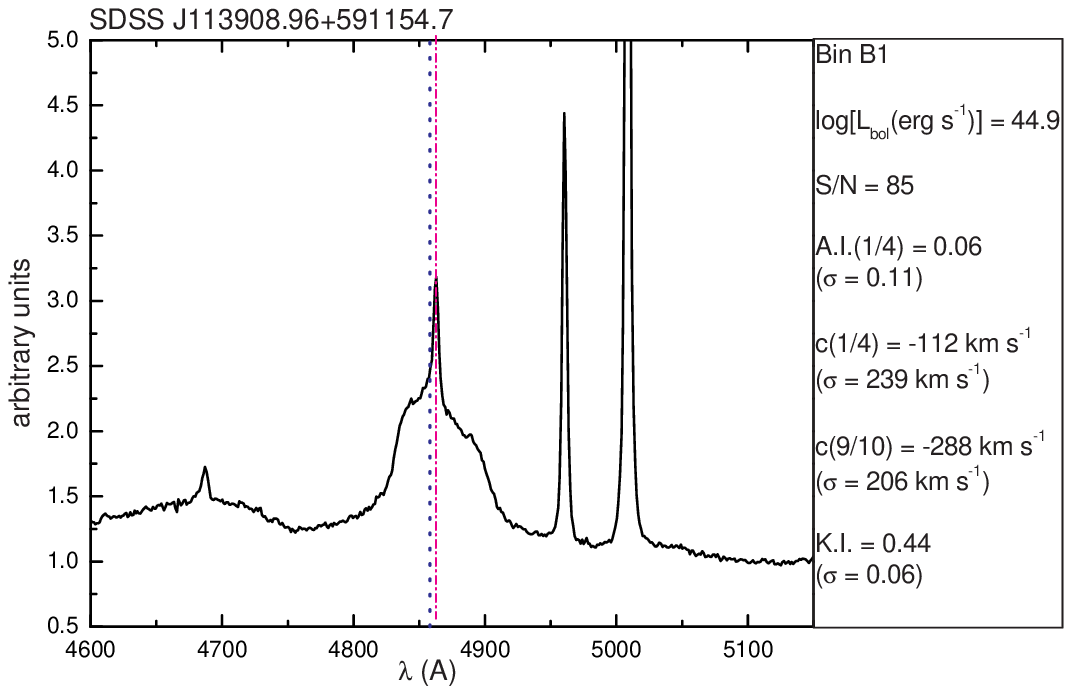}
\includegraphics[width=0.9\columnwidth,clip=true]{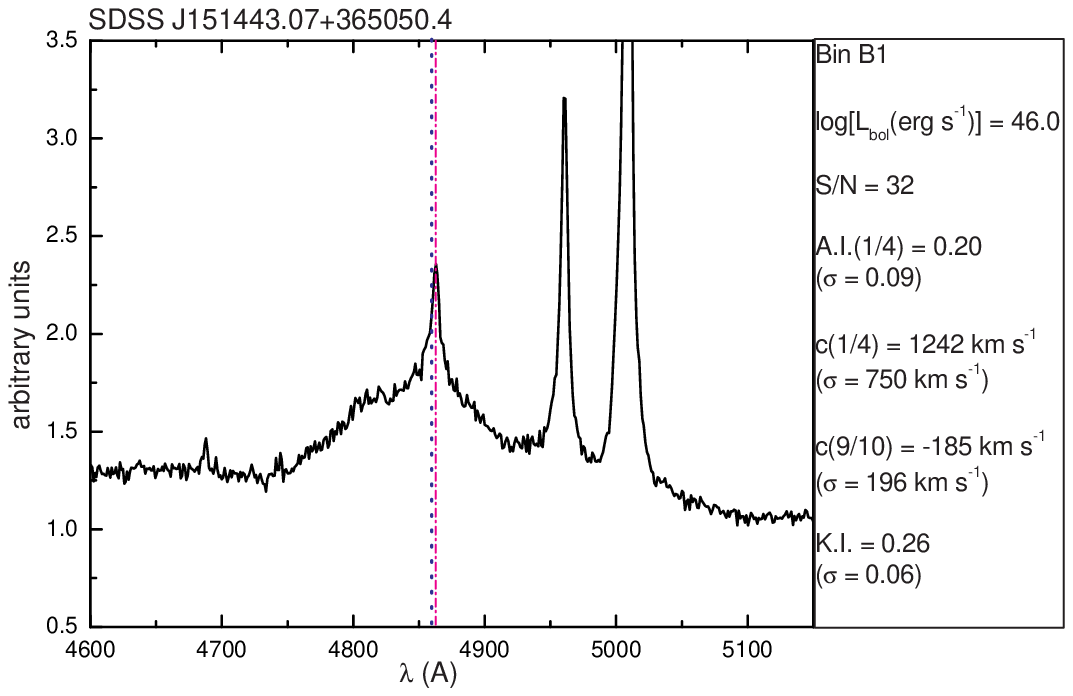}
\includegraphics[width=0.9\columnwidth,clip=true]{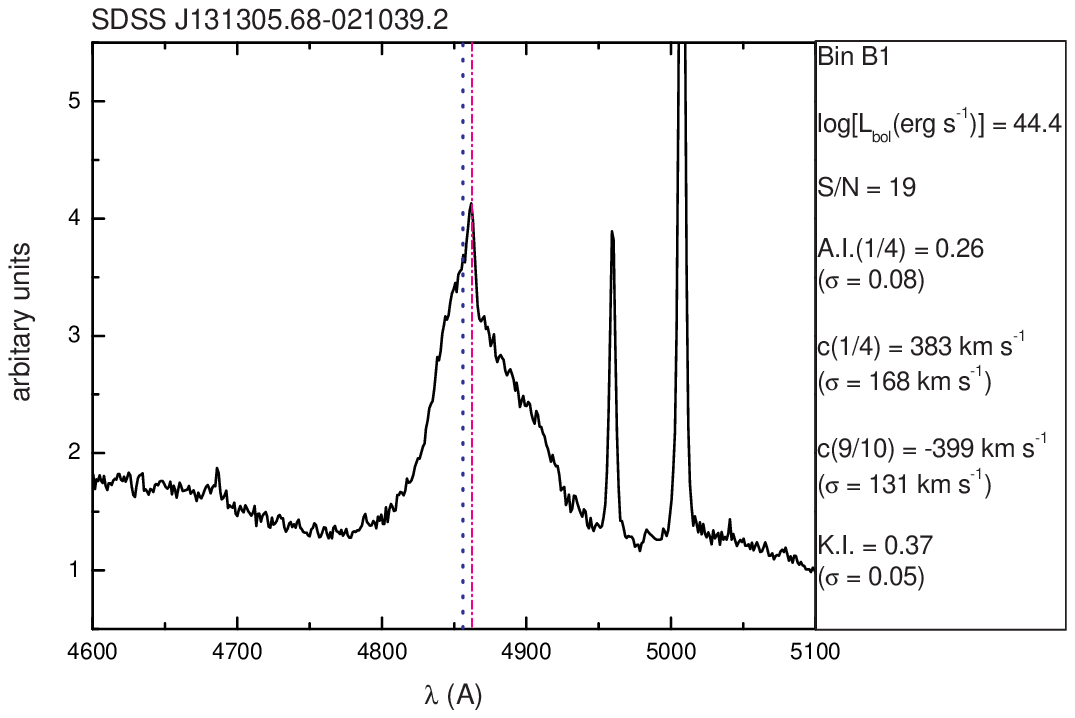}
\includegraphics[width=0.9\columnwidth,clip=true]{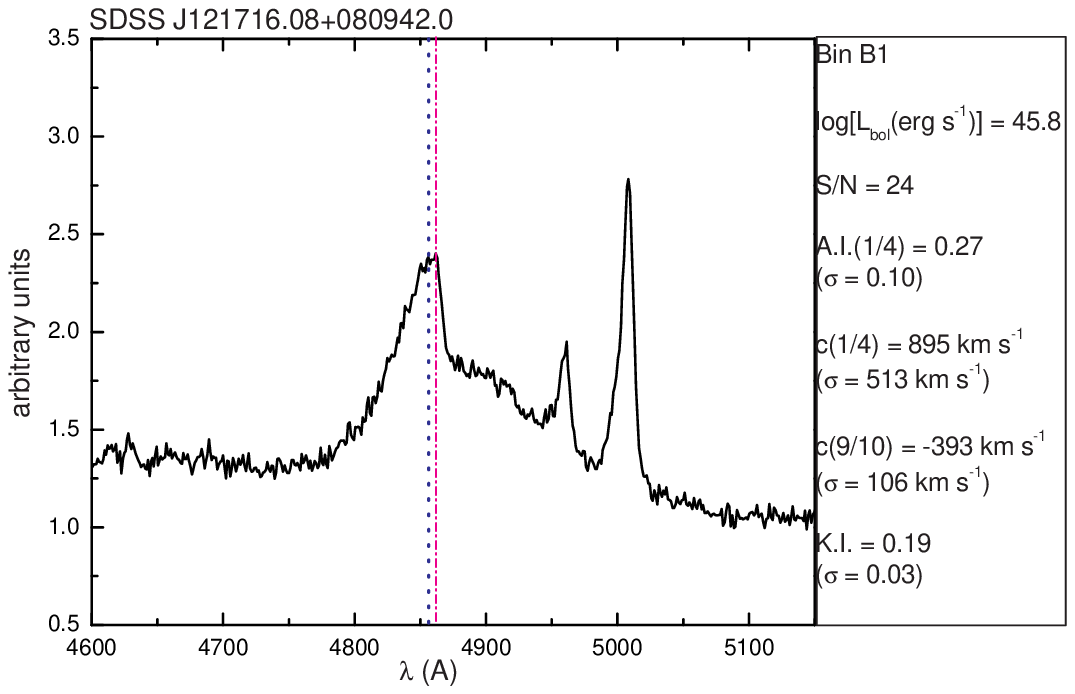}
\includegraphics[width=0.9\columnwidth,clip=true]{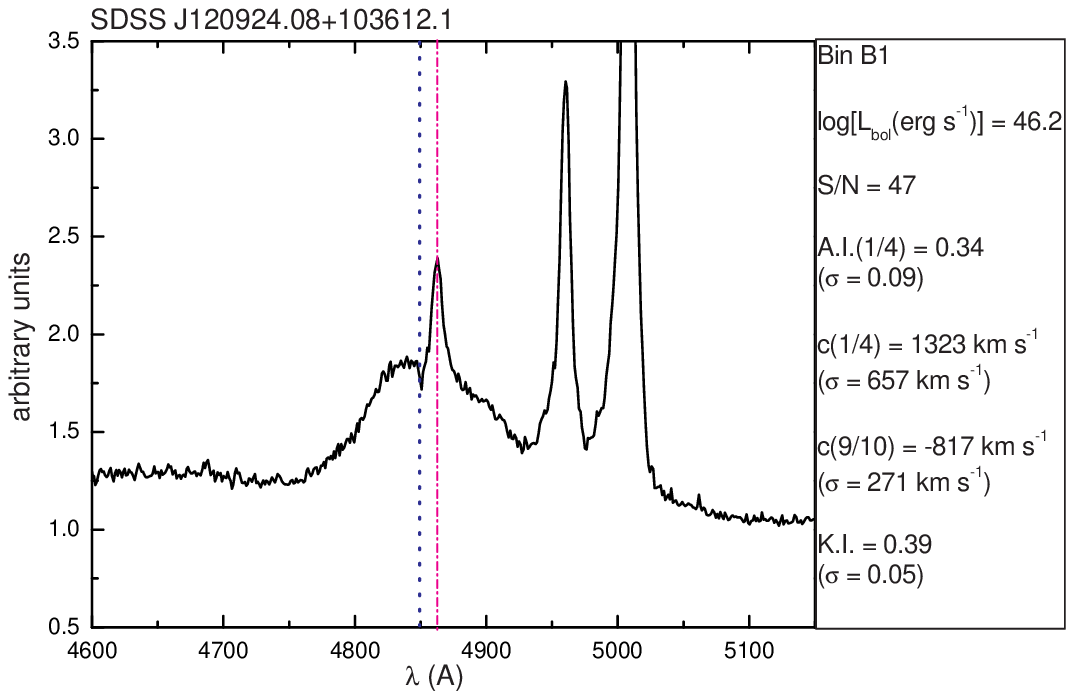}
\includegraphics[width=0.9\columnwidth,clip=true]{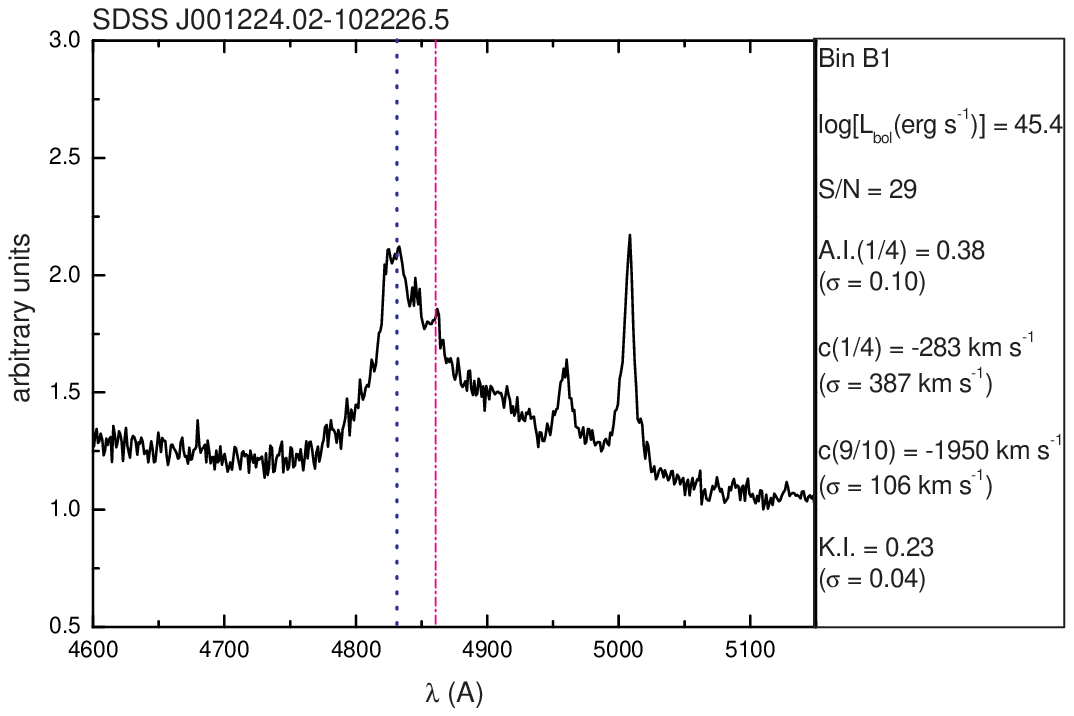}
\includegraphics[width=0.9\columnwidth,clip=true]{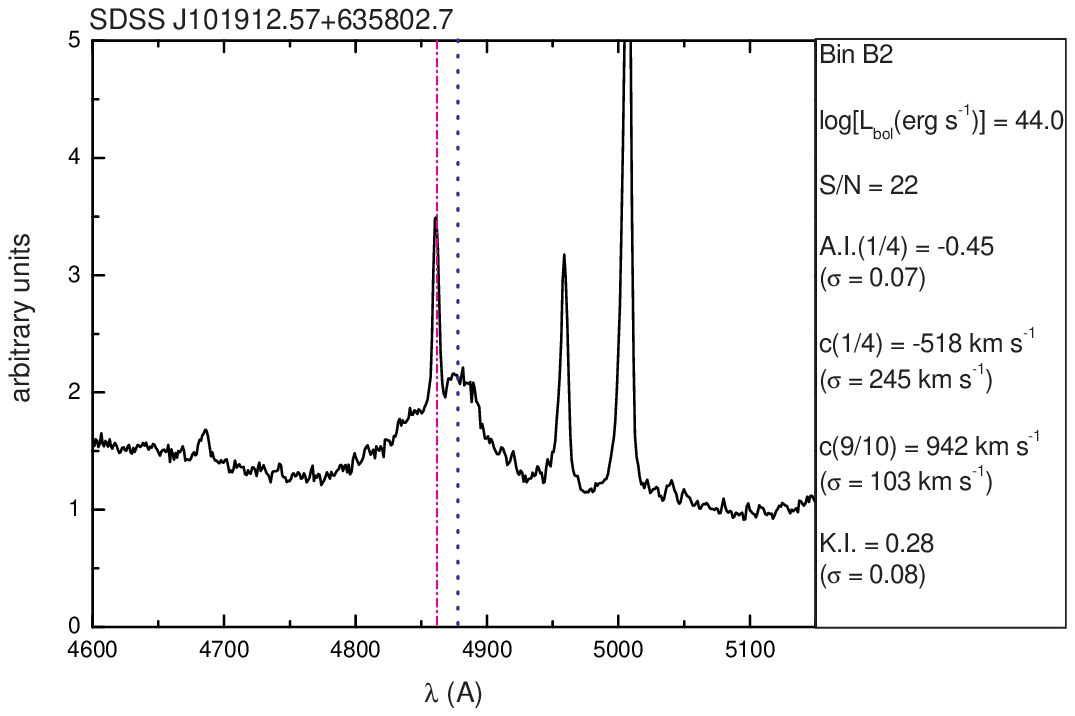}
\includegraphics[width=0.9\columnwidth,clip=true]{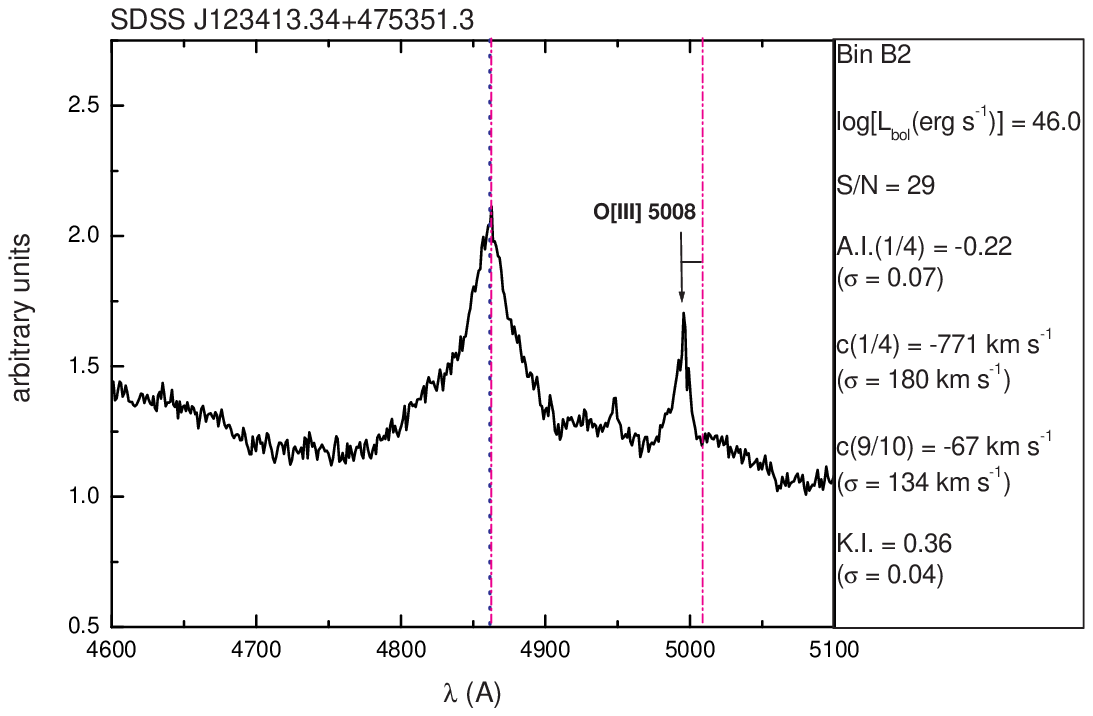}
\caption{Cont'd.}
\end{figure*}
\clearpage

\begin{figure*}
\centering \setcounter{figure}{4}
\includegraphics[width=0.9\columnwidth,clip=true]{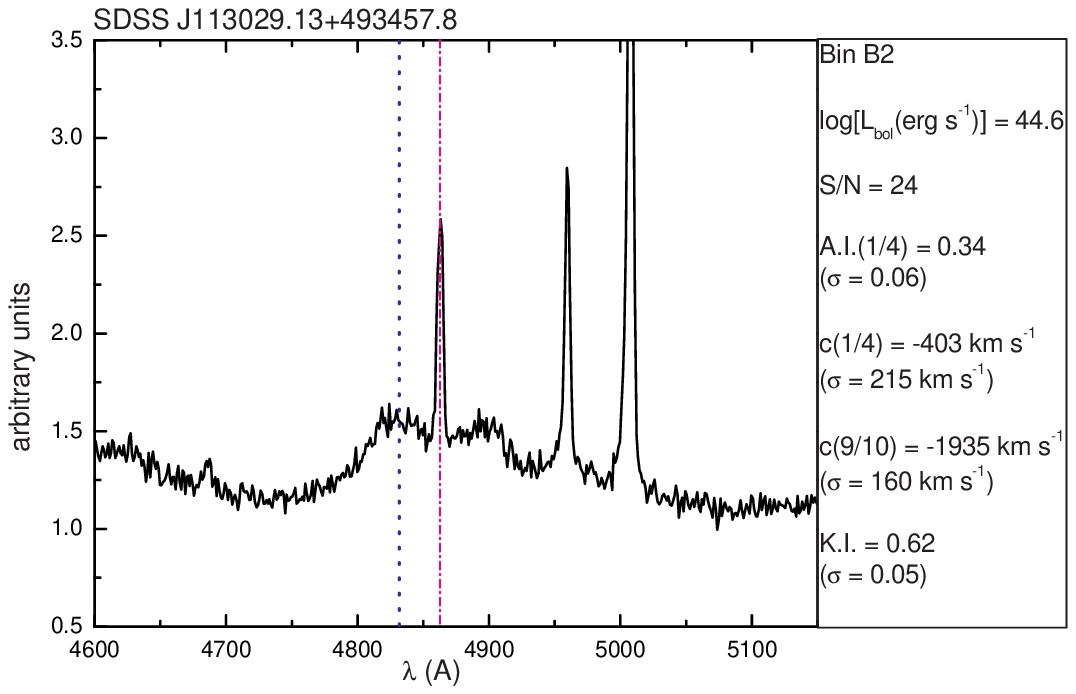}
\includegraphics[width=0.9\columnwidth,clip=true]{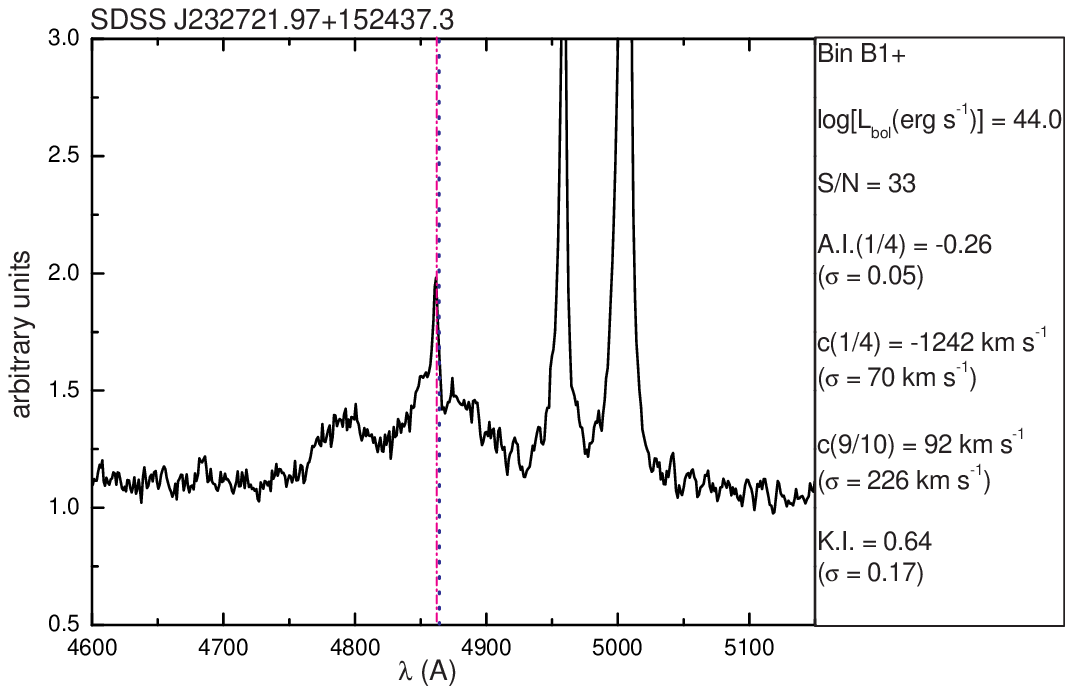}
\includegraphics[width=0.9\columnwidth,clip=true]{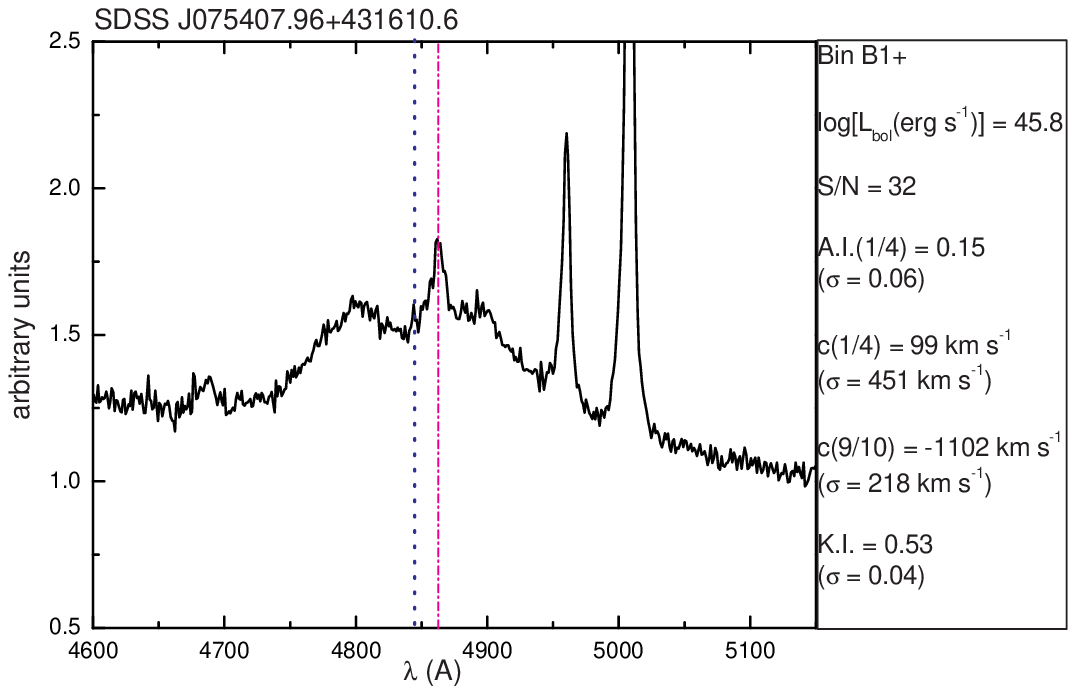}
\includegraphics[width=0.9\columnwidth,clip=true]{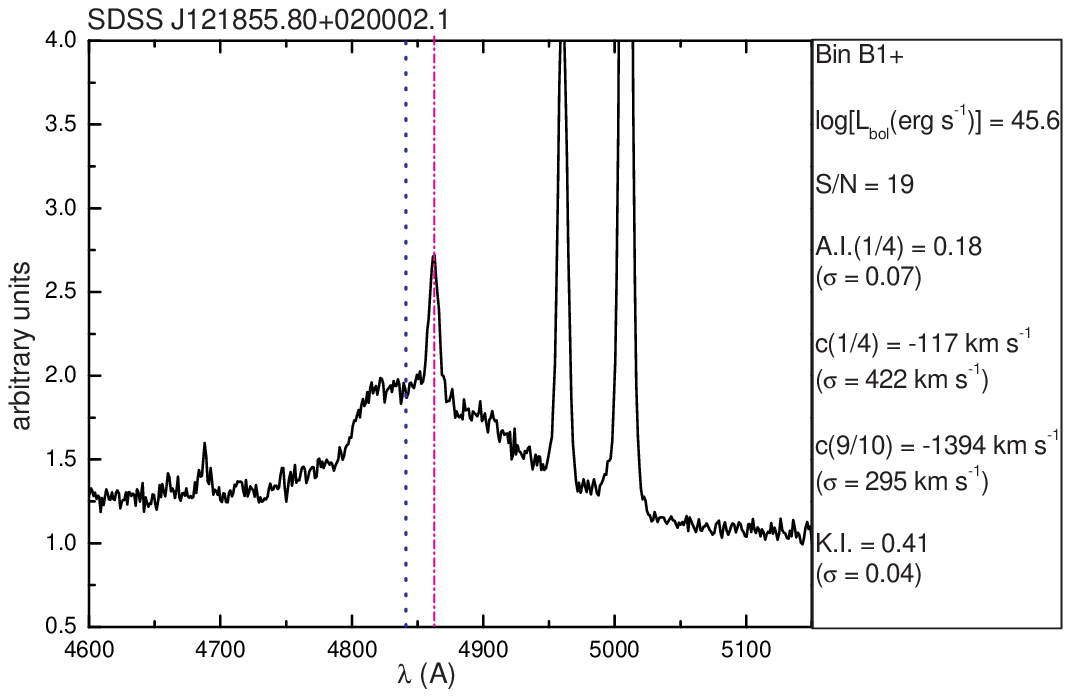}
\includegraphics[width=0.9\columnwidth,clip=true]{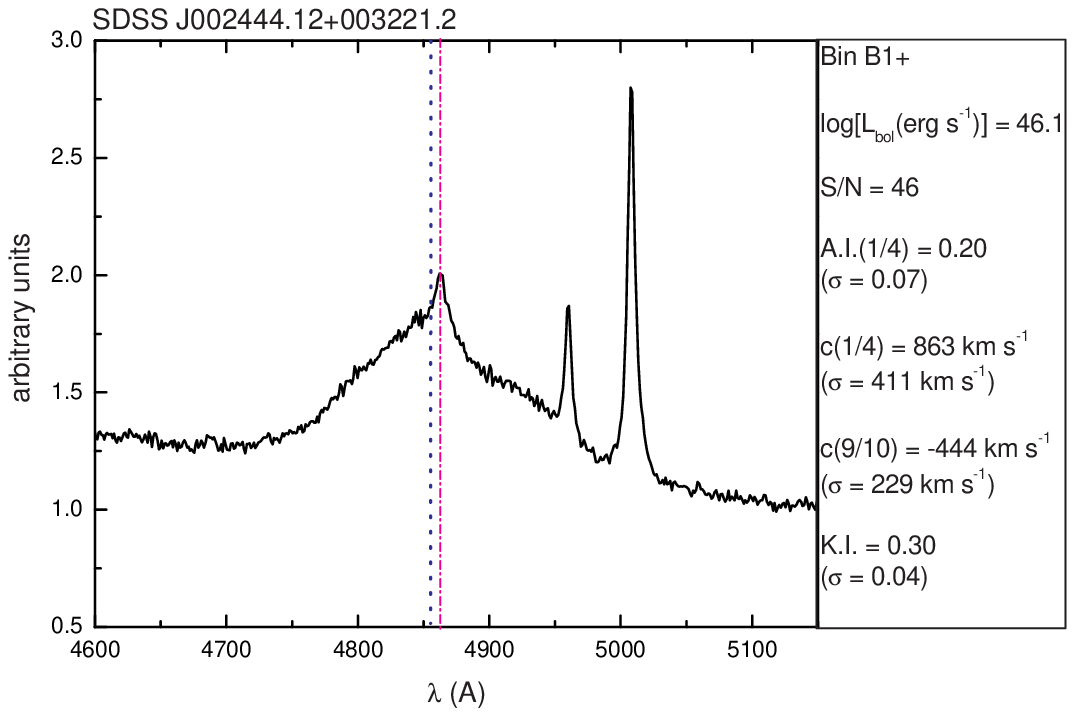}
\includegraphics[width=0.9\columnwidth,clip=true]{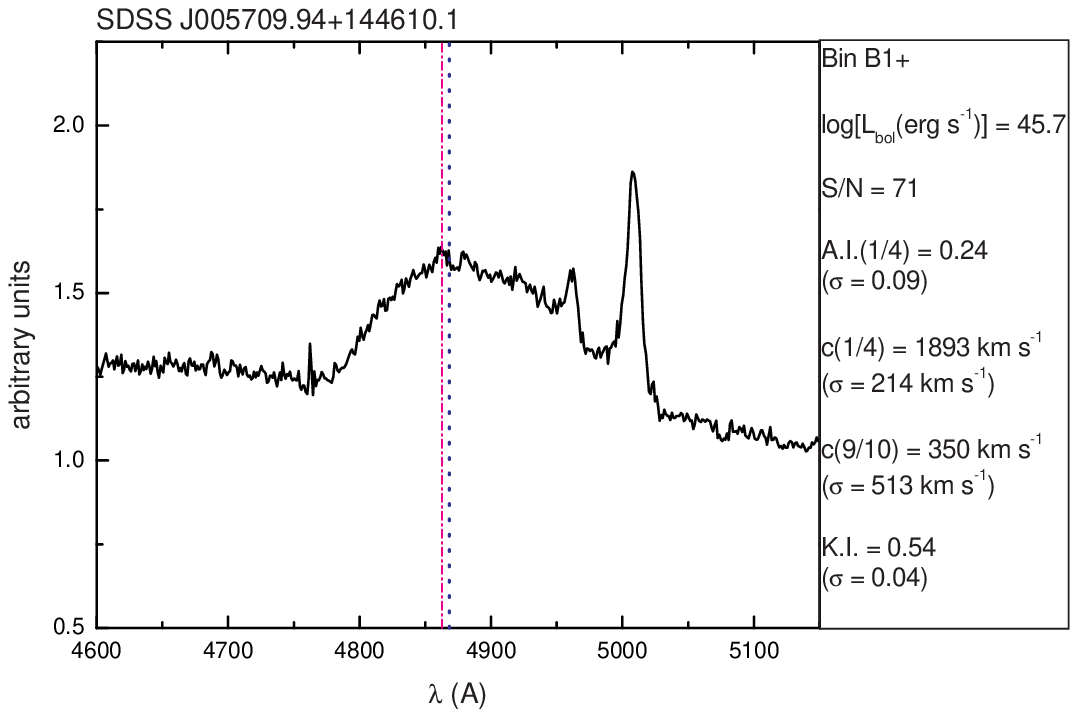}
\includegraphics[width=0.9\columnwidth,clip=true]{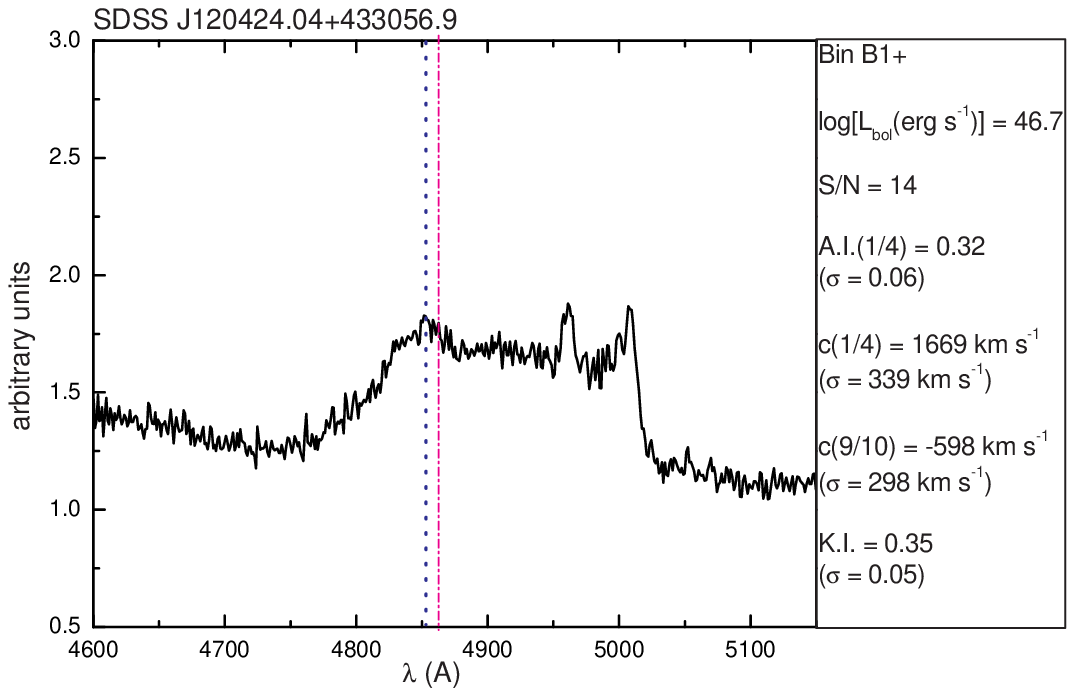}
\includegraphics[width=0.9\columnwidth,clip=true]{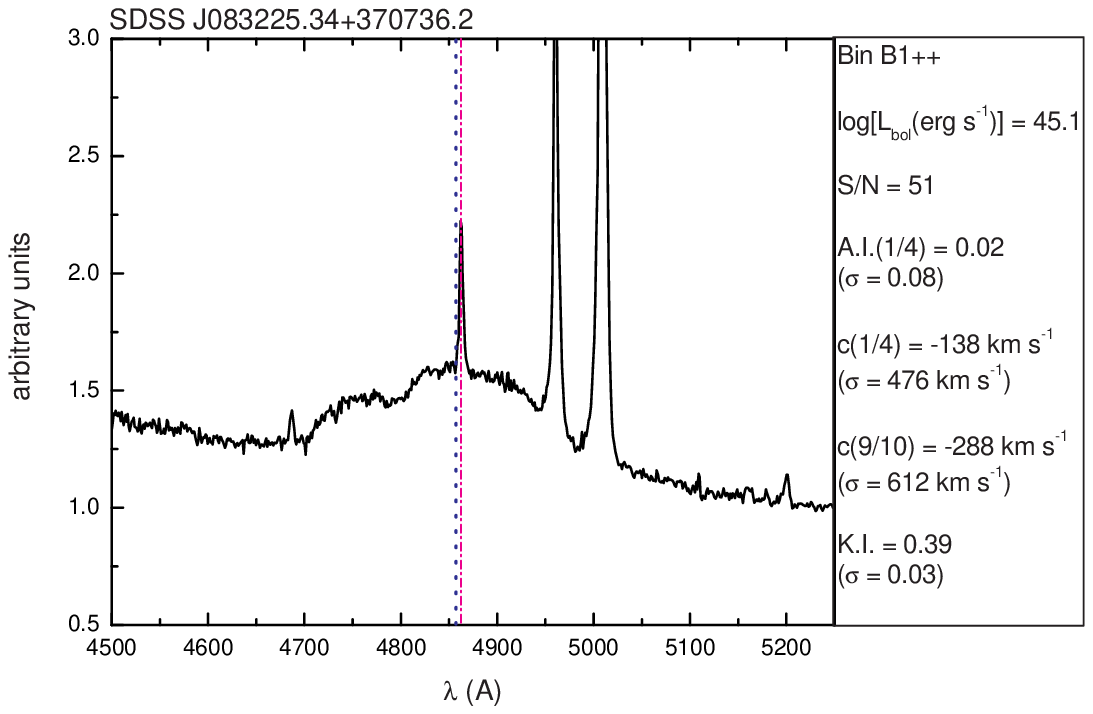}
\caption{Cont'd.}
\end{figure*}
\clearpage

\begin{figure*}
\centering \setcounter{figure}{5}
\includegraphics[width=\columnwidth,clip=true]{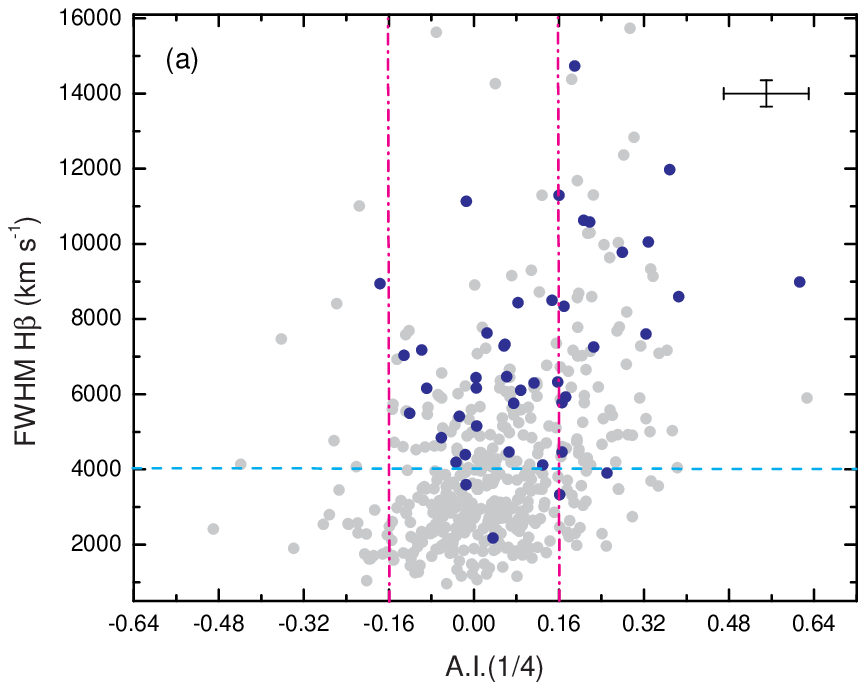}
\hspace{0.5cm}
\includegraphics[width=\columnwidth,clip=true]{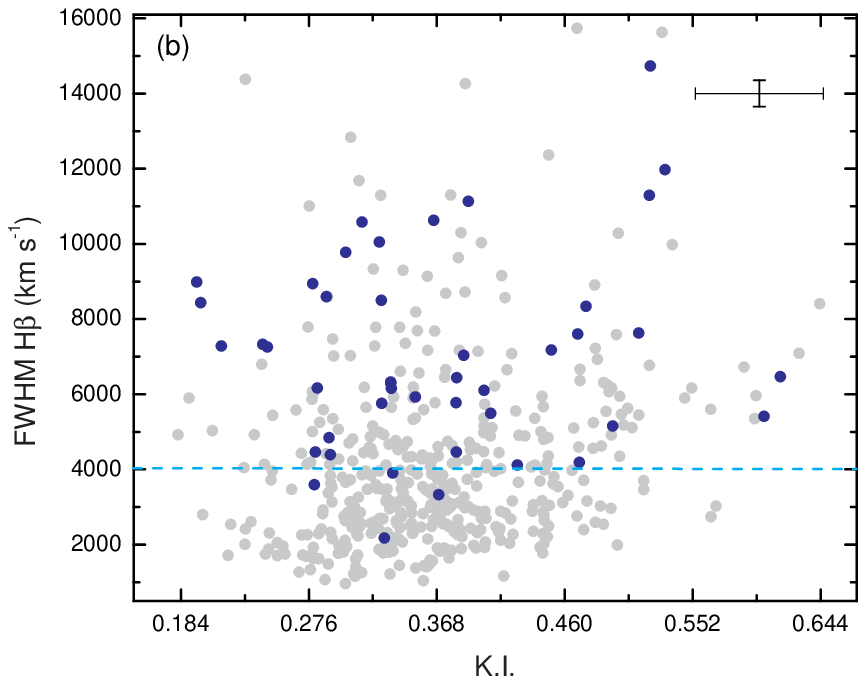}
\includegraphics[width=\columnwidth,clip=true]{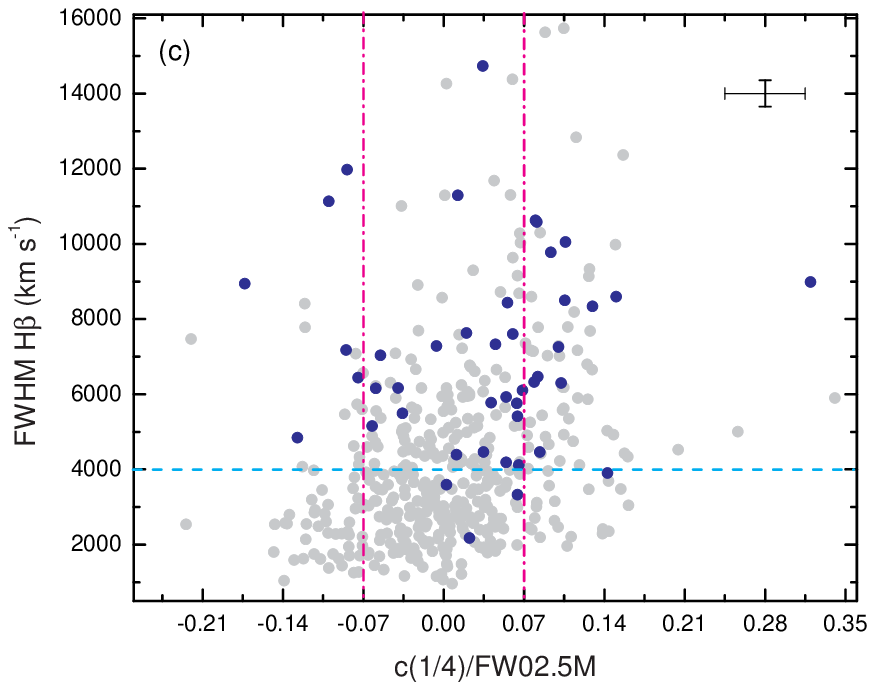}
\hspace{0.5cm}
\includegraphics[width=\columnwidth,clip=true]{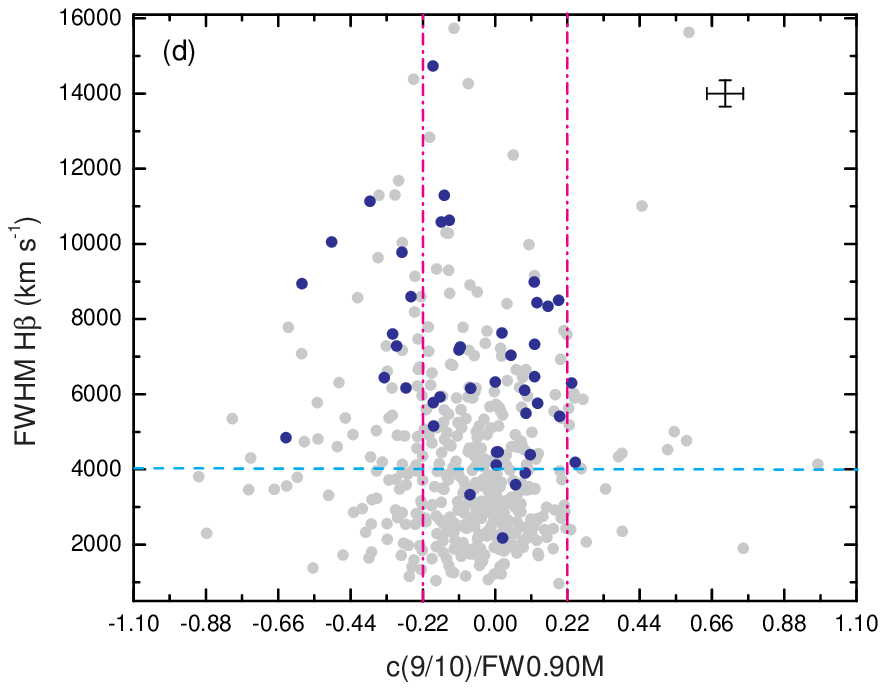}
\caption{Plots of FWHM(H$\beta$) versus (a) line asymmetry, (b)
shape measure kurtosis, (c) ``base'' normalized shift c(1/4)/FW0.25M
and (d) ``peak'' normalized shift c(9/10)/FW0.90M. The vertical
lines indicate the 2$\sigma$ uncertainty on either side of zero in
all but panel (b). In case we also show the median typical 2$\sigma$
error bars. The horizontal line at 4000 km s$^{-1}$ is the
Population A/B boundary  proposed in the context of the 4DE1 space.
The FRII sources are shown with distinct blue symbols in each case.}
\end{figure*}
\clearpage

\begin{figure*}
\centering
\includegraphics[width=0.9\columnwidth,clip=true]{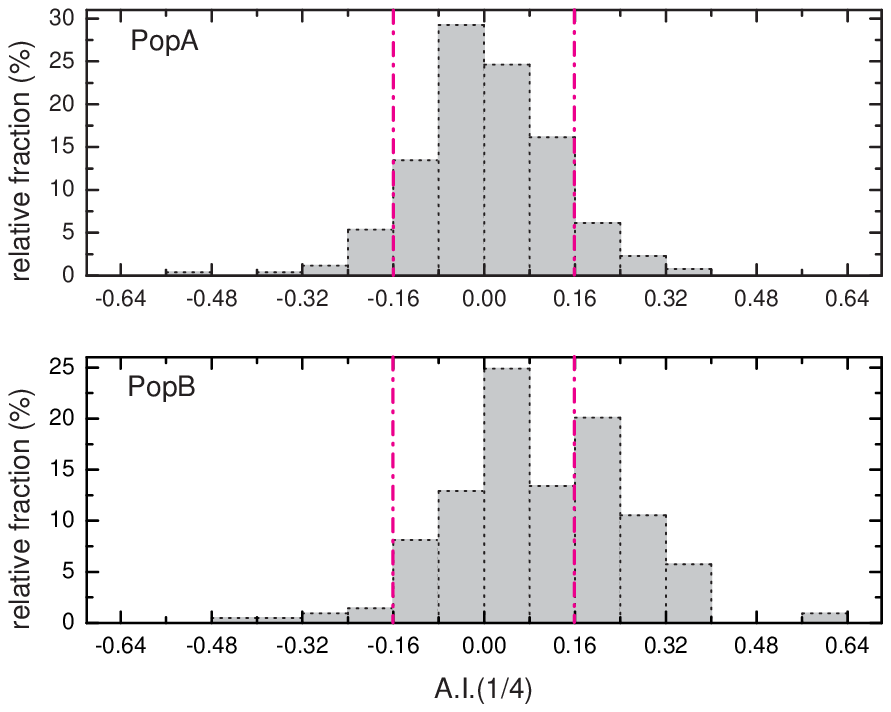}
\hspace{0.5cm}
\includegraphics[width=0.9\columnwidth,clip=true]{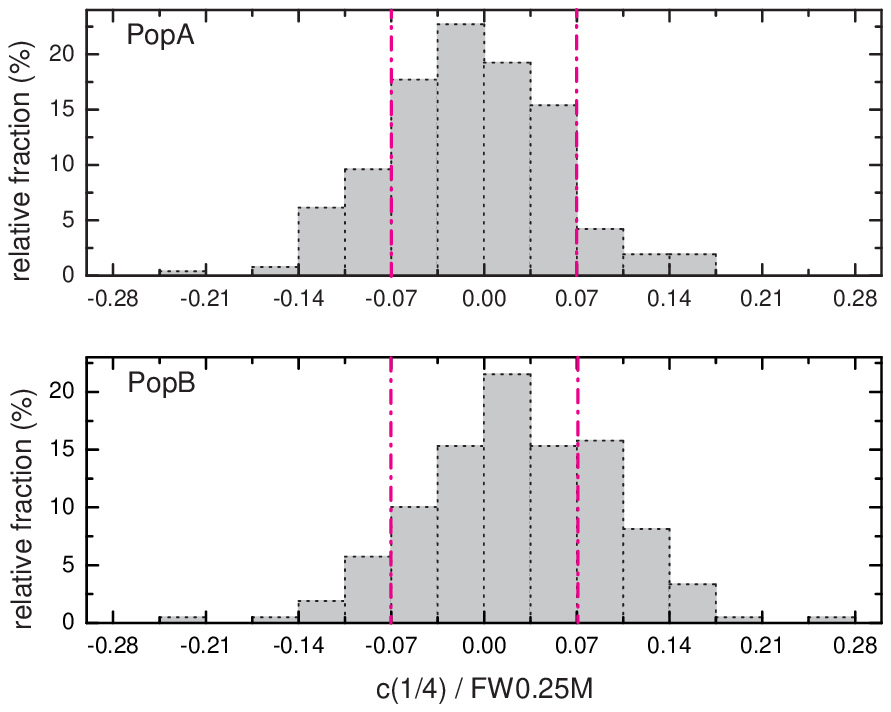}
\includegraphics[width=0.9\columnwidth,clip=true]{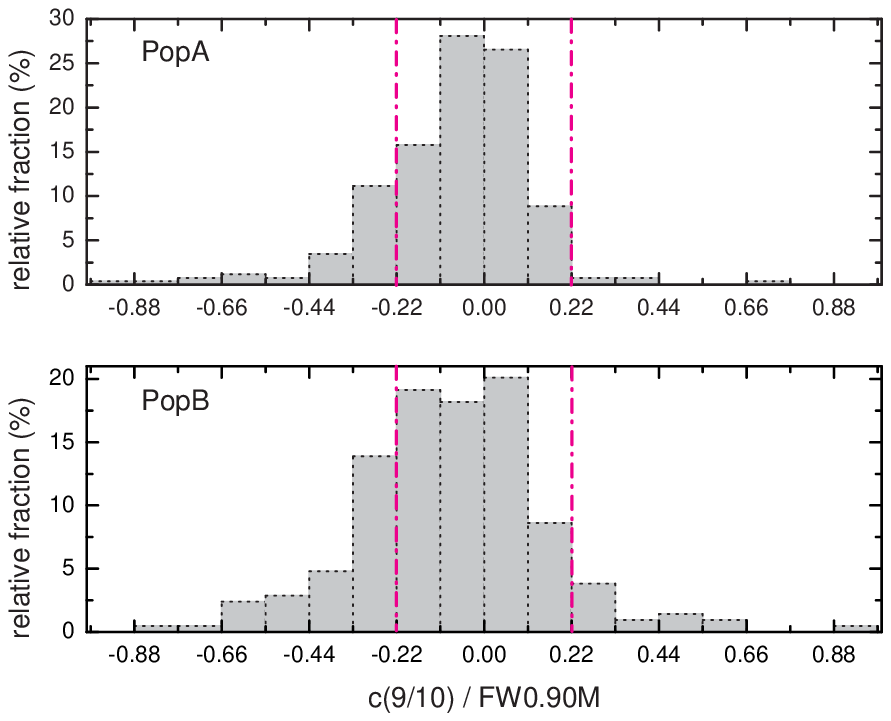}
\hspace{0.5cm}
\includegraphics[width=0.9\columnwidth,clip=true]{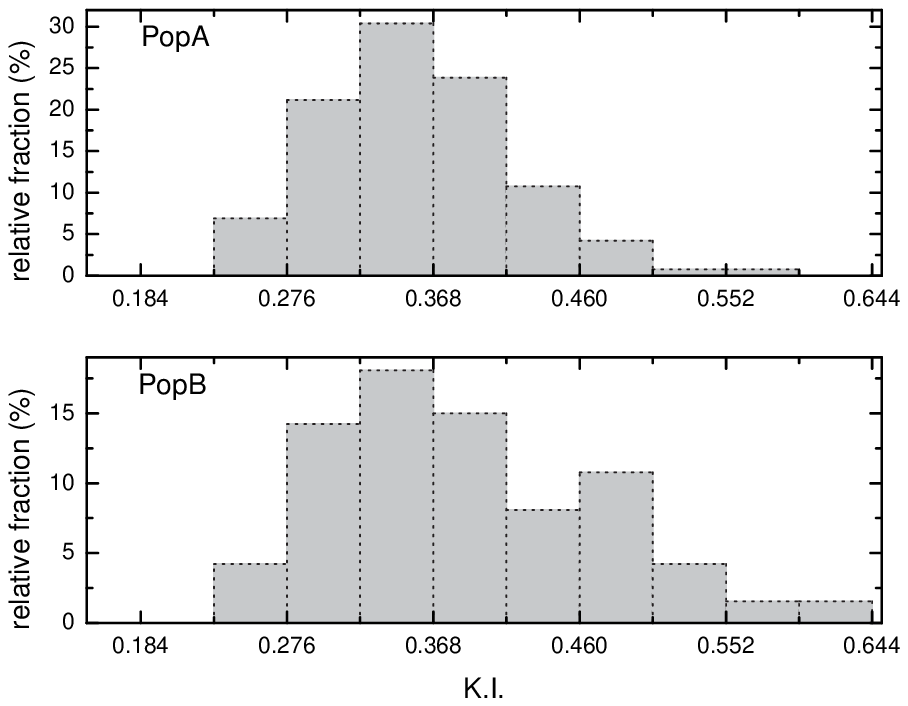}
\caption{Histogram distributions of the asymmetry index A.I.(1/4) -
upper left, base centroid shift c(1/4) normalized to FW0.25M - upper
right, peak centroid shift c(9/10) normalized to FW0.90M - lower
left and line shape kurtosis index K.I. - lower right. We show
separately the distributions for Population A and B sources for each
measure. The width of the bins is equal to $\sigma$ estimated
typical error. The vertical lines indicate a $\pm2\sigma$ interval
on either side of 0 for the first three parameters.}
\end{figure*}
\clearpage

\begin{figure*}
\centering
\includegraphics[width=0.9\columnwidth,clip=true]{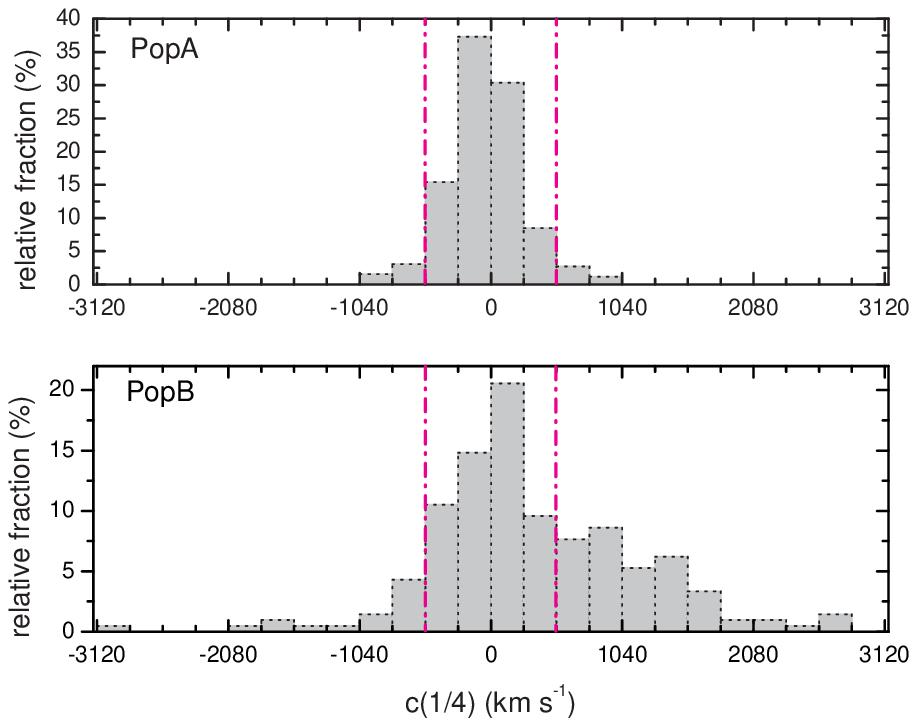}
\hspace{0.5cm}
\includegraphics[width=0.9\columnwidth,clip=true]{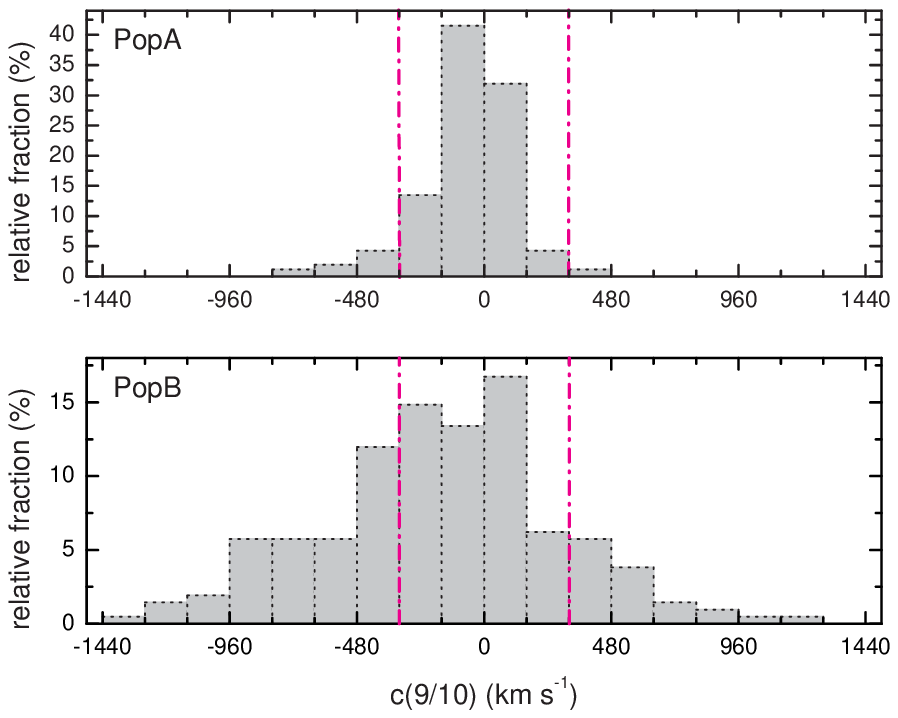}
\caption{Histogram distributions for the centroid shifts at the base
(1/4 fractional intensity) and peak (9/10 fractional intensity)
without normalization to the width of the profile. Population A and
B are illustrated separately. The width of the bins is equal to
$\sigma$ estimated typical error. The vertical lines indicate a
$\pm2\sigma$ interval on either side of 0 in each case.}
\end{figure*}
\clearpage

\begin{figure*}
\centering
\includegraphics[width=0.79\columnwidth,clip=true]{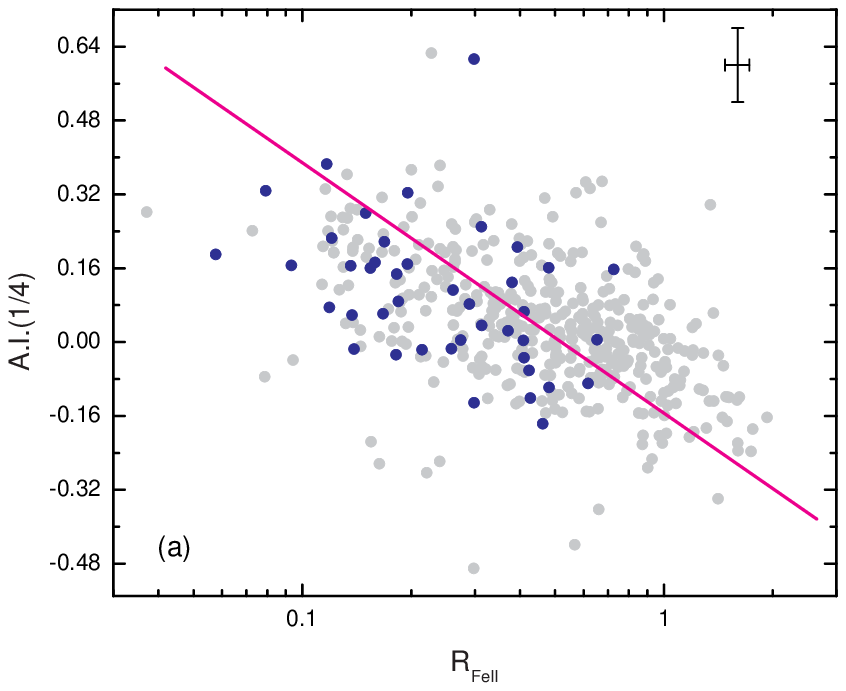}
\hspace{0.3cm}
\includegraphics[width=0.79\columnwidth,clip=true]{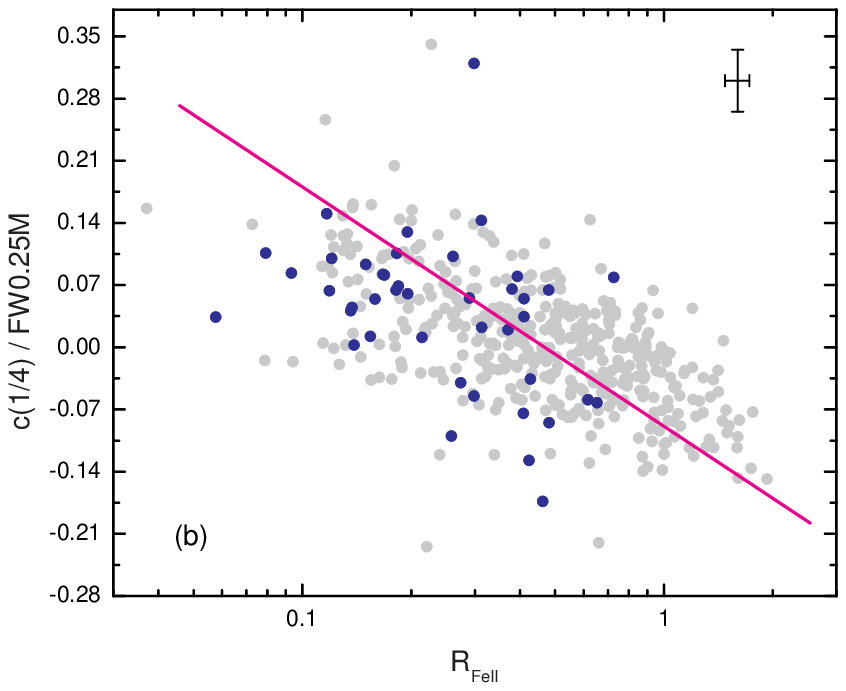}
\includegraphics[width=0.79\columnwidth,clip=true]{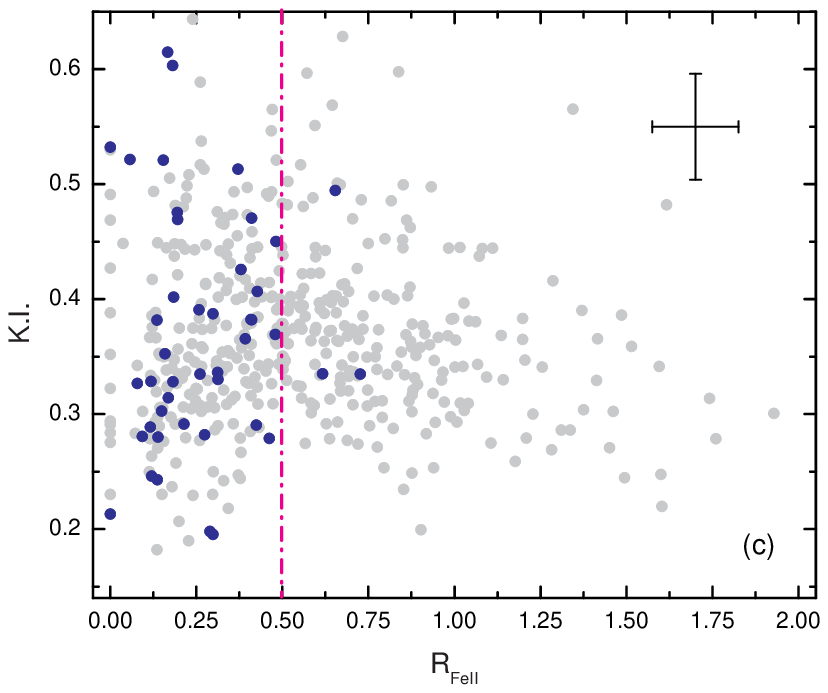}
\hspace{0.3cm}
\includegraphics[width=0.79\columnwidth,clip=true]{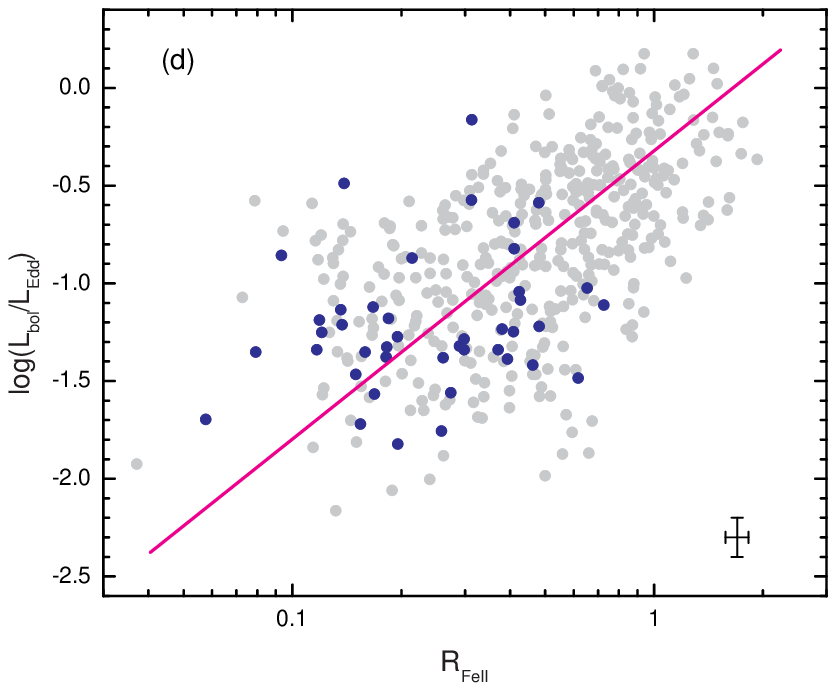}
\includegraphics[width=0.79\columnwidth,clip=true]{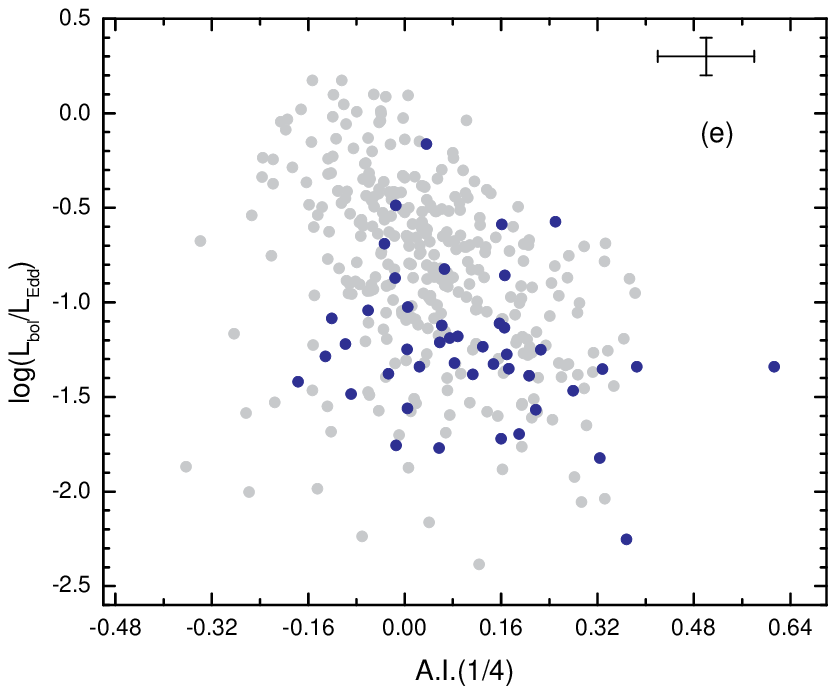}
\hspace{0.3cm}
\includegraphics[width=0.79\columnwidth,clip=true]{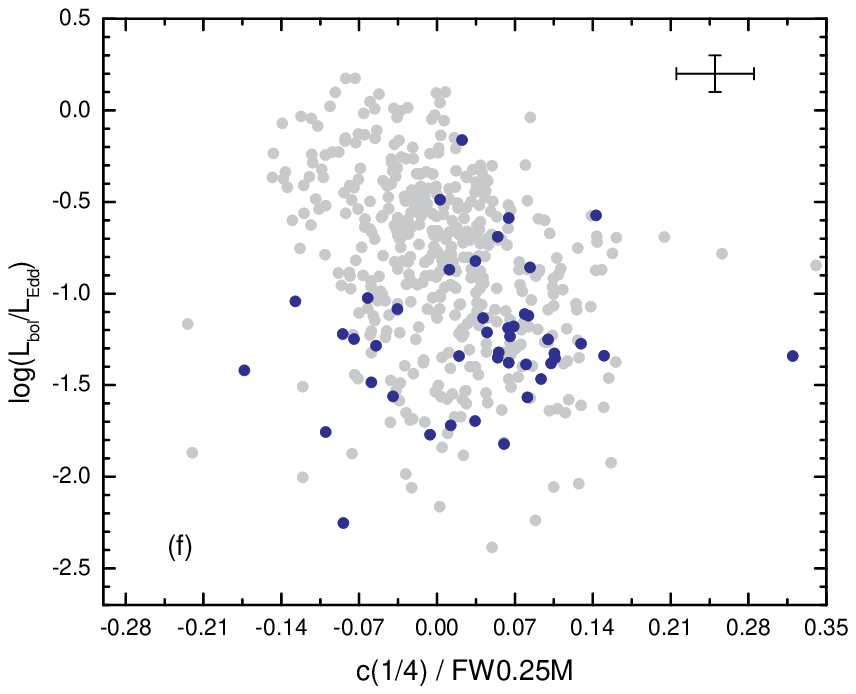}
\includegraphics[width=0.79\columnwidth,clip=true]{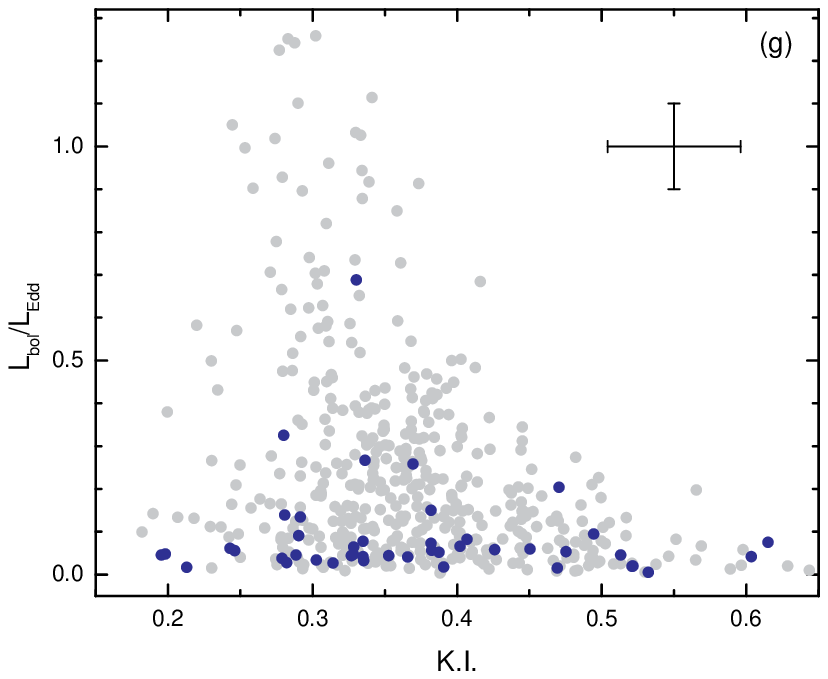}
\caption{The connection between the line diagnostic measures and the
Eigenvector 1 optical parameters and its chief driver, Eddington
ratio. Darker symbols indicate (in each panel) the FRII sources of
our sample, assumed to be the parent population of RL quasars along
with typical 2$\sigma$ error bars in the corner of the panel. (a)
The asymmetry index is plotted as a function of R$_{FeII}$. The
bisector fit is also indicated. (b) The normalized base centroid
shift is plotted as a function of R$_{FeII}$ and the bisector fit is
also shown. (c) The line shape measure kurtosis index K.I. is
plotted as against R$_{FeII}$. The vertical line suggests a possible
separation of two different trends around 0.50 abscissa value. (d)
Eddington ratio is shown as a function of R$_{FeII}$ along with a
bisector fit. Panels e-f show the relation between base measures
A.I.(1/4) and normalized centroid shift c(1/4)/FW0.25M and the
Eddington ratio, the main driver of the 4DE1 diversity. Panel g
shows the plot of L$_{bol}$/L$_{Edd}$ vs. K.I.}
\end{figure*}
\clearpage

\begin{figure*}
\centering
\includegraphics[width=0.9\columnwidth,clip=true]{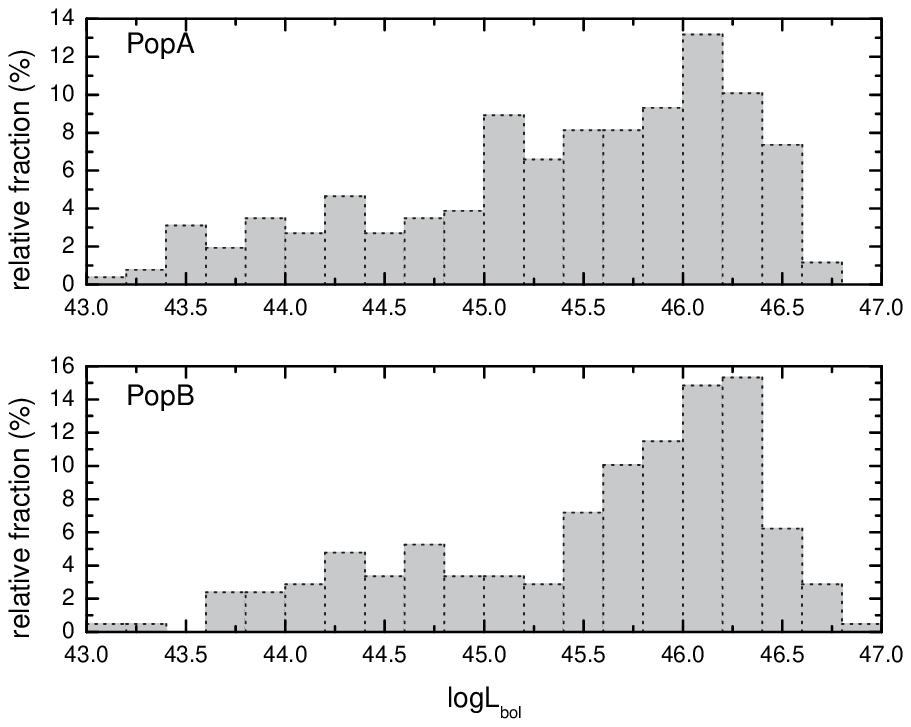}\\
\includegraphics[width=0.9\columnwidth,clip=true]{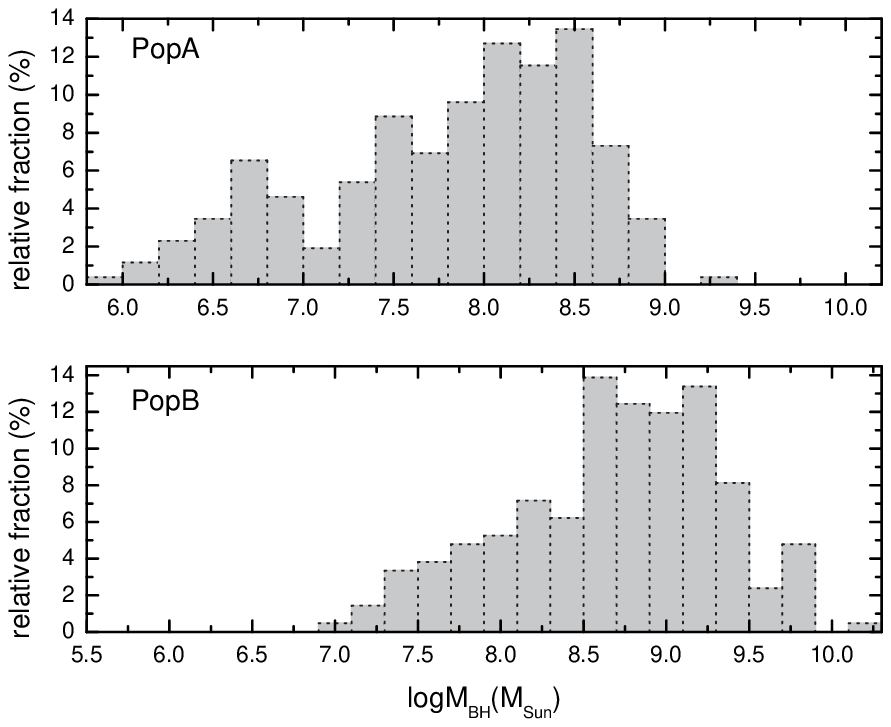}
\hspace{0.2cm}
\includegraphics[width=0.9\columnwidth,clip=true]{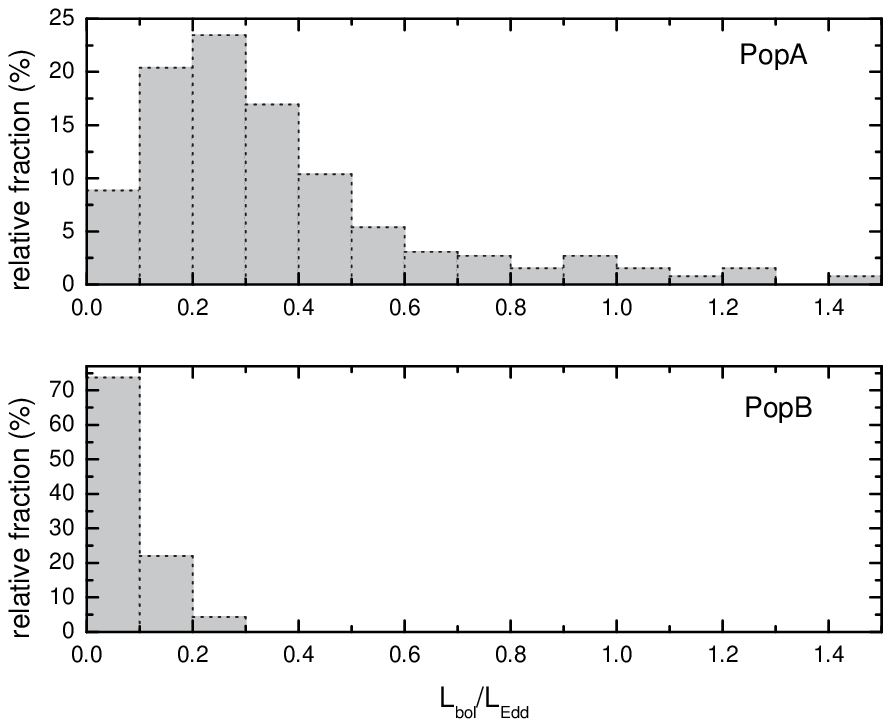}
\caption{Population A and B quasars show similar luminosity
distributions, but strikingly different distributions in terms of BH
mass and Eddington ratios, Population A typically showing smaller BH
masses and higher accretion rates and population B harboring more
massive BH and accreting much slower.}
\end{figure*}
\clearpage

\begin{figure*}
\centering
\includegraphics[width=0.9\columnwidth,clip=true]{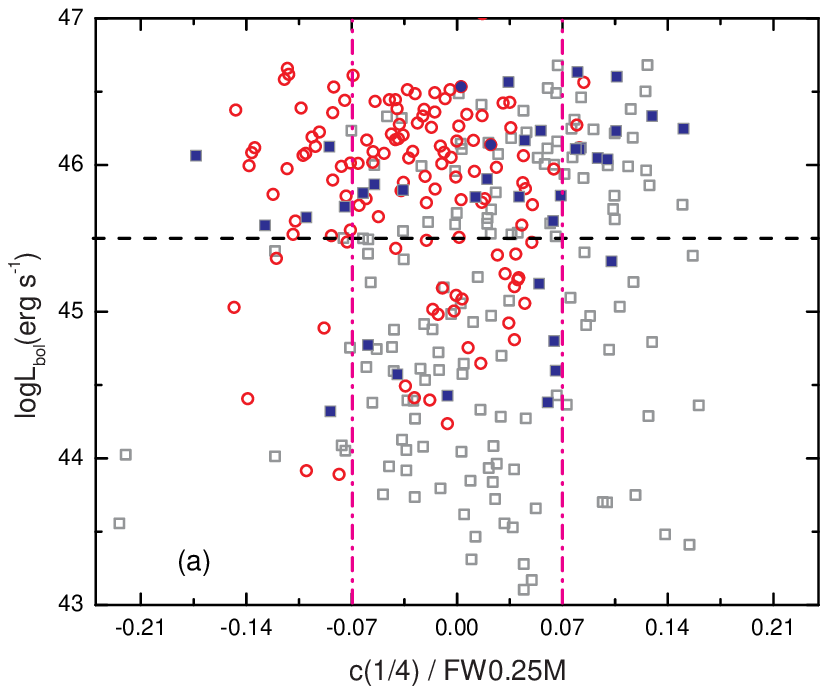}
\hspace{0.2cm}
\includegraphics[width=0.9\columnwidth,clip=true]{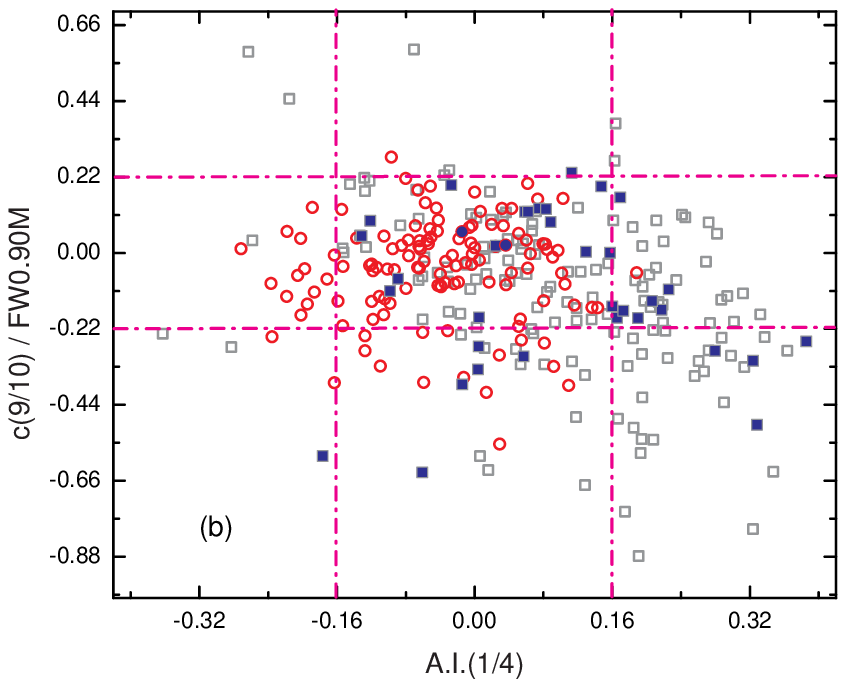}
\includegraphics[width=0.9\columnwidth,clip=true]{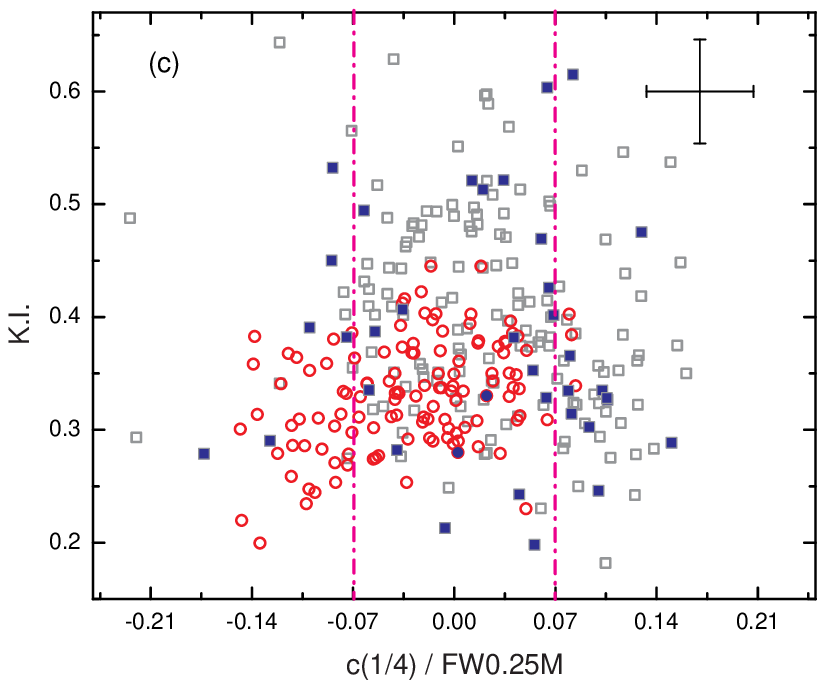}
\hspace{0.2cm}
\includegraphics[width=0.9\columnwidth,clip=true]{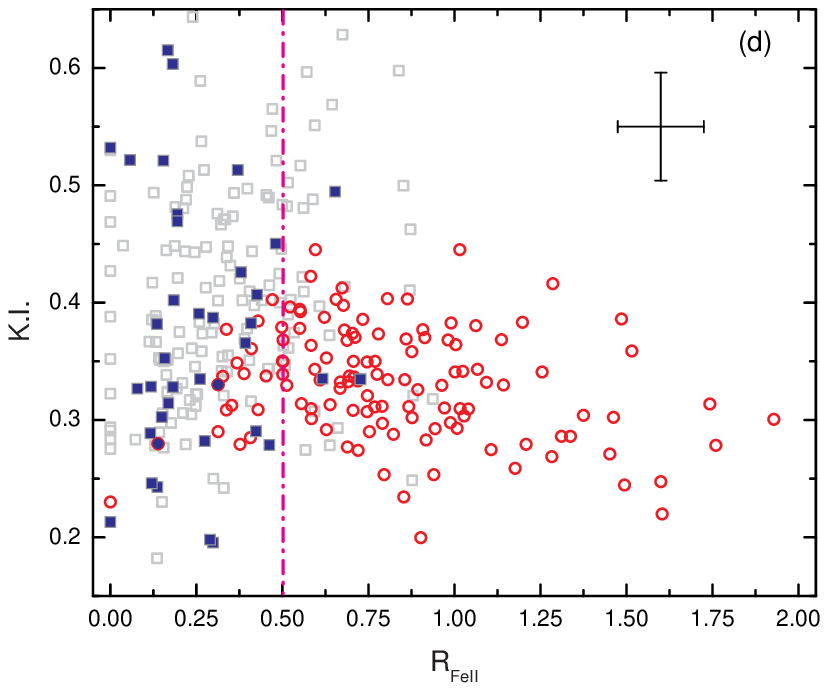}
\caption{Two subsamples of N=177 low (L$_{bol}$/L$_{Edd}$ $<$ 0.1)
and N=123 highly accreting (L$_{bol}$/L$_{Edd}$ $>$ 0.3) sources are
considered, the squares representing the former and the circles
showing the latter. The filled symbols indicate the FRII sources.
Pairs of vertical and horizontal lines on either side of zero show
the typical $\pm2\sigma$ median uncertainty. (a) The bolometric
luminosity versus the base (1/4 fractional intensity) normalized
shift. The horizontal line at 45.5 ordinate indicates the values
above which both samples are well represented numerically (see
text). (b) The normalized peak shift versus normalized base centroid
shift. (c) Kurtosis index versus normalized base centroid shift.
Typical median $2\sigma$ errors are shown in the upper corner of the
panel. (d) Kurtosis index versus R$_{FeII}$. Typical median
$2\sigma$ errors are shown in the upper corner of the panel.}
\end{figure*}
\clearpage

\begin{figure*}
\centering
\includegraphics[width=0.9\columnwidth,clip=true]{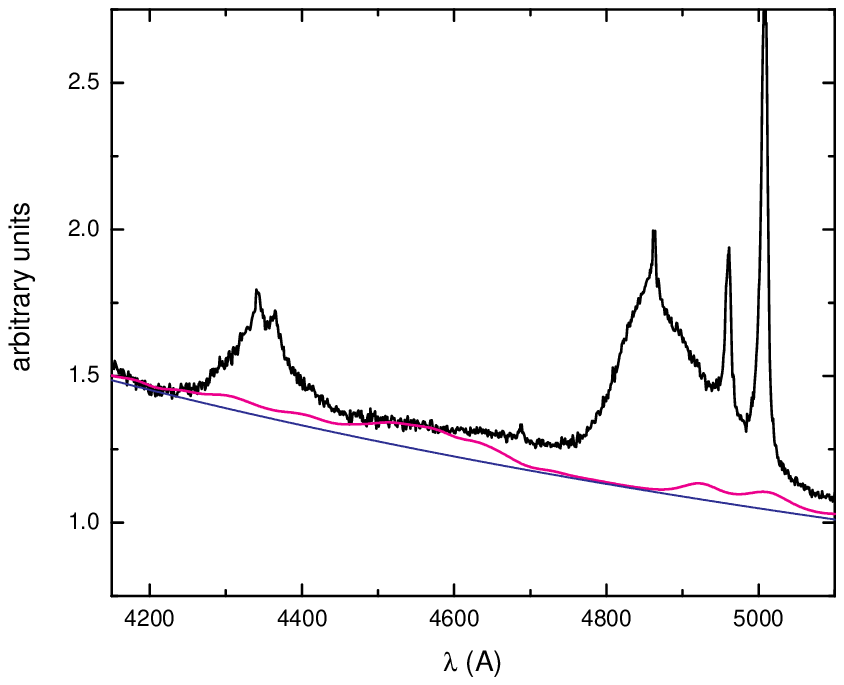}\\
\includegraphics[width=0.9\columnwidth,clip=true]{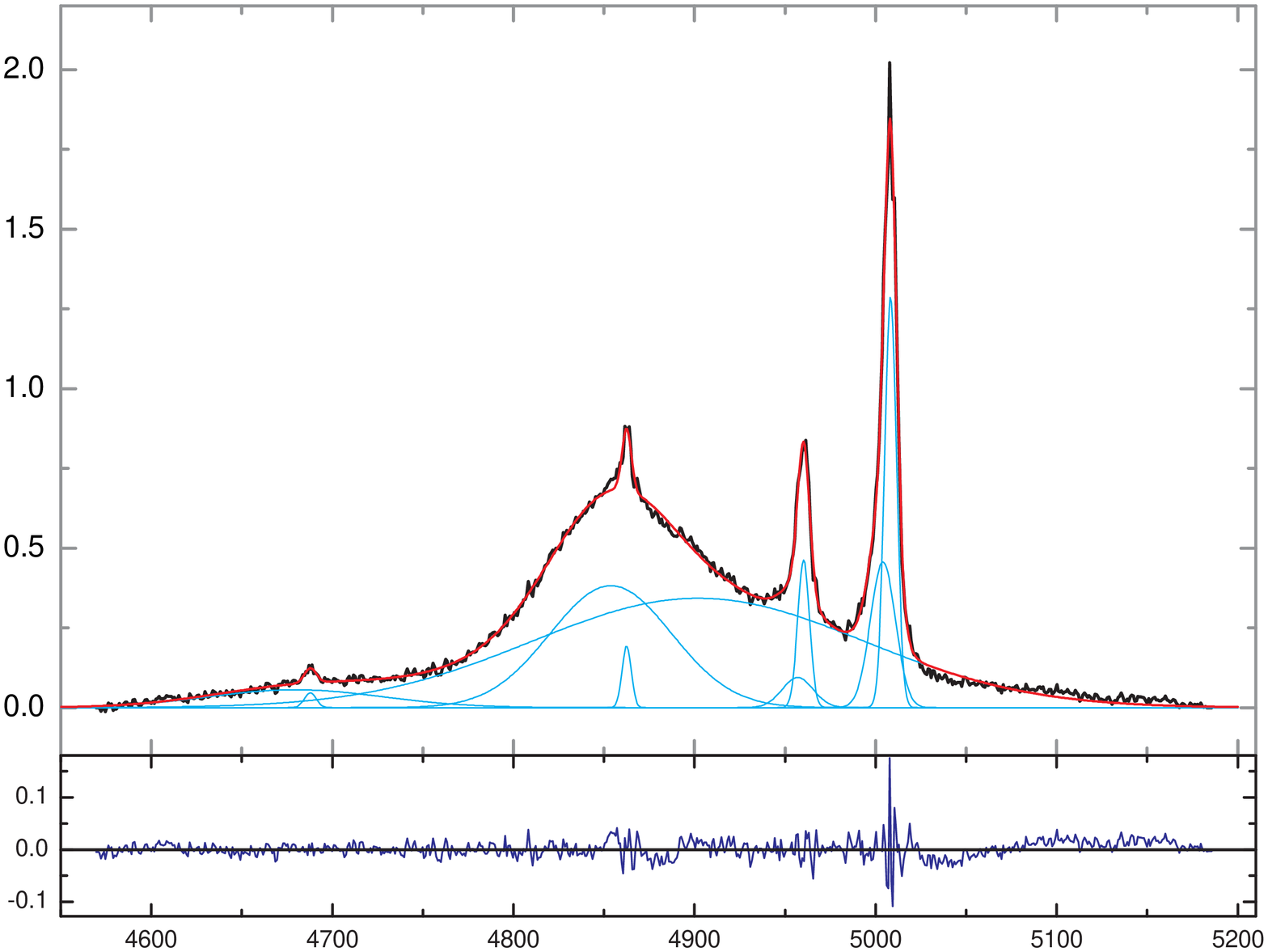}
\hspace{0.2cm}
\includegraphics[width=0.9\columnwidth,clip=true]{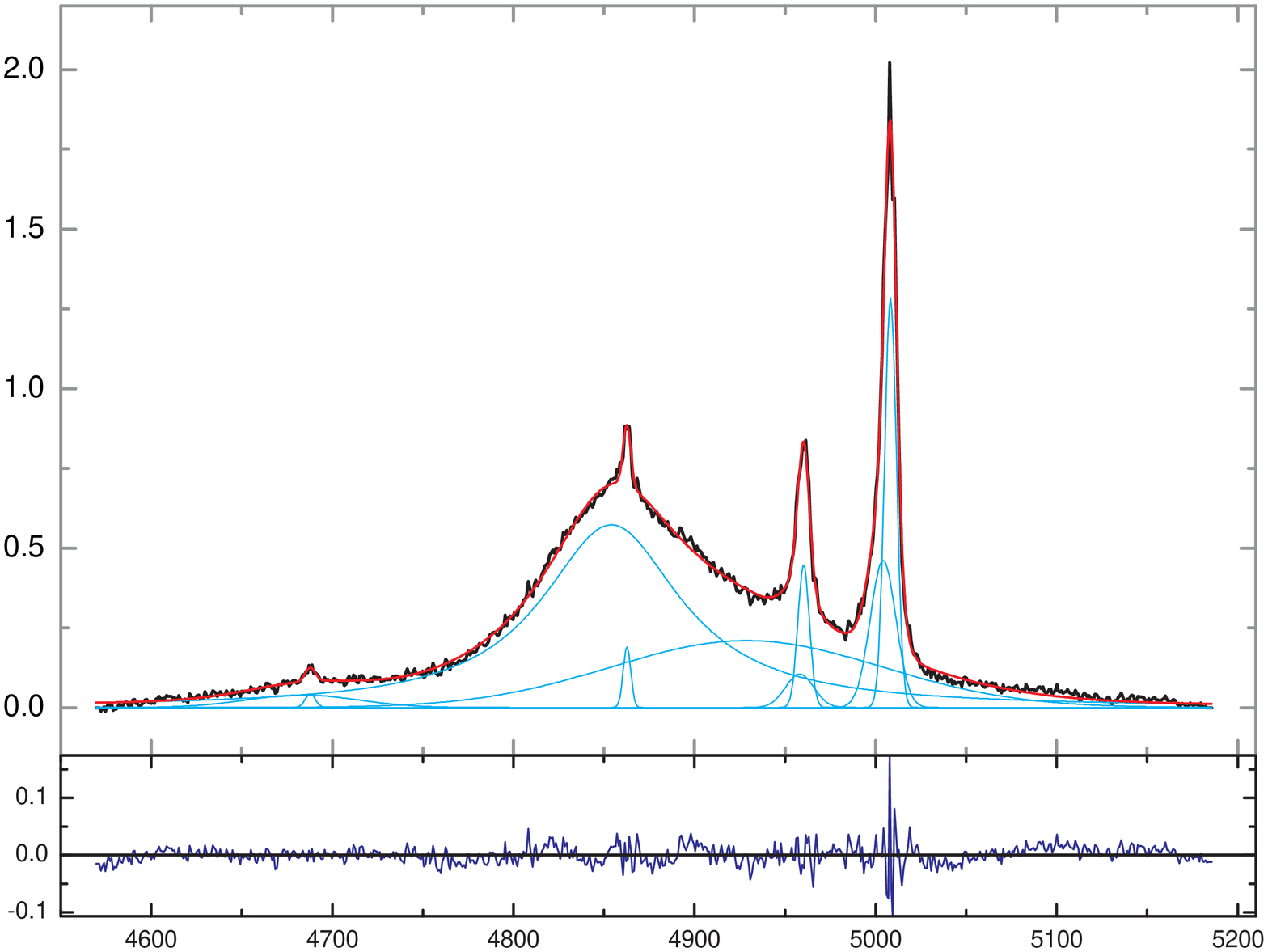}
\caption{Upper-panel: Composite median spectrum for N=31 sources
with logL$_{bol}$[erg s$^{-1}$] $>$ 45.5 and c(1/4)/FW0.25M $>$
0.07. The best model of FeII is also shown along with the underlying
continuum. Lower-left and lower-right panels show the best fitting
solution for the emission lines in the interval 4600-5150 \AA. The
left panel shows a Gauss-Gauss combination for the H$\beta$ profile,
the right panel shows a Lorentz-Gauss model of H$\beta$. All other
lines/components are modeled with Gaussian functions. In both panels
we also show the residuals.}
\end{figure*}
\clearpage

\begin{figure*}
\centering
\includegraphics[width=0.9\columnwidth,clip=true]{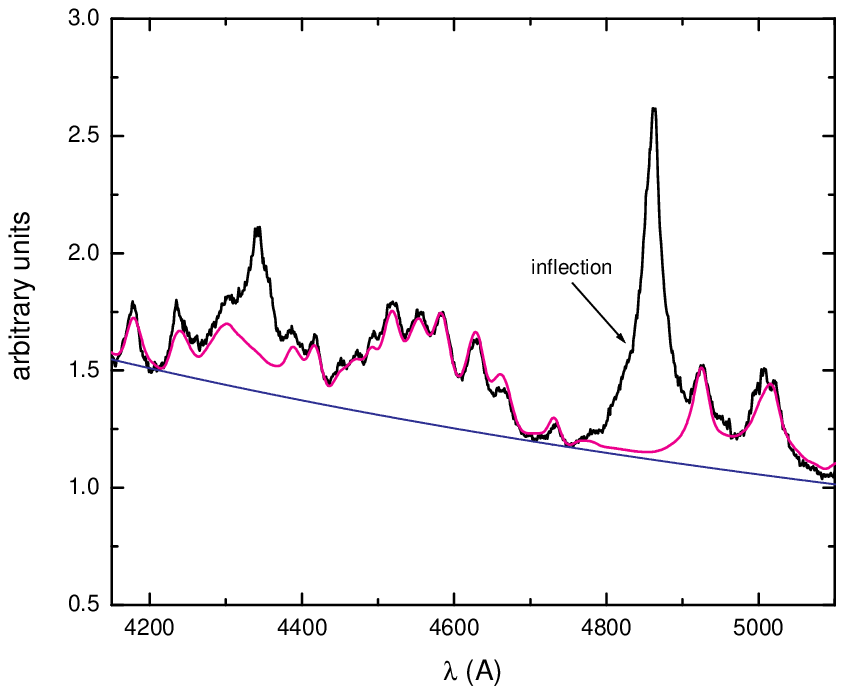}\\
\includegraphics[width=0.9\columnwidth,clip=true]{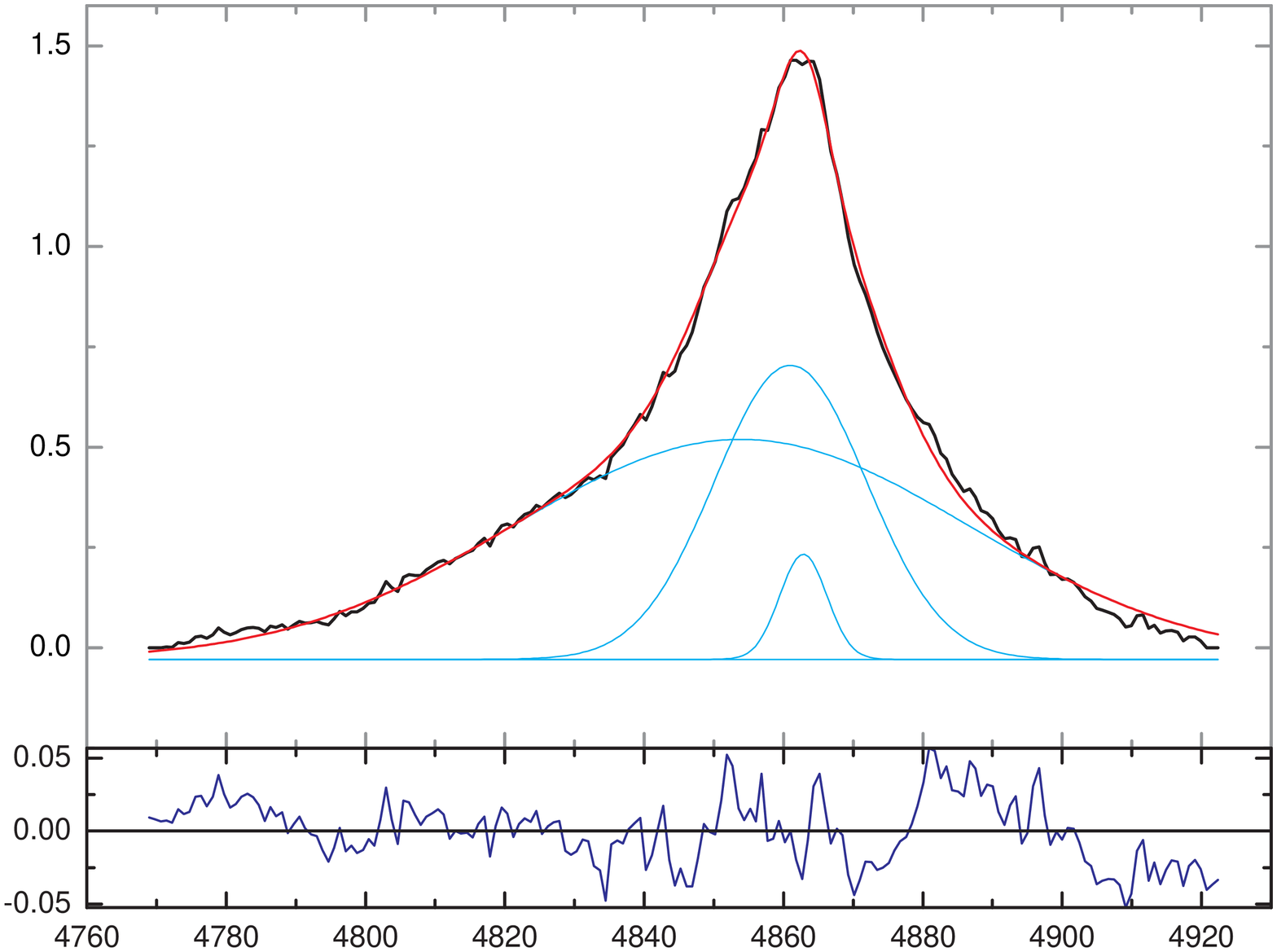}
\hspace{0.2cm}
\includegraphics[width=0.9\columnwidth,clip=true]{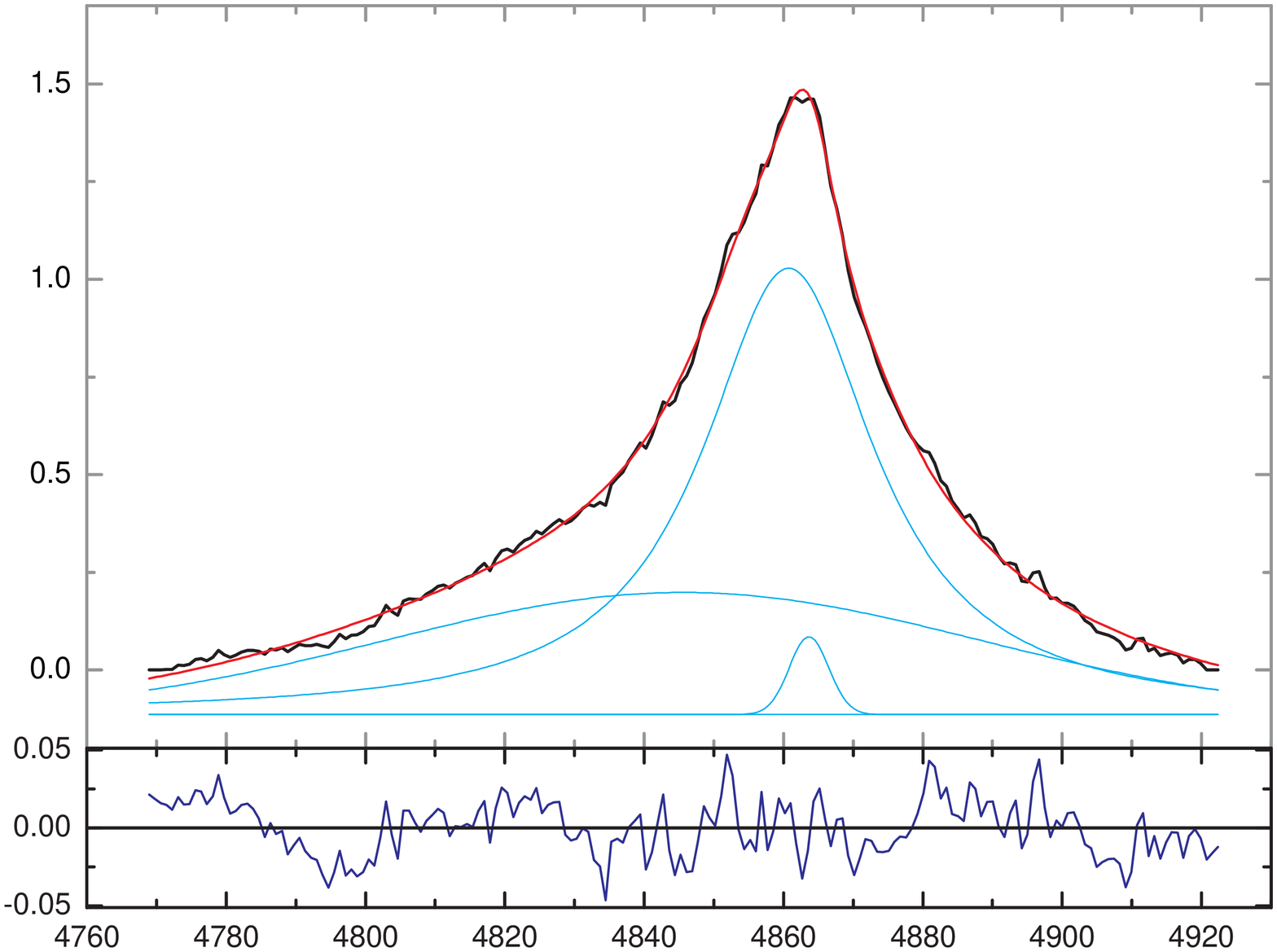}
\caption{Upper-panel: Composite median spectrum for N=26 sources
with logL$_{bol}$[erg s$^{-1}$] $>$ 45.5 and c(1/4)/FW0.25M $<$ -
0.07. The best model of FeII is also shown along with the underlying
continuum. Lower-left and lower-right panels show the best fitting
solution for the emission lines in the interval 4600-5150 \AA. The
left panel shows a Gauss-Gauss combination for the H$\beta$ profile,
the right panel shows a Lorentz-Gauss model of H$\beta$. In both
panels we also show the residuals.}
\end{figure*}
\clearpage

\begin{figure*}
\centering
\includegraphics[width=0.9\columnwidth,clip=true]{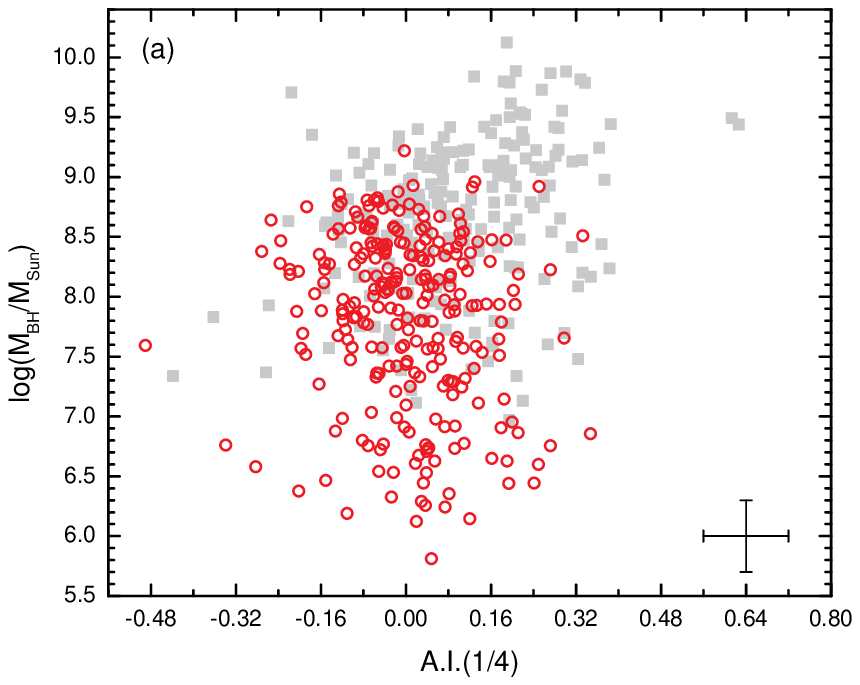}
\includegraphics[width=0.9\columnwidth,clip=true]{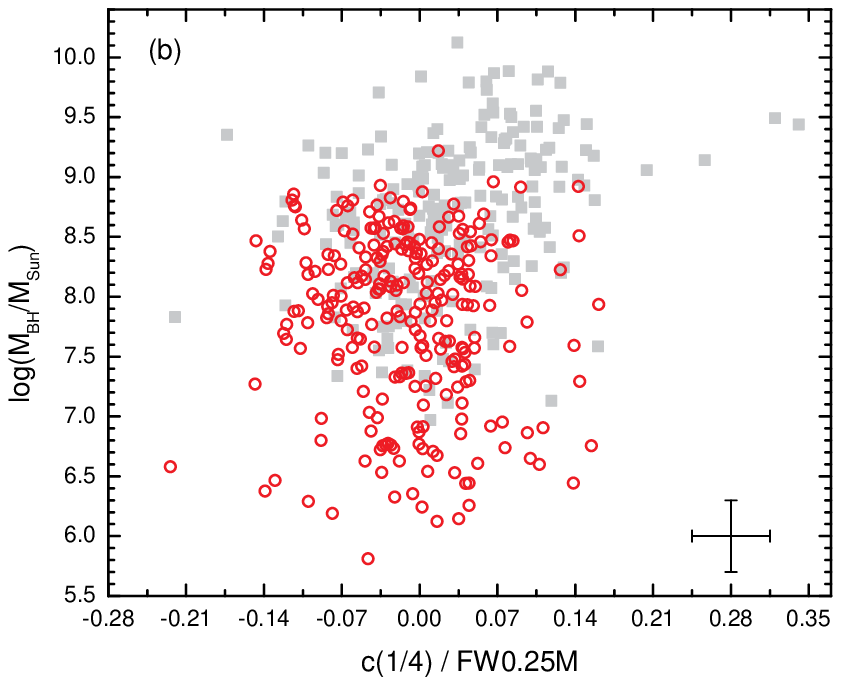}
\caption{Population A and B sources (open circles and solid squares,
respectively) are shown in plots of BH mass versus asymmetry index
A.I.(1/4) (panel a) and normalized centroid shift c(1/4)/FW0.25M
(panel b). In each panel we also indicate typical 2$\sigma$ error
bars.}
\end{figure*}
\clearpage

\begin{figure*}
\centering
\includegraphics[width=1.8\columnwidth,clip=true]{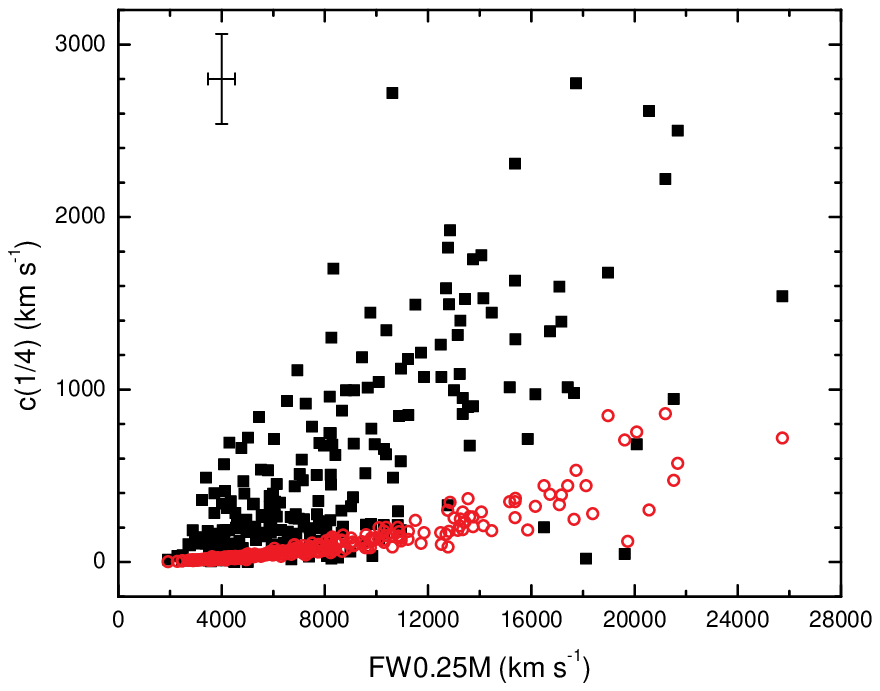}
\caption{For N=249 sources that show a positive centroid shift
c(1/4) we plot that measured shift (solid squares) and the predicted
gravitational redshift (open circles) versus the FW0.25M. We also
indicate the typical 2$\sigma$ error bars for the measured values of
c(1/4).}
\end{figure*}
\clearpage

\begin{figure*}
\centering
\includegraphics[width=0.9\columnwidth,clip=true]{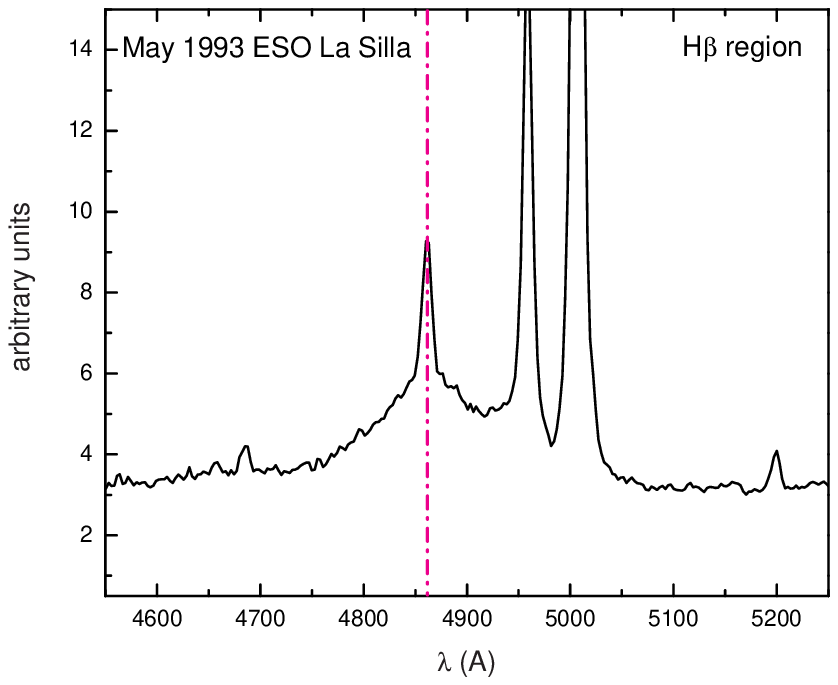}\\
\includegraphics[width=0.9\columnwidth,clip=true]{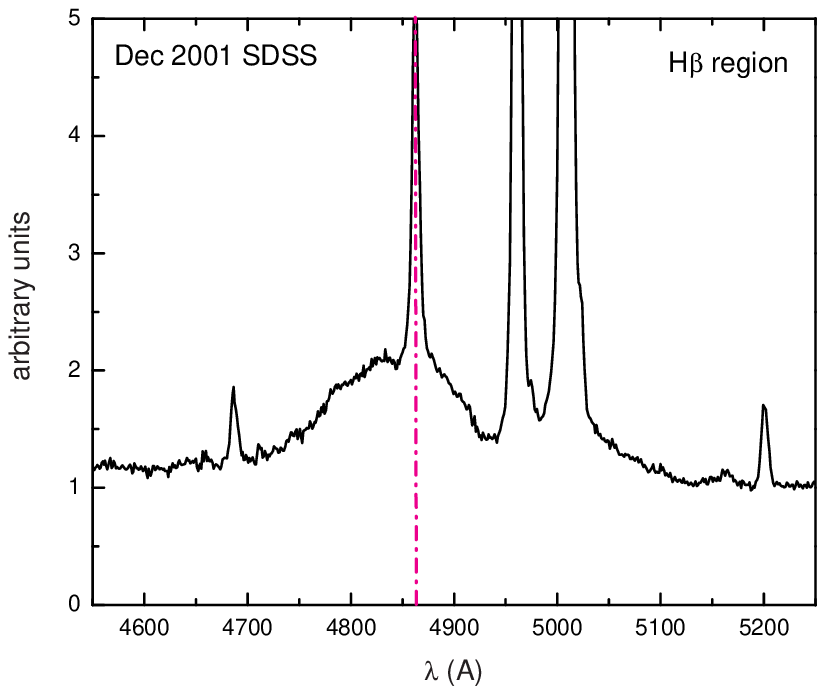}
\hspace{0.17cm}
\includegraphics[width=0.94\columnwidth,clip=true]{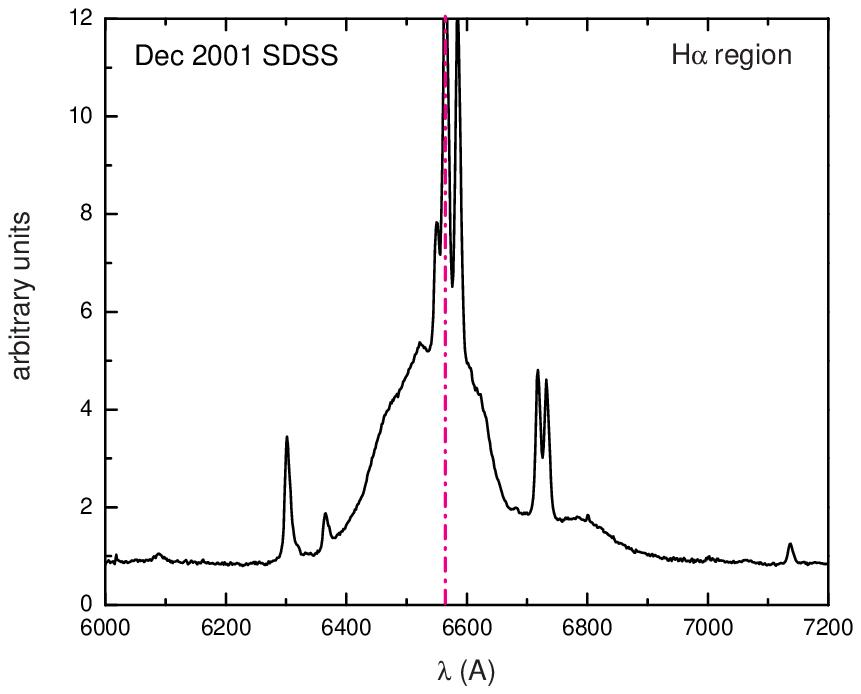}
\caption{SDSS J230443.47-084108.6 observed at two different epochs.
We note the change in the functional shape of the Balmer H$\beta$
line wings. For the more recent records (lower panel) we show both
the H$\beta$ and the H$\alpha$ regions. The vertical lines indicate
the rest-frame Balmer wavelength positions. The ``redshelf'' is
clear in the later observation, especially in the H$\alpha$ region.
An inflection along the red wing of the Balmer lines may be a
transient feature due to the variability of the emitting source.}
\end{figure*}
\clearpage

\begin{figure*}
\centering
\includegraphics[width=1.8\columnwidth,clip=true]{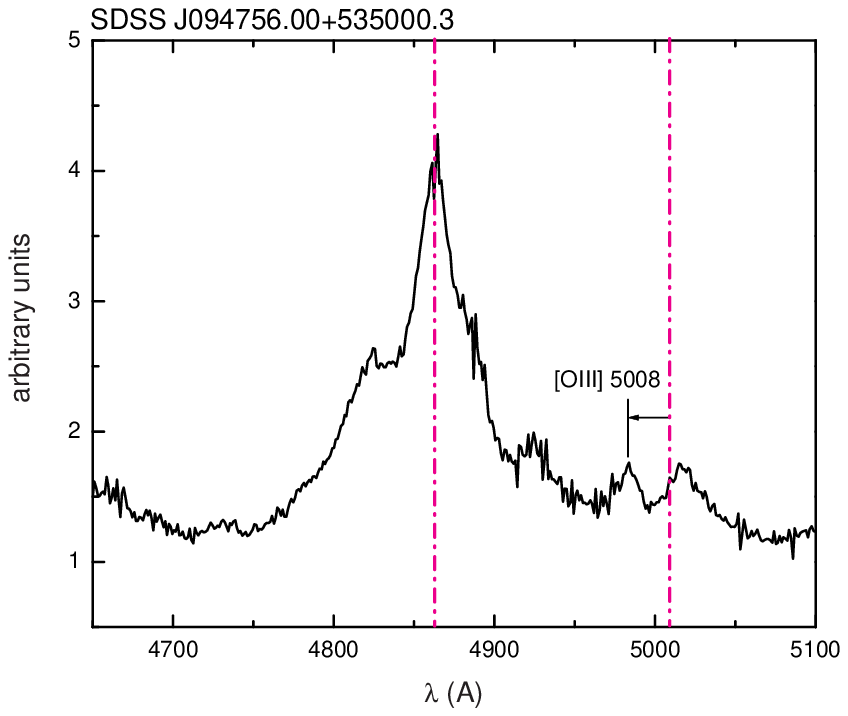}
\caption{An extreme case that shows a clear inflection on the blue
wing of H$\beta$ accompanied by a -1500 km s$^{-1}$ blueshift of
[OIII] $\lambda$5008\AA. The vertical lines indicate the restframe
positions for H$\beta$ 4862.7\AA\ and [OIII] 5008.2\AA.}
\end{figure*}
\clearpage

\begin{figure*}
\centering
\includegraphics[width=1.8\columnwidth,clip=true]{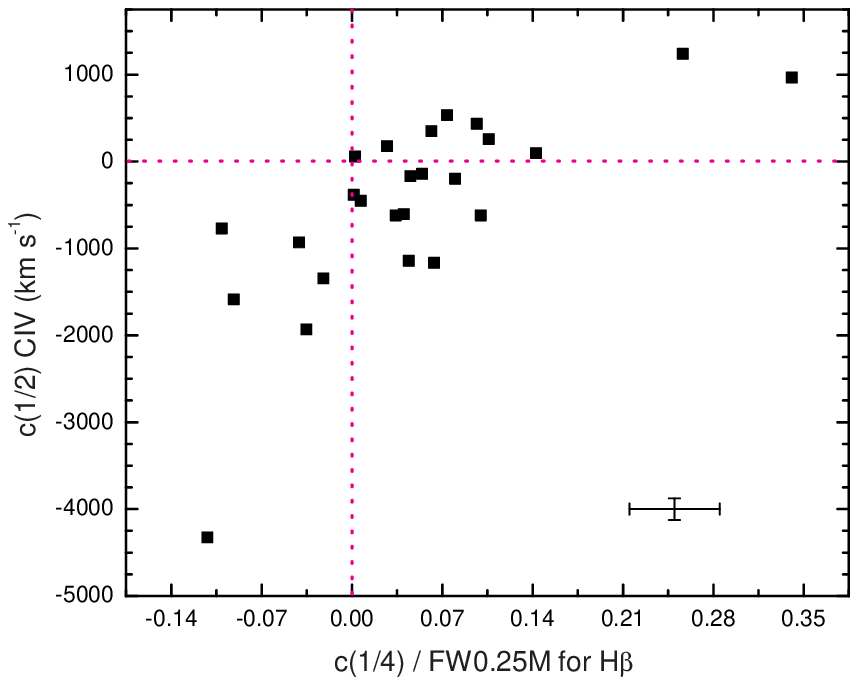}
\caption{A comparison between the CIV $\lambda$1549\AA\ and Balmer
H$\beta$ shift. Data shown represents the overlap between the
current sample and the sample of quasars examined in
\citet{Sulentic07}. The dotted lines indicate the zero shift for CIV
(vertical axis) and H$\beta$ (horizontal axis). Typical 2$\sigma$
uncertainties are indicated.}
\end{figure*}
\clearpage

\begin{figure*}
\centering
\includegraphics[width=1.8\columnwidth,clip=true]{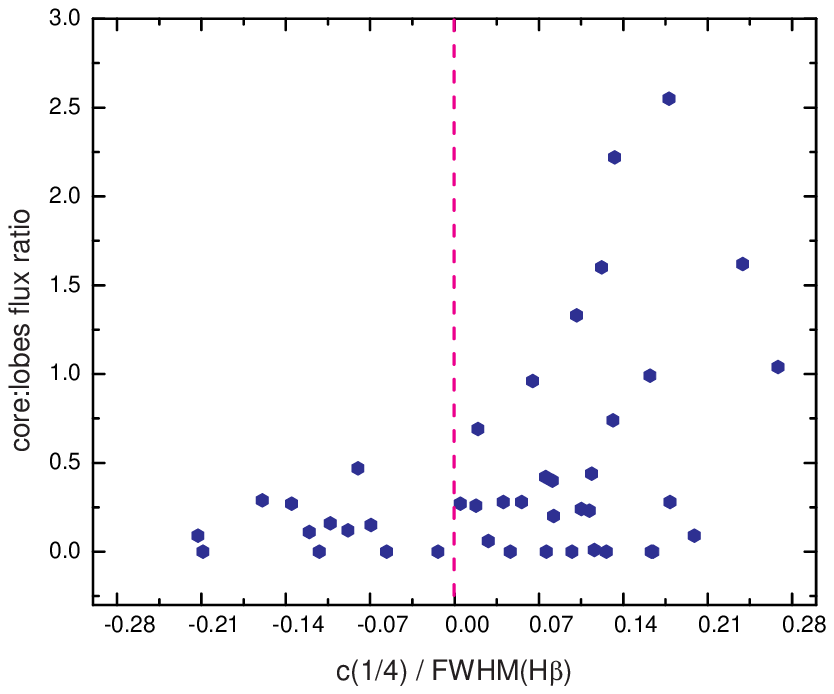}
\caption{The FRII sample is shown here in terms of core:lobes flux
density ratio at 1.4GHz, based on FIRST and the normalized H$\beta$
emission line base shift. The vertical line indicates an unshifted
measure.}
\end{figure*}

\end{document}